\documentclass[12pt]{cmuthesis}

\usepackage{xcolor}
\usepackage{multicol}
\usepackage{dblfloatfix}
\usepackage{soul}
\usepackage{color, colortbl}
\definecolor{Gray}{gray}{0.95}
\usepackage{times}
\usepackage{float}
\usepackage{fullpage}
\usepackage{graphicx}

\usepackage{booktabs, makecell, tabularx}

\definecolor{brown}{rgb}{0.59, 0.29, 0.0}
\definecolor{darkgray}{rgb}{0.59, 0.59, 0.59}
\definecolor{tablegray}{gray}{.9}

\newcommand\revision[1]{\textcolor{blue}}

\usepackage[utf8]{inputenc}
\usepackage{diagbox}
\usepackage{colortbl}

\usepackage{tabularx}

\usepackage{nameref}

\usepackage{xspace}
\usepackage{enumitem}
\usepackage{mathtools}
\usepackage{commath}

\usepackage{amssymb}
\usepackage{pifont}
\newcommand{\cmark}{\ding{51}}%
\newcommand{\xmark}{\ding{53}}%

\newcommand{\customtilde}{{\raise.17ex\hbox{$\scriptstyle\sim$}}}

\newcommand{\etal}{et~al.\xspace}
\newcommand{\eg}{e.g.,\xspace}

\usepackage{amsmath}
\usepackage{gensymb}
\usepackage{bbding}
\usepackage{multirow}
\usepackage[numbers,sort]{natbib}
\usepackage[backref,pageanchor=true,plainpages=false, pdfpagelabels, bookmarks,bookmarksnumbered,
]{hyperref}
\usepackage{subfigure}

\usepackage{url}

\usepackage[letterpaper,twoside,vscale=.8,hscale=.75,nomarginpar]{geometry}

\begin {document} 
\frontmatter

\pagestyle{empty}

\title{ {\bf Practical and Rich User Digitization}}
\author{Karan Ahuja}
\date{August 2023}
\Year{2023}
\trnumber{CMU-HCII-23-103}

\committee{
Dr. Chris Harrison, Co-Chair, CMU\\
Dr. Mayank Goel, Co-Chair, CMU\\
Dr. Nikolas Martelaro, CMU\\
Dr. Andrew D. Wilson, Microsoft Research
}

\support{}
\disclaimer{}



\maketitle


\pagestyle{plain} 

\begin{abstract}

A long-standing vision in computer science has been to evolve computing devices into proactive assistants that enhance our productivity, health and wellness, and many other facets of our lives. User digitization is crucial in achieving this vision as it allows computers to intimately understand their users, capturing activity, pose, routine, and behavior. Today’s consumer devices – like smartphones and smartwatches – provide a glimpse of this potential, offering coarse digital representations of users with metrics such as step count, heart rate, and a handful of human activities like running and biking. Even these very low-dimensional representations are already bringing value to millions of people’s lives, but there is significant potential for improvement. On the other end, professional, high-fidelity comprehensive user digitization systems exist. For example, motion capture suits and multi-camera rigs that digitize our full body and appearance, and scanning machines such as MRI capture our detailed anatomy. However, these carry significant user practicality burdens, such as financial, privacy, ergonomic, aesthetic, and instrumentation considerations, that preclude consumer use. In general, the higher the fidelity of capture, the lower the user's practicality. Most conventional approaches strike a balance between user practicality and digitization fidelity. 

My research aims to break this trend, developing sensing systems that increase user digitization fidelity to create new and powerful computing experiences while retaining or even improving user practicality and accessibility, allowing such technologies to have a societal impact. Armed with such knowledge, our future devices could offer longitudinal health tracking, more productive work environments, full-body avatars in extended reality, and embodied telepresence experiences, to name just a few domains. 

\end{abstract}

\begin{acknowledgments}

At the culmination of this challenging and rewarding journey, I am profoundly grateful to the individuals who have been my pillars of strength. To my parents, Reshma and Deepak Ahuja, your unwavering support fueled my determination. My achievements are a reflection of the values you instilled in me. To my grandparents, your wisdom and stories have been a guiding light, shaping my path with resilience and curiosity. 

My heartfelt appreciation extends to my friends and collaborators, your collective enthusiasm has enriched my academic journey. To my advisors and thesis committee members, thanks for your constant support, encouragement, mentorship and insights that have elevated my work.

\end{acknowledgments}

\tableofcontents
\listoffigures
\listoftables

\mainmatter

\normalsize

\part{Introduction}

\chapter{User Digitization}
\label{user_digitization}

\section{Defining User Digitization}

The desire to digitize the human body to glean insights about the various aspects of the user dates back to 1780 when Abraham-Louis Perrelet created the first mechanical pedometer to measure the steps and distance while walking. Since then, digitization technologies have matured to capture a user's every motion and facial expression to create their digital replicas for movies, drive avatars in games or create doppelgangers for holograms. In the modern era of computing, with the growing proliferation of smart devices, each and everyone of us is surrounded by technologies that digitize our bodies. 

``User digitization" refers to the process of creating a digital representation of an individual. It can encompass various aspects of their identity, attributes, behaviors, and interactions. This can include personal information, preferences, activities, and even physiological or biometric data. User digitization often involves the sensing, collection, analysis, and storage of data to create a comprehensive digital model of a person, which can be used for various purposes such as personalization, health monitoring, improved productivity, virtual avatars, and embodied telepresence experiences, to name just a few domains.

\section{Spectrum of User Digitization}

User Digitization technologies can span across a wide spectrum of granularity. It can be something as simple as a touchscreen that can capture the position of our fingertips. Or a wearable like a smartwatch that can estimate metrics such as heart rate, step count, and a handful of activities such as running or biking. Even these low-dimensional representations are already immensely popular. Millions of people around the world use such devices for tracking their productivity, health, and fitness. On the other end of the spectrum, we have specialized Motion Capture systems that capture a high-fidelity digital representation of the user comprising of their full-body pose and mesh. These have made them very popular to drive life-like avatars in movies and games, and many human-computer interaction tasks. Further, motion capture technology is at the cutting-edge of clinical movement research, from helping stroke and cerebral palsy patients with rehabilitation, to understanding how to better treat those with Parkinson’s and Arthritis. Looking at these technologies, it is evident that user digitization is a continuum. Let’s look at this schematically by focusing on two very important dimensions of user digitization.

\begin{figure}
\centering
  \includegraphics[width=\columnwidth]{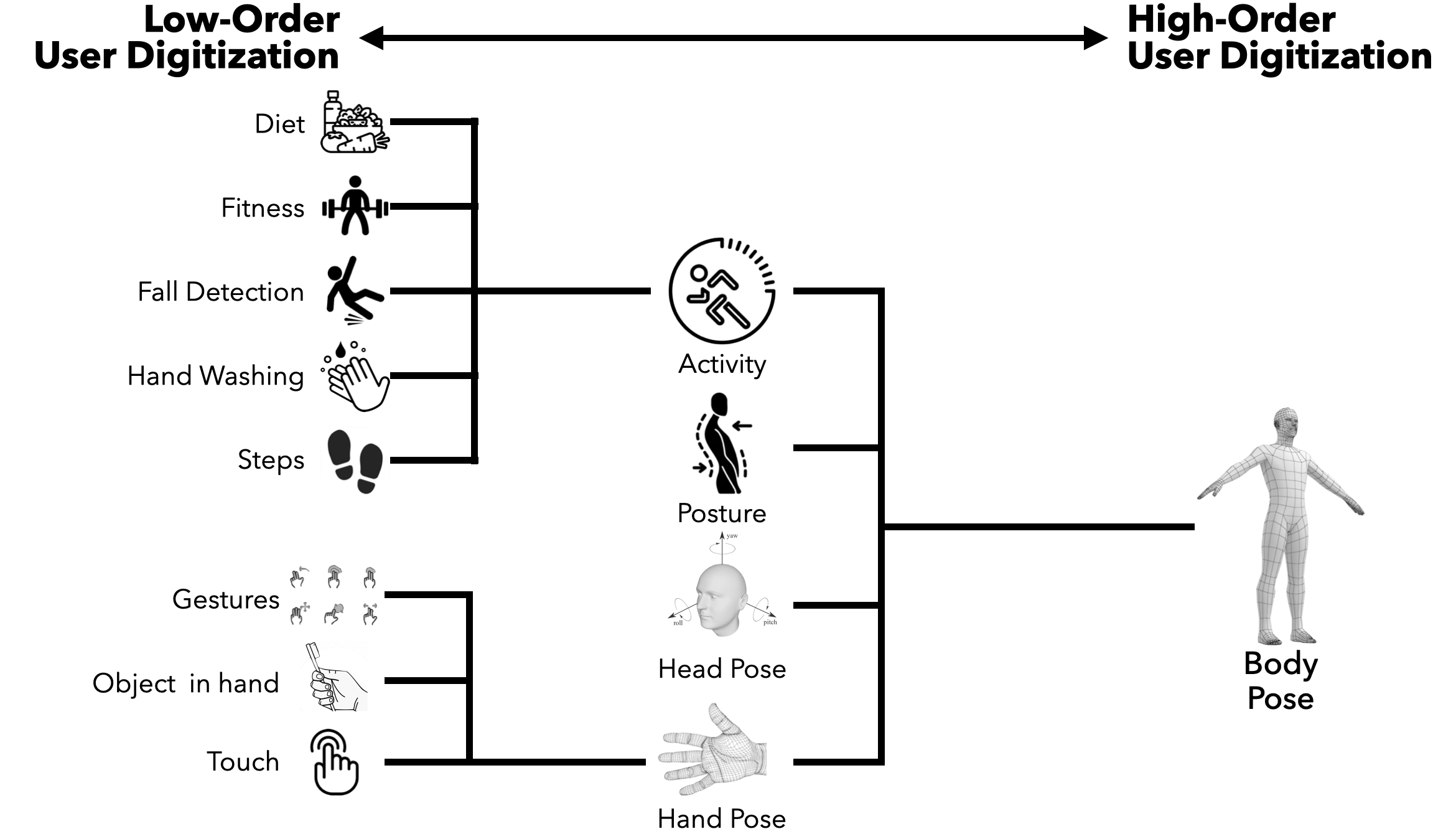}
  \caption[Digitization richness]{Spectrum of user digitization richness going from lower-order user digitization (on the left) to higher order digitization (on the right).}
  \label{fig:richness_spectrum}
\end{figure}

\subsection{Digitization Richness}

The first dimension is Digitization Richness or Fidelity, which encapsulates the degree of capturing an exact and faithful representation of our physical selves. Figure~\ref{fig:richness_spectrum} conveys the richness of different technologies on the user digitization spectrum. We have low-order digitization such as step counts on one end to higher-order digitization such as full-body pose and mesh on the other end. 

\subsection{User Practicality}

The second dimension is User Practicality. Importantly, practicality is not one thing, it is multifaceted. A practical system would have the following characteristics:

\begin{itemize}
    \item \textbf{Low invasiveness}: A minimally invasive system would require practically zero bootstrapping from the user. It would have no calibration or initialization steps, and would have a form factor that is not discomforting to the user.  
    \item \textbf{Privacy-aware}: The digitization paradigm should be designed with privacy as one of its core tenets. An ideal system would be one that captures no user privacy compromising data, processes data on edge devices and denatures any collected features.  
    \item \textbf{Minimal instrumentation}: We want systems to have as few points of instrumentation as possible (ideally no instrumentation at all).
    \item \textbf{Works on-the-go}: Typically, a lot of the sensing systems work in highly constrained, lab conditions. Ideally, we want these to work when people are mobile and outside of controlled settings.
    \item \textbf{Low hardware cost}: We ideally want approaches that have a low hardware cost - something that can be a consumer product in the near future.
\end{itemize}

\subsection{Relationship between Digitization Richness and User Practicality}

\begin{figure}
\centering
  \includegraphics[width=0.5\columnwidth]{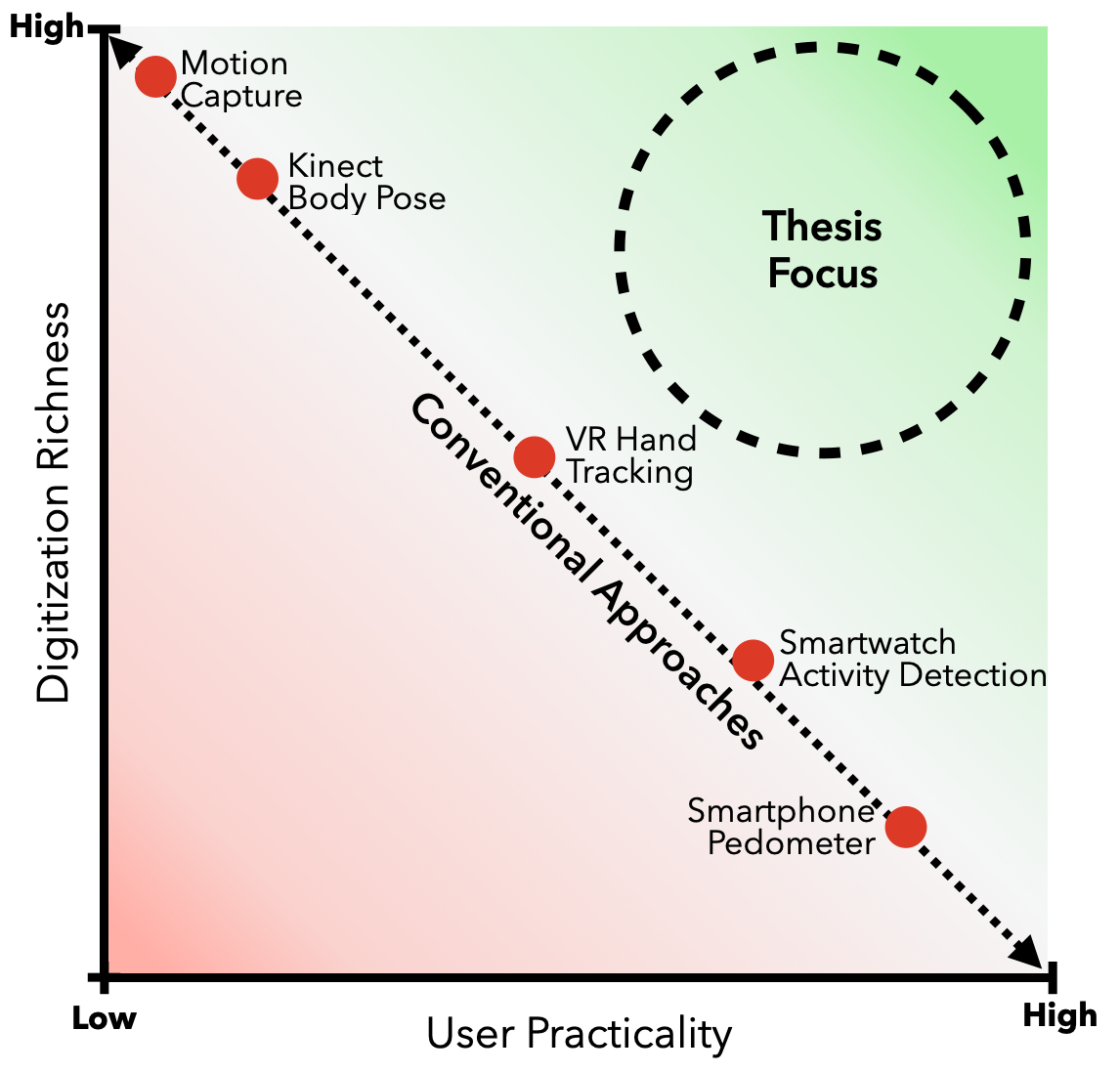}
  \caption[Digitization richness vs. user practicality]{A design space plotting digitization richness vs. user practicality. Most approaches lie along the diagonal, where practicality decreases as richness increases. Ideally, we want systems that offer higher richness without commensurate practicality trade-offs.}
  \label{fig:intro_fig}
\end{figure}

Both  richness and practicality are key dimensions of user digitization. Usually, as the richness of the sensing system increases, so does the number of dimensions it can capture and the more aspects of our life it can support. Ideally, we want these rich user digitization systems without commensurate practicality trade-offs. Unfortunately, in general, human sensing practicality decreases in lockstep (Figure~\ref{fig:intro_fig}, diagonal line), to the point where they have limited consumer adoption. For example, the gold standard of full-body user digitization such as Vicon and OptiTrack are generally associated with blockbuster films and games, using expensive multi-camera rigs and special suits with markers. On the other end of the spectrum, we have technologies such as the pedometer which can only track steps but are non-invasive and can be easily implemented on ubiquitous devices such as smartphones and smartwatches. 

We can plot these two dimensions – digitization richness vs. user practicality – as a high-level design space (Figure~\ref{fig:intro_fig}). At a high level, we see an inversely proportional trend that the higher you go in richness, the less practical it is for users. The dashed line denotes this conventional axis occupied by conventional approaches (Figure \ref{fig:intro_fig} red dots), balancing user practicality and digitization richness. But what we ideally want are systems that are practical and rich. Such techniques would lie above the conventional approaches diagonal line and occupy the upper left quadrant of our design space (Figure 1, green region). In this thesis, I will describe my efforts of such sensing systems that start to pull away from the line and towards this idealized domain. 

This would enable consumer devices to calculate a continuous comprehensive representation of the user. Such a digital representation can go beyond a simple step count and capture a user’s gait cycle, dexterity, balance, posture, and active range of motions of joints, providing contextualized behavior and pose insights in every day practical scenarios that even a once a day visit to a Motion Capture studio would fail to provide. Using this holistic continuous information, a user can get fine-grained wellness data and even enable doctors to remotely monitor their patients, for example, tracking a user’s recovery after surgery, and even chances of injury and relapse. It will also be able to inform the design of future consumer devices that enable full-body avatars for telepresence, interactions in extended reality and to create more productive and collaborative work environments.

\section{Glossary}

In order to enhance the clarity and comprehension of this document, a glossary of key terms and their definitions has been included. This glossary serves as a reference guide to assist readers in understanding the terminology and concepts discussed within the text.

\begin{itemize}
    \item \textbf{User Digitization:} The process of creating a digital representation of an individual, encompassing various aspects of activity, intent, behavior, and interactions.
    \item \textbf{Motion Capture:} Technology to record and translate human movements into digital data for applications like films, games, and simulations.
    \item \textbf{Virtual Reality (VR):} Immersive technology creating computer-generated virtual environments for users to interact with.
    \item \textbf{Augmented Reality (AR):} Technology overlaying digital information onto the real world data modalities.
    \item \textbf{Extended Reality (XR):} Encompasses VR and AR, blending physical and virtual environments.
    \item \textbf{Inertial Measurement Unit (IMU):} A device that combines multiple sensors to measure orientation, angular velocity, and acceleration.
    \item \textbf{Avatar:} Digital representation of a user or character in virtual or digital environments.
    \item \textbf{Human Digital Twin:} A virtual replica of a person, capturing physiological and behavioral aspects for analysis.
    \item \textbf{Activity Recognition:} Identifying and categorizing activities performed by users. 
    \item \textbf{Pose:} Position and orientation of a body or body parts.
    \item \textbf{Design Space:} A conceptual framework that visualizes relationships and trade-offs between dimensions, such as digitization richness and user practicality, to guide system design and innovation.
    \item \textbf{Telepresence:} The technology enabling a user to feel present or interact in a remote location as if they were physically there, often involving avatars or digital representations.
    \item \textbf{Digitization Richness:} A dimension that measures the degree of capturing a faithful representation of physical attributes and behaviors in user digitization technologies.
    \item \textbf{User Practicality:} The multidimensional characteristic of a user digitization system, encompassing but not limited to factors such as non-invasiveness, privacy awareness, minimal instrumentation, mobility, and low hardware cost.
\end{itemize}

\section{Organization of Thesis}

In this thesis, I will focus on my research towards making rich and practical user digitization solutions. Chapter~\ref{chap:rel_work} provides an overview of the literature and prior work domains that intersect with my thesis work. 

In Part \ref{part:improving_richness}, I describe my works that improve the digitization richness of user digitization systems while maintaining user practicality. I start by exploring techniques that advance activity recognition to create a vocabulary to digitize and express a users behavior routine. I then delve into pose sensing, focusing on head pose, eye-gaze, hands and finally full-body pose capture systems that are mobile and provide a holistic representation of the user on-the-go. A common thread across all these works is that they make use of existing ubiquitous devices (such as smartphones) and require no additional hardware while enriching user digitization capabilities.

Part \ref{part:improving_practicality} even further advances the field, with the focus on user digitization systems that further improve upon the user practicality while retaining or even advancing digitization richness, for instance by making them less invasive or more privacy aware. Finally, I summarize my key insights, takeaways and contributions.

\chapter{Background}
\label{chap:rel_work}

My  work on user digitization intersects with many diverse areas of human-computer interaction research, including activity detection, touch sensing, context-aware computing, pose estimation, computer vision, wearables and ubiquitous computing. Here I focus on the most highly related work that was influential.

\section{Human Activity Recognition}

There has been extensive research done in detecting human activities from sensors such as microphones \cite{ubicoustics, privacyMic, ericCoughSensing}, IMUs \cite{viband, laputHandActivities, BaoARAccel, accelHAR, KwapiszARcellPhone}, cameras \cite{cameraHAR}, powerline sensors \cite{electriSense}, pressure-sensors \cite{waterPressureSensors} and other multimodal sensing approaches \cite{syntheticSensors, MMDL4HAR}. Refer to \cite{dl4HARsurvey, subetha2016survey, GuptaSurveyHAR} for a detailed survey. In this section, I focus on video, audio, motion and multimodal methods, that most closely resemble my efforts.

\subsection{Camera-Based Human Activity Recognition}

Computing devices equipped with cameras are becoming increasingly prevalent in our daily lives. The high fidelity afforded by them make them a powerful sensor for activity recognition.Researchers have explored different motion and temporal feature representations of videos learned by 3D Convolutional Neural Networks (CNNs)~\cite{carreira2017quo, hussein2019timeception, wang2018videos}. In addition to CNNs, long short-term memory (LSTM) models are popular, taking advantage of dependencies across video frames \cite{donahue2015long, li2018videolstm, yue2015beyond, sharma2015action}. 

Instead of using raw visual information, researchers have also explored the idea of relying on high-level semantic representations for human activity. For example, using body pose~\cite{jhuang2013towards, xiaohan2015joint, wang2013approach, baradel2018human, du2017rpan, zolfaghari2017chained}, motion of semantic keypoints~\cite{choutas2018potion}, and joint representations~\cite{du2015hierarchical, hou2016skeleton, liu2016spatio, shahroudy2016ntu, song2016end}.

\subsection{Audio-Based Human Activity Recognition}

Researchers have leveraged the ubiquity of microphones to build numerous audio based human activity recognition systems. In the past, these systems were designed to recognize a limited set of activities.
For instance, in SoundSense \cite{soundSense}, Lu et al. made use of a smartphone microphone and a scalable classification architecture to classify between speech, music and ambient sound. Thomaz et al. \cite{thomazEating} showed that eating activities could be effectively detected using wrist worn acoustic sensing. Along similar lines, BodyScope \cite{BodyScope} trained a support vector machine with data collected from a wearable acoustic sensor which recorded throat sounds, in order to distinguish between user activities such as eating, drinking, speaking, laughing and coughing.

Recently, with the advent of more advanced machine learning techniques, there has been widespread success in building more general purpose human activity recognition systems. Stork et al. \cite{StorkNonMarkovian} used non-markovian ensemble voting on Mel-frequency cepstral coefficients to classify activities performed in the bathroom and kitchen contexts, such as using a blender, microwave, hairdryer or a toothbrush. DeepEar \cite{lane2015deepear} proposed a deep learning based approach to acoustic sensing, supporting tasks such as ambient scene analysis, stress detection, emotion recognition and speaker identification. 
Laput et al. \cite{ubicoustics} further built a plug and play acoustic activity recognition system to detect thirty different daily activities, employing a convolutional neural network trained with augmented data from sound effect libraries. Liang et al. \cite{liangADL} and Wu et al. \cite{wu2020automated} further build upon these techniques to improve robustness. 

\subsection{IMU-Based Human Activity Recognition}

A large number of activity recognition systems rely on inertial sensors such as accelerometers, gyroscopes and magnetometers present in smartphones, smartwatches and other wearables to detect human activity. Similar to acoustic activity recognition, in the past, motion based activity recognition systems were generally constrained to detect coarse human activities, such as walking, running and biking \cite{accelHAR, KwapiszARcellPhone}. Refer to \cite{GuptaSurveyHAR, sousa2019human} for a detailed survey. With the recent widespread adoption of wearables, especially wrist-worn smartwatches, researchers have developed more general purpose activity \cite{kwon2020imutube, mollyn2022samosa} and gesture recognition \cite{viband} systems. 

These include multi-device systems such as that of Shoaib et al. \cite{shoaibBadHabits} that use the accelerometer from a smartphone in conjunction with one from a smartwatch to detect 13 activities such as smoking and drinking coffee. Single wearable approaches have also seen renewed interest with the advent of deep learning. For instance, by using a wrist-worn smartwatch accelerometer sampled at 4~kHz, Laput et al. \cite{laputHandActivities} detected fine grained hand activities. 

\subsection{Multimodal Human Activity Recognition}

Multimodal deep learning systems have seen widespread adoption in other communities for tasks such as image captioning \cite{karpathy2015deep, antol2015vqa, mao2014explain, donahue2015long}, pose estimation \cite{ahuja2021pose,rogez20143d} and autonomous driving \cite{bojarski2016end, caesar2020nuscenes}. While this field is much more nascent for human activity recognition, a number of systems have adopted similar techniques.

Synthetic Sensors \cite{syntheticSensors} used a sensor board containing 9 different sensors (such as an accelerometer, magnetometer, microphone, etc.) to create sensory abstractions that could be combined together to create context-sensitive applications. In \cite{MMDL4HAR}, Radu et al. presented a case study of different challenges and approaches to multimodal deep learning for human activity recognition. GestEar \cite{beckerGestEar} proposes the use of sound and motion from a smartwatch for gesture classification (e.g. snapping, knocking, clapping). 

\section{Body Pose Capture}
\label{sec_pose_rel_work}

Full-body motion capture has a long history, dating back to at least 1878 with Muybridge’s “The Horse in Motion” \cite{eadweard1878horse}. This film is widely regarded as the start of chronophotography, a photographic technique for the scientific study of movement and especially locomotion. Later pioneers, such as Max Fleischer, used rotoscoping as a way to capture and then transform complex movements, such as locomotion, for use in early 20th century animated films. Today, digital technologies have enabled a wide variety of highly-automated and precise techniques for human motion capture. I now review this literature, paying particular attention to systems that capture continuous full-body pose, as opposed to techniques for tracking just fingers or hands in free space.

\subsection{External Body Pose Capture}
There have been significant strides in capturing the user’s pose with external sensors. Commercial systems such as Vicon \cite{Vicon} and OptiTrack \cite{OptiTrack} make use of retroreflective markers tracked by high-speed external infrared cameras. These systems are considered the golden standard and have been extensively used for character animation, games and movies as they provide low-latency and accurate motion capture (MoCap). 

In recent years, with the advent of deep learning, computer vision based pose estimation systems have become increasingly popular. These include systems that make use of RGB cameras \cite{Cao:2017,alp2018densepose, Papandreou:2018}, depth sensors such as Kinect, OpenNI \cite{OpenNI} and Intel RealSense \cite{Intel_realSense}, and systems that combine the RGB and depth \cite{zimmermann20183d, michel2017markerless}. Researchers have also considered non-optical tracking systems, utilizing modalities such as RF \cite{zhao2018through}, capacitive sensing \cite{zhang2018wall++, shultz2022tribotouch}, magnetic fields \cite{Polhemus, trakSTAR}, mechanical linkages \cite{sutherland1968head} and sound-based sensing approaches \cite{dov} among others. These systems have limited range and precision when compared to their visual modality counterparts. 

In VR settings, the use of outside-in tracking (i.e., an external sensor looking at the user) has been explored in many systems such as the HTC Vive \cite{Vive}, Oculus Rift S \cite{Rift} and PlayStation VR \cite{PlayStationVR}. These systems typically track the headset and two-handheld controllers, and can roughly estimate the remainder of the joints with inverse kinematics \cite{root_motion_solver, parger2018human}. More faithful full-body tracking can be achieved by placing extra sensors on the limbs, such as HTC VIVE trackers, or bolstering capture with external cameras such as the Kinect.

\subsection{Worn Body Pose Capture}

Worn body capture systems generally offer greater ease-of-use and mobility to users, allowing for body digitization on-the-go and with less setup. While there are innumerable specialized systems that focus on digitizing the hands \cite{kim2012digits,zhang2015tomo, glauser2019interactive} and face \cite{iravantchi2019interferi, ahuja2018eyespyvr, thies2018facevr}, more related to my work are worn, self-contained approaches that attempt whole-body pose estimation. A wide variety of sensing techniques have been explored for whole-body, ranging from exoskeletons \cite{METAmotion}, magnetic trackers \cite{chen2016finexus} to ultrasonic beacons \cite{vlasic2007practical}. Distributed sensor based approaches are also prevalent that instrument multiple points of the body with the same sensor. IMU’s \cite{huang2018deep, xsens, ahuja2022activityposer}, cameras \cite{shiratori2011motion} and RFID’s \cite{jin2018rf} have been explored for these purposes. 

Researchers have also explored instrumenting various body parts with cameras, for instance, Back-Hand-Pose \cite{wu2020back} makes use of a wrist-worn fisheye camera for hand pose estimation. \cite{bailly2012shoesense} instruments the shoe with a depth sensor for hand gesture recognition. Similar camera arrangements have also been explored in a pendant \cite{starner2000gesture} and ring \cite{chan2015cyclopsring, nanayakkara2013eyering, yang2012magic} form factor for hand and finger tracking. 

In terms of body tracking, there are handful of self-contained, worn systems able to capture full body pose. The first is EgoCap~\cite{Rhodin:2016}, which used a pair of downward-facing fisheye cameras cantilevered from the head, proving a sufficient view for a computer vision pose model to extract a body skeleton. Mo2Cap2~\cite{xu2019mo} is virtually identical, except it uses just a single fisheye camera. Instead of requiring extra cameras, MeCap~\cite{Ahuja:2019} shows that the rear camera of a smartphone placed into a low-cost VR headset can capture full body pose with a clip-on mirror accessory. Finally, and perhaps most practical, are~\cite{tome2019xr, ahuja2022controllerpose}, which demonstrate that the wide-angle RGB cameras contained in VR devices could be used to capture a wearer’s body skeleton. Equally practical is \cite{chen2014around}, which uses a smartphone's user-facing camera and IMU for 3-DOF localization of the device with respect to the user's body for around body interaction. Later,~\cite{oddeyecam} extended this idea and added a depth camera to improve tracking accuracy and capture the user's shoulders. Finally,~\cite{simo} uses two smartphones in a 3D printed case to provide tracking of a user’s hand holding the device, along with the user's head position,  relative to a distant display using inside-out tracking.

\part{Improving Digitization Richness while maintaining Practicality}
\label{part:improving_richness}

\chapter{Overview}

\section{Utility of Higher-order User Digitization}

The advancement of digital representations that can faithfully mirror the intricate nuances of human behavior, from minute gestures to complex movements, has led to a plethora of innovative technologies. The traditional way to improve digitization richness has been to add more points of instrumentation either on the user or in the environment. Along these lines, the gold standard of user digitization still remains motion capture techniques such as Vicon and OptiTrak that make use of multiple cameras, markers, and specialized suits. While they have been instrumental in capturing high-fidelity data, they are however cumbersome, restrictive, and resource-intensive.

Thus, the pursuit of improving digitization richness has always been met with a critical challenge: the delicate balance between improving fidelity and ensuring user practicality. My research shows that full-body pose is not exclusive to the realm of specialized and external devices, but can now be enabled on mobile consumer devices. This affords my approach with superior practicality and wide applicability to work in unconstrained settings in the wild, over methods that require special hardware or external instrumentation. To achieve this I add layers of intelligence to increase the fidelity of existing practical and ubiquitous sensing platforms. 

I further note that higher-order digitization have a trickle down effect to other lower-order digitization representations. For example, if I know the pose of a person, I can use that information to improve the precision of fall detection systems. Similarly, I can also inform new insights, rather than just predicting whether the user has fallen, I can detect what led to the fall, estimate the severity of it and also track how the user will recover from such an injury. Let's take a look at the following three scenarios wherein modeling a higher-order digitization unlocks new new applications and uplifts other lower-order representations.

\subsection{Objective Measurement of Hyperactivity in Children}
\label{lemurdx}

Globally, Attention-Deficit Hyperactivity Disorder (ADHD) affects approximately 5\% children and adolescents~\cite{song2021prevalence,edition2013diagnostic}. 
ADHD is a neurodevelopmental syndrome that often leads to increased inattentiveness, impulsivity, and hyperactivity.
Children with ADHD start showing these signs as early as four years of age.
In school-age children, 55\% of all ADHD cases show hyperactivity symptoms~\cite{willcutt2012prevalence}.
The current standard of measurement of hyperactivity in children depends on subjective reports via questionnaires from parents or teachers.
These questionnaires are convenient as they save time, money, and effort.
However, research has also shown very low inter-rater reliability for these surveys between parents and teachers ($\kappa$~=~0.11)~\cite{wolraich2004assessing}.
The inherent subjectivity of these tests and physicians' lack of contextual awareness often lead to misdiagnoses.
This is problematic as overdiagnosis leads to unnecessary treatment, and underdiagnosis can lead to delayed treatment.
Thus, there is a need to add some objectivity to the diagnosis and measurement of hyperactivity. 

Researchers have used commodity sensors to measure the amount of body motion and correlate it with hyperactivity.
For example, Lin~\etal\cite{Lin2020Quantitative} tracked children's arm movement using smartwatches, comparing children with and without hyperactivity.
They found a significant difference in the measured acceleration signals between the two conditions.
Several other works have developed similar machine-learning (ML) models to detect hyperactivity in children from passively sensed acceleration data.
However, these works assumed the measurements happened while children were in specific activities, such as a 1-hour session at a clinic~\cite{OMahony2014Objective} or taking a test~\cite{Gilbert2016Aiding}.
Finding such settings for children is often not practical, especially given that the condition is chronic and requires frequent measurement and management.
Moreover, a clinical examination does not cover a full spectrum of a child's behavior and condition during a typical day or how their behavior changes on different days (\eg weekends \textit{vs.} school days).

While all prior approaches only looked at acceleration data of the wearers arm (a very low-level digitization of user), adding activity context enabled me to identify hyperactivity symptoms in children without putting any behavioral constraints on children. For this, I ran a preliminary study \cite{lemurdx}, wherein we gave smartwatches to 61 children out of which 25 had a hyperactivity diagnosis. Analyzing this pilot data across participants, showcased that hyperactivity is dependent on the activity context. In particular, the different in hyperactivity between ADHD and control group manifests itself only in certain activity classes throughout the day. Our ML pipeline to estimate the risk of hyperactivity achieves a detection accuracy of 85.2\% (F1-score~=~81.6\%) when using sensor data from activity moments labeled as ``quiet'' or ``sitting''. Without such contextualization, the model accuracy drops to 67.2\% (F1-score~=~63.0\%). Further clinicians stressed the need to contextualize hyperactivity. Here, the use of a higher-order digitization (activity context) augmented the capabilities of a lower-order digitization (average motion), wherein both were sensed from the same acceleration sensor. Further, as the activity context limits the search space, we need to collect less data for training, which is great as collecting medical data is very challenging. 

\subsection{Speaker Head-Pose Estimation for Augmenting Speech Interaction}
\label{dov}

\begin{figure}
\centering
  \includegraphics[width=\columnwidth]{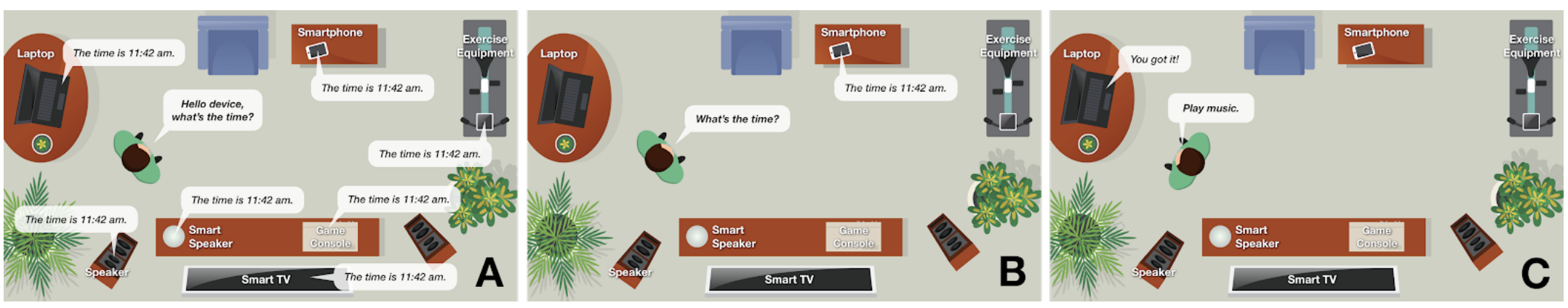}
  \caption[Direction-of-Voice Estimation for addressability]{Future smart homes and offices are envisioned to contain many “smart” devices able to respond to voice commands. However, without device-specific wakewords, multiple devices may try to respond to generic queries (left). Ideally, users would be able to face and speak to a device, more akin to human-human interaction (center and right). Thus, there is a need for Speaker Head-Pose estimation approaches, especially those than can run locally on self-contained devices, without having to install extra sensors in the environment or rely on multi-device interoperability, which does not appear to be forthcoming in the near future.}
  \label{fig:dov}
\end{figure}

Where a person is looking is an important social cue in 
human-human interaction, allowing someone to address a particular person in conversation or denote an area of  interest. For several decades, human-computer interaction researchers have looked at using gaze data to ease and enhance interactions with computing systems, ranging from social robots to smart environments. However, to capture gaze direction, special sensors must either be worn on the head (unlikely for consumer adoption) or external cameras are used (which can be privacy invasive). 

In this research, I explored a practical approach - utilizing speech as a directional communication channel. In addition to receiving and processing spoken content, I propose that devices also infer the Direction of Voice (DoV) with is akin to modeling the user's head pose with respect to the target device of interaction. Note this is different from Direction of Arrival (DoA) algorithms, which calculate from where a voice originated. In contrast, DoV calculates the direction along which a voice was projected. 

This helps devices with microphones not only infer the activity of the user but also infer their head orientation to the device. Such estimation innately enables voice commands with addressability (Figure \ref{fig:dov}), in a similar way to gaze, but without the need for cameras. This allows users to easily and naturally interact with diverse ecosystems of voice-enabled devices, whereas today’s voice interactions suffer from multi-device  confusion. With DoV estimation providing a disambiguation mechanism, a user can speak to a particular device and have it respond; e.g., a user could ask their smartphone for the time, laptop to play music , smartspeaker for the weather, and TV to play a show. Another benefit of DoV estimation is the  potential to dispense with wakewords (e.g., “Hey Siri”, “OK Google”) if devices are confident that they are the intended target for a command. This would also enable general  commands – e.g., “up” – to be innately device-context  specific (e.g., window blinds, thermostat, television).

DoV's approach relies on fundamental acoustic properties of both human speech and multipath effects in human environments. The machine learning model leverages features derived from these  phenomena to predict both angular direction of voice, and more coarsely, if a user is facing or not facing a device. The software is lightweight, able to run on a wide variety of consumer devices without having to send audio to the cloud for processing (thus helping to preserve privacy). Thus, modeling the speaker's head-pose unlocks new capabilities in current smart speakers and smart devices equiped with microphone arrays.  

\subsection{Pose-driven Activity Recognition for Practical Classroom Sensing}
\label{pose_drive_edusense}

Here I showcase how modeling the pose of a user unlocks new insights for lower-order representations such as discrete activity modeling. I explore this through the lens of classroom sensing by modeling the pose of instructors and students in the classroom. 

Providing university teachers with high-quality opportunities for professional development cannot happen without data about the classroom environment. Currently, the most effective mechanism is for an expert to observe one or more lectures and provide personalized formative feedback to the instructor. Of course, this is expensive and unscalable, and perhaps most critically, precludes a continuous learning feedback loop for the instructor. Here, I present EduSense, a comprehensive sensing system that produces a plethora of theoretically-motivated visual and audio features derived from a user's pose. They are correlated with effective instruction, which could feed professional development tools in much the same way as a Fitbit sensor reports step count to an end user app.

\begin{figure}
\centering
  \includegraphics[width=\columnwidth]{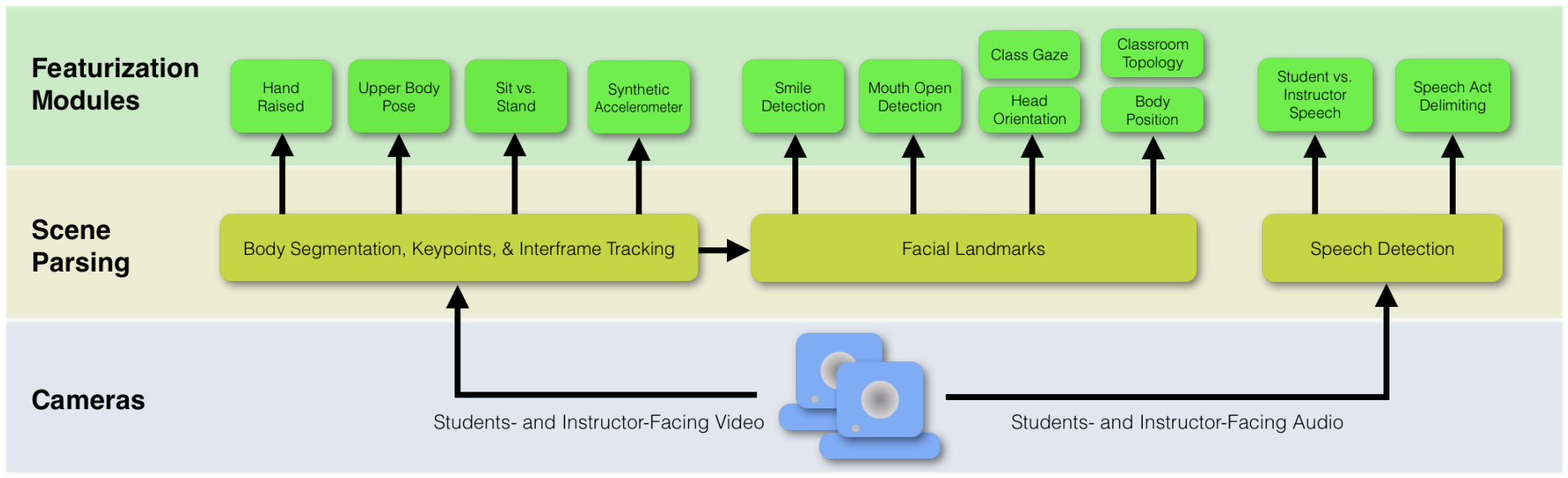}
  \caption[EduSense Featurization Pipeline]{Video and audio from classroom cameras first flows into a scene parsing layer which captures the user's pose, before being featurized by a series of specialized modules.}
  \label{fig:edusense_feat}
\end{figure}

\subsubsection{Augmenting Activity Recognition}

Pose can be used to derive insights for lower-order representations such as activities and posture. Here, I showcase how the user's pose can be used for such tasks. 

\begin{itemize}

\item \textbf{Sit vs. Stand Detection:} This featurization module uses body pose keypoints to predict if a person is sitting or standing. It requires seven keypoints to make an accurate prediction: neck (1), hips (2), knees (2), and feet (2). The relative geometry of these points is encoded by computing direction unit vectors between all pairs of these keypoints. To this feature vector, I also add the ratio of distances between the chest and foot, and chest and knee for both legs. This combined feature vector is passed to an MLP classifier. 

\item \textbf{Hand Raise Detection:} In addition to occlusion, hand raise detection is even more challenging due to the variation in the way students participate. This module ingests eight body keypoints per body: neck (1), chest (1), shoulder (2), elbow (2), and wrist (2). I compute direction unit vectors between all pairs of these points. I also compute the distance between all pairs of points, normalized by the distance between the nose and neck points. Note these features are essentially scale invariant to both body distance from camera and physical body size. These values are used as input to an MLP classifier, which predicts either hand raised or not. 

\item \textbf{Upper Body Posture:} Body posture can be indicative of student affective and attentional state. To explore the feasibility of sensing similar attributes, but in a non-invasive computer-vision driven manner, I trained an upper body posture classifier. For this module, I utilize the same eight upper body keypoints we found to be successful in the hand raise detection module. As before, I compute direction unit vectors between all pairs of points, and distance between all pairs of points, normalized by the distance between the nose and neck points. These values are used as input to a multiclass MLP model (sklearn, default hyper parameters), which was trained during development to predict three proof-of-concept classes: arms at rest, arms closed (e.g., crossed), and hands on face.

\item \textbf{Smile Detection:} As a proof-of-concept of class-scale facial affect analysis, I built a module that detects smiles. For this, I use ten mouth landmarks on the outer lip and ten landmarks on the inner lip. I compute direction unit vectors from the left lip corner to all other points and use a SVM (sklearn, poly kernel with degree 3, default parameters) for binary classification. 

\item \textbf{Mouth State Detection:} Audio based speaker identification is challenging especially when multiple speakers are present. As a potential, future way to identify speakers, I developed a module that estimates if a mouth is open, the confidence of which could be tracked over many frames (per person) to produce a talking confidence. I use a binary SVM (sklearn, linear kernel, default parameters) and two highly descriptive features: the height of the mouth to the left and right of center, divided by the width of the mouth. 

\end{itemize}

\subsubsection{New applications based on Pose}

Apart from augmenting lower-order digitization, the pose itself can enable of a multitude of new features. Here are some listed below.

\begin{figure}
\centering
  \includegraphics[width=\columnwidth]{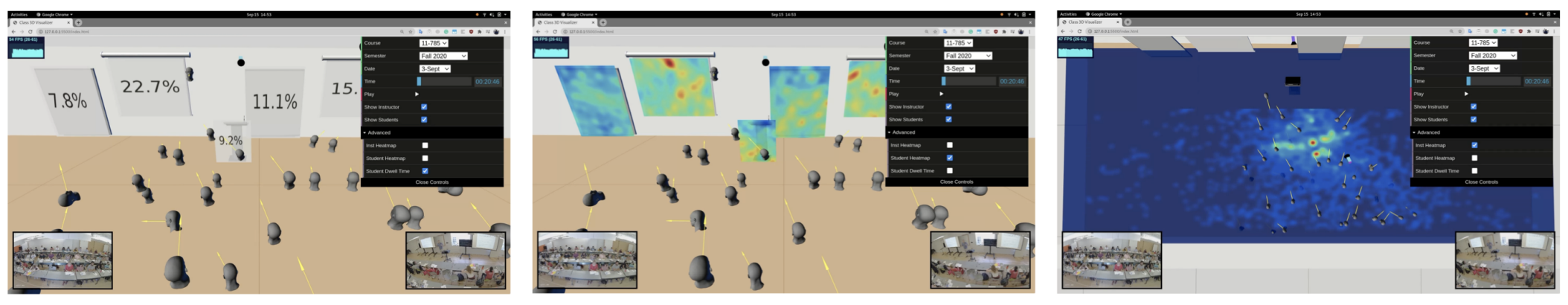}
  \caption[EduSense Gaze and Classroom Topology]{EduSense Gaze and Classroom Topology. Left: Percentage of student gaze across various classroom foci (whiteboards, projector screens, lectern) at the end of a class session. Center: Heatmaps of students gaze across the same foci. Right: Heatmap of the instructor gaze aggregated across a class session.}
  \label{fig:edusense_twin}
\end{figure}

\begin{itemize}

\item \textbf{Head and Gaze Estimation:} Head orientation can be used as a coarse proxy for gaze attention; e.g., toward the instructor and other classroom foci. Using a custom CNN \cite{ahuja2021classroom} that inputs the face and its landmarks, EduSense produces the 3D orientation of the head for each body. Once found, head orientations for individuals can be aggregated into a classroom histogram of foci, or even a combined mean gaze vector.   

\item \textbf{Body Position and Classroom Topology:} The gaze estimation module not only predicts the 3D orientation for each head, but also an estimated 3D position in real world coordinates (Figure \ref{fig:edusense_twin}). This can be used to reveal the classroom topologies (i.e., student layout), and in the future, help illuminate spatial patterns in the class (e.g., fewer hand raises in the back of class or where the instructor spends the most time). 

\item \textbf{Synthetic Accelerometer:} Worn accelerometers have been used previously to infer student engagement and affect. To achieve a similar result, but without the need for worn sensors, I simply track the motion of bodies across frames. Similar to the previous module, I use the 3D head position produced during scene parsing, and calculate a delta X/Y/Z normalized by the elapsed time since the previous frame. This affords a 3D motion and acceleration vector in real world units (e.g., m/s). 

\end{itemize}

\section{Organization}

In this part, I showcase my research projects that pave the path towards improving user digitization richness while maintaining user practicality. I work my way up the user digitization richness spectrum (Figure \ref{fig:richness_spectrum}) focusing on techniques that digitize a particular aspect of a user.

My interest in user digitization was informed and shaped by my initial explorations into activity recognition. Thus, I start with activity sensing \ref{chap:HAR} as a way to digitize and express a users behavior routine. I start off by exploring the  topic of acoustic activity recognition. I then expand my focus towards camera-based sensing systems that can detect activities across multiple people simultaneously and also profile them over time. 

I then delve into pose sensing \ref{chap:pose}, as it offers a continuous and more comprehensive representation of the user over discrete and pre-defined activities. I start with approaches that model individual body parts such as head-pose, eye-gaze, hands and upper body, building up to approaches that capture a holistic full-body representation of the user. A common theme that unites all these pose-sensing approaches is that they make use of commodity and ubiquitous consumer devices such as smartphones so that the advances in user digitization can be practical and accessible to all.

\chapter{Advancing Activity Recognition in Unconstrained Settings}
\label{chap:HAR}

\section{Introduction}

Activity recognition is an important cornerstone of user digitization. It not only helps glean insights about what the user is doing, but also contextualizes it with respect to the environment. It can enable a plethora of applications ranging from health sensing, informatics, smart assistants and assisted living to name a few.

Thus, a user's activities are the basic building blocks for digitizing and expressing their behavior routine. We are already seeing a glimpse of this in consumer devices. Take the Apple Watch for example - it is one of the best commercial devices available, already selling by the millions for its fitness and health features. At the time of writing this thesis (2023) it can automatically detect only a handful of exercises such as swimming, walking, running, and cycling. However, there are many many more activities of interest, that are more than just exercises, which occur in our daily lives, that these consumer devices around us are oblivious to.  

\section{Preliminary Work}
\label{ubicoustics}

\begin{figure}
\centering
  \includegraphics[width=\columnwidth]{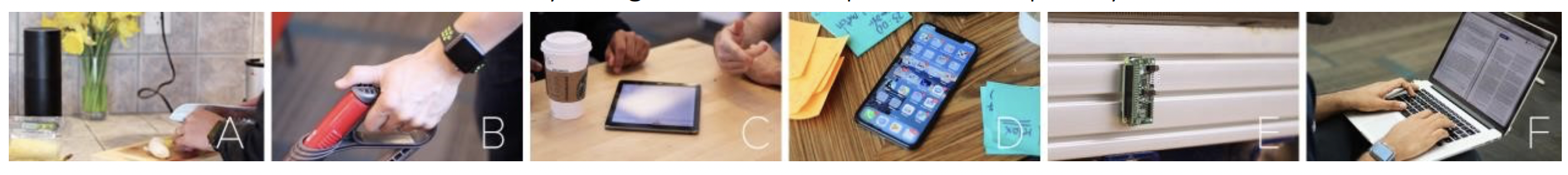}
  \caption[Ubicoustics for plug-and-play acoustic activity recognition]{Ubicoustics enables real-time activity recognition across diverse hardware platforms, including smart speakers (A), smartwatches (B), tablets (C), phones (D), IoT sensors (E) and laptops (F).}
  \label{fig:ubicoustics}
\end{figure}

Microphones are the most common sensor found in consumer electronics today, from smart speakers and phones to tablets and televisions. Despite sound being an incredibly rich information source, offering powerful insights about physical and social context, modern computing devices do not utilize their microphones to understand what is going on around them. For example, a smart speaker sitting on a kitchen counter-top cannot figure out if it is in a kitchen, let alone know what a user is doing in a kitchen. Likewise, a smartwatch worn on the wrist is oblivious to its user cooking or cleaning. This inability for “smart” devices to recognize what is happening around them in the physical world is a major impediment to them truly augmenting human activities.

Real-time, sound-based classification of activities and context is not new. There have been many previous application-specific efforts that focus on a constrained set of recognized classes. I sought to build a more general-purpose and flexible sound recognition pipeline – one that could be deployed to an existing device as a software update and work immediately, requiring no end-user or in situ data collection (i.e., no training or calibration). Such a system should be “plug-and-play” – e.g., plug in your Alexa, and it can immediately discern all of your kitchen appliances by sound. This is a challenging task, and very few sound-based recognition systems achieve usable end-user accuracies, despite offering pre-trained models that are meant to be integrated into applications.

Ubicoustics is a novel approach that brings the vision of plug-and-play activity recognition closer to reality across a myriad of practical form factors (Figure \ref{fig:ubicoustics}). The process starts by taking an existing, state-of-the-art sound labeling model and tuning it with high-quality data from professional sound effect libraries for specific contexts (e.g., a kitchen and its appliances). I found professional sound effect libraries to be a particularly rich source of high-quality, well-segmented, and accurately-labeled data for everyday events. These large databases are employed in the entertainment industry for post-production sound design (and to a lesser extent in live broadcast and digital games). 

Sound effects can also be easily transformed into hundreds of realistic variations (synthetically growing our dataset, as opposed to finding or recording more data) by adjusting key audio properties such as amplitude and persistence, as well mixing sounds with various background tracks. I show that models tuned on sound effects can achieve superior accuracy to those trained on internet-mined data alone. I also evaluate the robustness of our approach across different physical contexts and device categories. Results show that Ubicoustics can achieve human-level performance, both in terms of recognition accuracy and false positive rejection. 

Thus, Ubicoustics showcases the power of acoustic activity sensing across a myriad of smart devices. However, it is limited in its fidelity as it cannot detect simultaneous activities across a multitude of people in the environment. Camera-based activity recognition approaches can overcome this due to their wider field-of-view and thus ability to track multiple people. Further, they can provide a continuous capture of these activities and provide temporal insights. 

\section{Modeling Continuous and Temporal Activities}
\label{chaptergymcam}

\subsection{Introduction}

\begin{figure}
\centering
\includegraphics[width=0.6\textwidth]{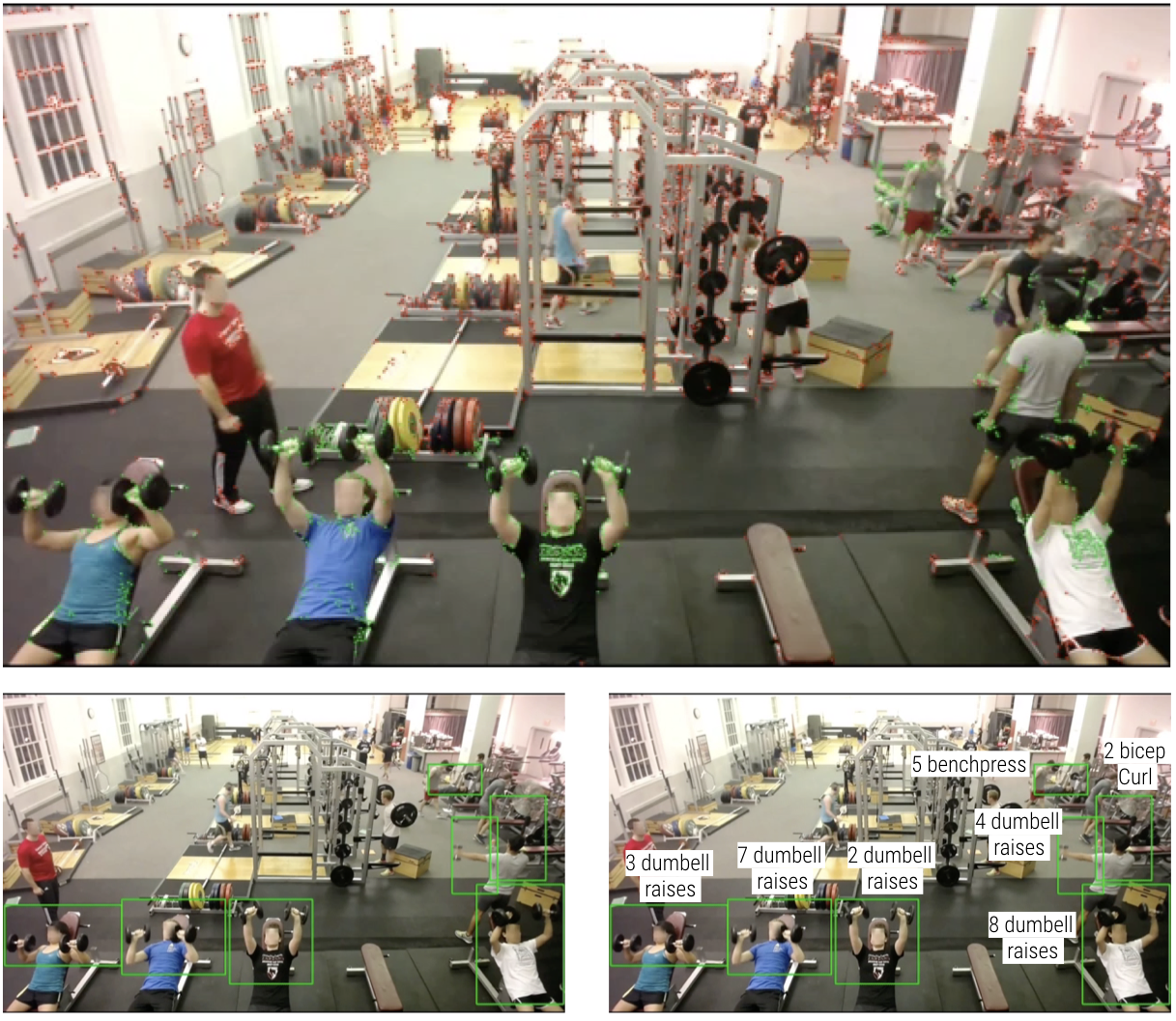}
\caption[GymCam system overview]{GymCam uses a camera to track exercises. Top: Optical flow tracking motion trajectories. Bottom Left: Exercises clustered. Bottom Right: Exercise Recognition.}
\label{hero}
\end{figure}

Detecting, recognizing and tracking simultaneous activities in unconstrained scenes is a challenging task. As an exemplar to showcase the utility of activity recognition I pick gyms as a setting to monitor exercises of multitudes of people. To this end, I present \textit{GymCam} (Figure~\ref{hero}), a vision-based system that uses off-the-shelf cameras to automate exercise tracking and provide high-fidelity activity analytics, such as repetition count, without any user- or environment-specific training or intervention. Instead of requiring each user in the gym to wear a sensor on their body, \textit{GymCam} is an external single-point sensing solution, \textit{i.e.,} a single camera placed in a gym can track \textbf{all} people and exercises simultaneously. 

While using cameras enables accurate exercise tracking that is not limited to certain kinds of motion, it of course also raises privacy concerns. These are important to address in order to attain rich digitization system which are also practical to the user. Thus, to mitigate these privacy concerns, the first step of the classification pipeline converts the raw video into optical flow trajectories. With this processed signal, GymCam can detect exercises, but sensitive user information is not easily recoverable. Indeed, with on-camera compute power, this could be the only data transmitted from the device, or perhaps the entire classification pipeline could be run locally. 

To develop and evaluate the machine learning algorithms, I collected data in the university's gym for five days. In total, I recorded 42 hours of video and annotated 597 different exercises. I did not record the number of gym users because the protocol required immediate anonymization of the data (\textit{i.e.,} faces blurred). Users of the gym were informed that a research team was recording video, but there was no other interaction with participants, minimizing observer effects (\textit{e.g.,} intentional or unintentional changes to their routine). I note this problem often affects research studies where users are aware they are part of an exercise tracking research study, and the evaluation setting is constrained~\cite{recofit}. I believe GymCam presents the first truly unconstrained evaluation of exercise tracking. 

The overall process of GymCam is as follows:

\begin{enumerate}
\item{Detect all exercise activities in the scene (acc. = 99.6\%), then}
\item{Disambiguate between simultaneous exercises (acc. = 84.6\%), then}
\item{Estimate repetition counts ($\pm$ 1.7 counts)}
\item{Recognize common exercise types (acc. = 93.6\% for 5 most common exercise types).}
\end{enumerate}

\subsection{Theory of Operation}

I now discuss the underlying premise behind GymCam that allows it to: (1) detect motion, (2) cluster motions into separate exercises, and (3) identify and track individual exercises. 

GymCam leverages the insight that almost all repetitive motion in a gym represents some form of exercise. Even if a camera cannot see an entire person, it is still often able to see a small part of the body exhibiting repetitive motion, and can track that body part, linking it to an exercise later. However, when multiple users are exercising and potentially overlap in a video, it can be hard for camera-based systems to delineate the exact boundaries between the exercises -- an issue worn sensors do not have to handle. Fortunately, I found it is extremely rare for two users to perform their exercises at exactly same time, speed, and phase. Thus, by calculating features that capture these dimensions, GymCam is able to differentiate between simultaneous exercises without any supervised training data.

Apart from distinguishing different users, there are other challenges when relying solely on repetitive motion tracking. Foremost, periodicity can be exhibited by a user's gait or warm up before starting an exercise. Secondly, when placed in an unconstrained environment, users tends to be less deliberate with between-exercise moments (\textit{e.g.} fidgeting, stretching, walking). These interludes can be quite periodic, and thus indistinct from exercises. Moreover, in the unconstrained environment of a gym, users may challenge themselves (\textit{e.g.} lift challenging weights). Morris~\textit{et al.}~\cite{recofit} observed that ``\textit{self-similarity [or periodicity] may break down in intensive strength-training scenarios. For this reason, more validation of intensive weightlifting is important future work.}" I believe that the only viable approach to solve the problem of variations in exercise and noisy human behavior, is to collect extensive training data in the user's actual workout environment without significant observer effect. 

\subsection{Algorithm}
\begin{figure}
\includegraphics[width=0.9\textwidth]{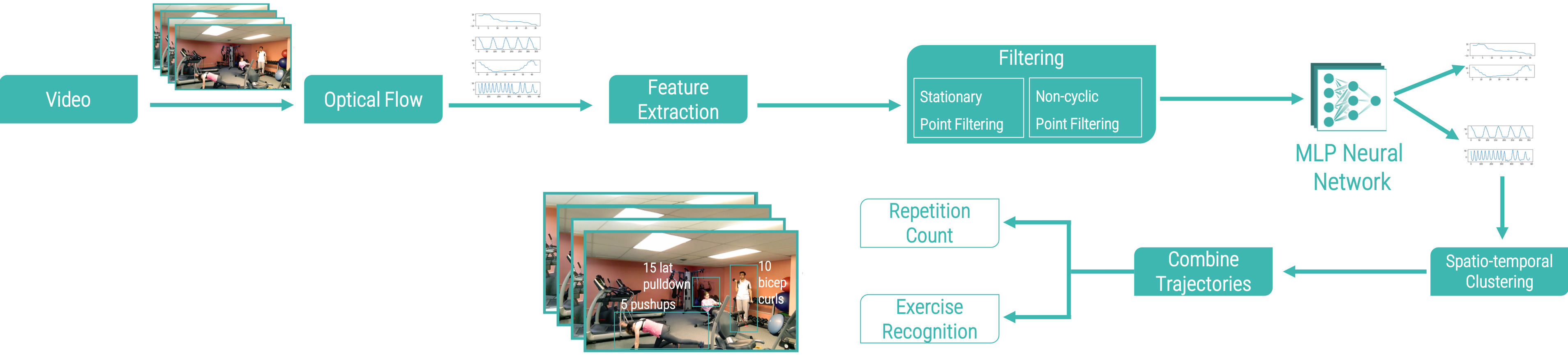}
\caption[GymCam system architecture]{GymCam system architecture.}
\label{pipeline}
\end{figure}

The goal of the work is to detect, identify, and track exercises, including when people are only partially visible. In fact, the real test of the approach is when the user is \textit{barely} visible, but the camera can merely see a weight or a handlebar moving. Thus, GymCam starts by identifying all movements, and classifying them as repetitive or not. There could be several movements in a video that belong to the same exercise (\textit{e.g.,} movement of different limbs, weights, and handlebar), so I combine similar repetitive exercise movements into exercise clusters. Next, for all motion trajectories in each identified cluster, I derive a combined trajectory to recognize the exercise type and estimate repetition count for that exercise (cluster). I will now describe the pipeline (Figure~\ref{pipeline}) in detail.

\subsubsection{Detecting Exercise Trajectories}
\label{sec:track_ex}
To detect movement, I start by extracting optical flow trajectories from the video. I initially investigated OpenCV's implementation of Lucas-Kanade sparse optical flow \cite{bouguet2001pyramidal}. However, the algorithm failed to track large, sudden movements and I switched to Wang~\textit{et al.}'s~\cite{densetraj} dense optical trajectory extraction method to process all motion captured by the camera. For every video frame, the algorithm generates new keypoints, which are tracked continuously across frames to produce a motion trajectory. I found a keypoint max lifespan of 11~seconds was ideal for capturing several exercise repetitions, while also managing the processing time needed to track thousands of points in a video stream.

These motion trajectories are then converted into features and passed to a classifier. To limit the number of data points, I trim motion trajectories by removing stationary points (\textit{i.e.}, any keypoint that moved less than 4-pixels between frames). I then normalize motion trajectories by their maximum translation and calculate a feature vector over an (empirically determined) sliding window of five seconds, with a stride of one second. The feature vector consists of 27 features, a subset of which have also been used in prior work (see~\cite{recofit,bedri2017}).

\begin{itemize}
	\item \textbf{Frequency-based features:} The working principle is that exercises are more periodic than non-exercises. I use frequency-based features to encode this property:
    
    \begin{itemize}
    \item \textbf{Number of zero crossings:} I calculate the number of zero crossings of the keypoint motion trajectory, only in the \textit{x}-axis, and only in the \textit{y}-axis.
    
    \item \textbf{Variance in zero crossings:} Exercises will be more periodic and have a lower variance in zero crossings than non-exercises.
    
    \item \textbf{Dominant Frequency:} The dominant frequency of the signal calculated by frequency transformation.
    
    \item \textbf{Autocorrelation:} Autocorrelation characterizes the periodicity of a signal. 
   
   \item \textbf{Maximum autocorrelation peak:}  Higher value indicates higher periodicity. 
   
    \item \textbf{Frequency via autocorrelation:} The dominant frequency of signal determined via autocorrelation.
   
   \item \textbf{Number of autocorrelation peaks:} Unusually high number of peaks indicate noisy signals, which are more likely to be non-exercises. 
   
   \item {\textbf{Number of prominent peaks:}} Represents the number of peaks higher than their neighboring peaks by a threshold (25\%). A greater number of prominent peaks indicates higher periodicity.
   
   \item{\textbf{Number of weak peaks:}} Similarly, I calculate the number of peaks smaller than their neighboring peaks by a threshold (25\%). A greater number of weak peaks represents noisy and less periodic motion.
   
   \item \textbf{Height of first autocorrelation peak after first zero crossing.}  The height of the first peak after a zerolag provides an estimate of the signal's periodicity.
   
    \end{itemize}
    
    \item \textbf{Non-frequency-based features:} Apart from the frequency-based features, I also calculate some non-frequency based features: 
    
    \begin{itemize}
    
    \item \textbf{RMS:} The root-mean-square amplitude of the signal.
    
    \item \textbf{Span:} The span of the motion helps to characterize the intensity of the motion. I use overall span, and span in both \textit{x}- and \textit{y}-axes as features.
    
    \item \textbf{Displacement Vector:} Displacement helps us distinguish between exercises and other periodic motions such as walking. Non-exercise motions (such as walking) often have a higher displacement than exercise motions. I use the coefficients of the overall displacement vector, and displacement in both \textit{x-} and \textit{y}-axes, for a total of 9 features.
    
    \item \textbf{Decay:} Decay signifies the loss of intensity over time, a characteristic of exercise motions. I fit a line to the observed trajectory and use its coefficients as features.
    
\end{itemize}

\end{itemize}

I filter motion trajectories to bias the classifier to minimize false positives, at the cost of lower precision. This is because when an person is exercising, not all body parts may be involved in the motion. For example, legs do not move during a bicep curl, so a keypoint on a person's leg may be inside the bounding box created by an annotator, but would not be periodic. Similarly, improper form may cause a point to move while performing an exercise. Thus, not every motion trajectory inside an ``exercise" bounding box is indicative of actual exercise motion. To protect the classifier from inaccurate training data, I filter motion trajectories with aggressive thresholds on frequency-based features. By filtering, I only provide the strongest and most representative examples of exercise trajectories to train the classifier. However, I do not perform any such filtering while validating the algorithm.

I use a multilayer perceptron (Scikit-Learn implementation with default hyperparameters) to classify every 5~second window segment of each keypoint trajectory as an exercise or not. The neural network optimizes the log-loss function using stochastic gradient descent. To smooth the output, I take a majority vote of three consecutive classifications and assign that as the output for each of those three classifications. Finally, I combine all consecutive \textit{positive} classifications to construct a motion trajectory that was predicted as an exercise.

\subsubsection{Clustering Points for Each Exercise}
\label{sec:cluster}
Exercises are often captured by many keypoint motion trajectories. Thus, the next step is to cluster keypoint motion trajectories into exercise groups. I perform clustering in two steps: (1) use spatio-temporal distribution of motion trajectories, and (2) use phase-differences between motion trajectories. 

Given an exercise, the motion trajectories of its encapsulating keypoints will likely be close to one another in space \textit{and} time. For space, I bootstrap the clustering algorithm by drawing bounding boxes next to each workout machine and station. Note, this only needs to happen once at the start of the system deployment (assuming machine and stations do not move). These boxes are non-overlapping and are representative of the exercise areas of the gym. 

Apart from spatial distribution, I also investigated the temporal separation between exercises. The exercise keypoints that overlap temporally as well as spatially are assigned to the same cluster. However, there is still a chance that exercises that are close to one another and occur together will be wrongly combined. To seperate such clusters, I also use phase information. For each cluster, I compute a phase-based similarity score between each trajectory-pair. For a pair of points that are not temporally co-located, the similarity is set to zero, and for others, the similarity is equal to the phase difference. I then threshold the phase difference (15~degrees) to assign a binary similarity score. In the end, I have a complete $N\times N$ adjacency matrix, where $N$ denotes the number of motion trajectory points classified as an exercise. Given such a matrix, I calculate all connected graphs. Each graph denotes one exercise cluster associated with the nearest bounding box. 

At the end of clustering, I combine the trajectories of all keypoints within a cluster to create a representative, average trajectory for further analysis. More specifically, I take the average of all the points within the cluster, accounting for the duration of each point, and smoothing it with a Hann window (size=1~second). This trajectory is used in the next process: exercise recognition and repetition count.

\subsubsection{Repetition Count}
\label{sec:repcount}
Once a representative, average trajectory for each cluster is obtained from the previous step, I calculate the repetition count. To objectively disambiguate actual exercises from warm ups, I disregard any exercises that have less than five repetitions in the ground truth annotations. I train a multilayer perceptron regressor (Scikit-Learn; default hyperparameters) that uses the frequency-based features for each combined trajectory (as detailed in section~\ref{sec:track_ex}), and outputs an estimate for the repetition count.

\subsubsection{Exercise Recognition}
\label{sec:ex_id}
Similar to repetition count, I leverage the cluster-average trajectory to infer the exercise type. I first quantize the trajectory into fixed-length segments as input to the classifier. I then run a sliding window (length five seconds, stride of one second) over this motion trajectory. Each window is passed to a multi-layer perceptron classifier (Scikit-Learn; default hyperparameters) to predict the exercise label, and I take a probability-based majority vote over all windows in the trajectory.

\subsection{Results}

\subsubsection{Detecting Exercise Trajectories}
I first report the results for distinguishing keypoint motion trajectories as \textit{exercise} or \textit{non-exercise}. For this, I performed a leave-one-day-out-cross-validation, which yielded a per-day, mean cross-validation exercise detection accuracy of 99.86\%, with a mean false positive rate of 0.001\% and precision of 23\%. Again, I optimized the algorithm to reduce false positives at the expense of precision. 

\subsubsection{Clustering Points for Each Exercise}
There are 597 distinct exercises in the ground truth annotated data. GymCam was able to accurately track 84.6\% of these exercises. It also had a false positive rate of 13.5\%, with most errors due to miscellaneous cyclic non-exercise motion such as warm-ups, rocking while seated, and walking.

\subsubsection{Repetition Count}
Repetition count accuracy helps in objective assessment of the time overlap between a predicted cluster and its corresponding ground truth match. I used 5-minute folds for cross-validation and achieved an accuracy of $\pm$1.7 for counting repetitions with a standard deviation of 2.64.

\begin{figure}
\includegraphics[width=\textwidth]{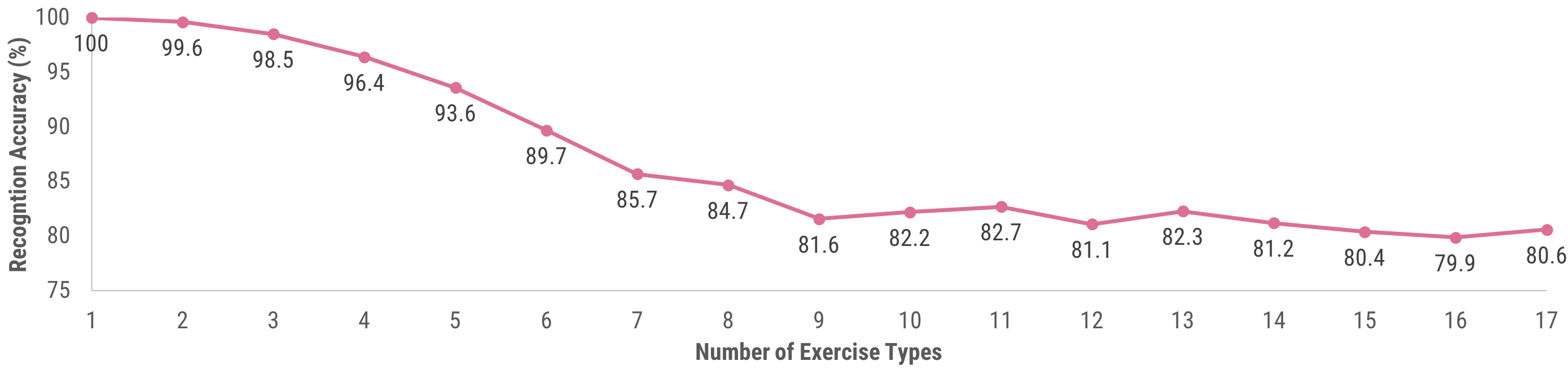}
\caption[GymCam Exercise Recognition Accuracy]{Plot showing the accuracy of exercise recognition \textit{vs.} number of exercise types}
\label{fig:ex-reco}
\end{figure}

\subsubsection{Exercise Recognition}

As discussed previously, the data was collected in an uncontrolled environment where participants were not instructed to perform a specific set of exercises, and so the distribution of exercise types was not uniform. Participants performed numerous atypical exercises and curating a balanced training set of conventional exercises from the data was challenging. I identified 18 common gym exercises and annotated their instances in the dataset. I decided to disregard warm-up exercises because the annotator labeled many different exercises as ``warm-up". The remaining 17 exercise types were classified with an accuracy of 80.6\% with cross-validation across 5-minutes folds. The five most frequently performed exercise types constituted roughly 69\% of the data. I noticed that a lack of training data caused the less frequently seen exercises to be misclassified. Thus, if I only focus on the most frequent exercises, GymCam recognition accuracy increases to 93.6\%. 

Figure~\ref{fig:ex-reco} shows the average identification accuracy as the number of recognized exercise types increase. This result indicates that the approach has the potential to differentiate between exercises based on the feature set, but a larger annotated dataset is needed. 

\subsection{Conclusion}
\textit{GymCam} showcases the power of detecting, recognizing and tracking simultaneous activities in unconstrained scenes. It is a system that leverages a single camera to track a multitude of simultaneous exercises. GymCam relies on tracking motion and assumes that most repetitive motion in a gym are exercises in progress. To develop and evaluate the machine-learning algorithms, I collected data in a varsity gym for five days. I segmented \textit{all} concurrently occurring exercises from other activities in the video with an accuracy of 84.6\%; recognized the type of exercise (acc.=92.6\%) and counted the number of repetitions ($\pm$ 1.7 counts). GymCam advances the field of real-time exercise tracking by filling some crucial gaps, such as tracking whole body motion, handling occlusion, and enabling single-point sensing for a multitude of users. The efficacy of GymCam showcases the use of only motion trajectories for activity recognition. This helps preserve the high fidelity afforded by cameras while decreasing their invasiveness.

\chapter{Towards Full-Body Pose Capture from Ubiquitous Consumer Devices}
\label{chap:pose}

\section{Introduction}

Delving into the intricate details of human activities yields substantial insights, but these endeavors remain constrained by their predetermined and discrete nature. In contrast, the concept of ``pose" offers a continuous and all-encompassing portrayal of an individual's physical orientation and alignment. This holistic viewpoint not only opens the door to novel applications but also enhances the capabilities of existing ones. Consider its application in the context of physical therapy. Instead of being confined to predetermined exercises, a comprehensive understanding of a patient's pose allows for tailored rehabilitation programs. By precisely capturing their posture and movements, therapists can design interventions that dynamically adjust to the patient's progress, offering more effective and personalized recovery journeys. This demonstrates the transformative impact of higher-order pose-based representations, which consequently permeate and uplift lower-level representations like activity classifications.

However, the prevailing pose capture landscape predominantly features commercial systems focused on capturing full-body pose through specialized suits or elaborate camera arrays and depth-sensing technology integrated into the environment. Estimating full-body pose from ubiquitous consumer devices remains a challenging task. Let's start by understanding the key over-arching challenges in achieving this.

\subsection{Key Challenges}

First, estimating full-body pose from ubiquitous consumer devices such as smartphones is a very under-constrained problem. Such devices don't have that many specialized sensors designed for the purpose of pose capture. Realistically speaking, you can only expect to have the sensors already in consumer devices. Second, these few sensors that these devices do have are not very good.  Consumer-grade sensors tend to be pretty noisy, low frame rate, and are at sub-optimal locations. 

Second, we are not likely to convince users to wear more sensors or adopt new devices. Further, users want high-resolution output, like fine-grained health data, but they don’t want to input high fidelity data, like continuous camera or microphone data that is privacy invasiveness. Thus, we don’t have many sensors, can’t add more and the ones that are there are pretty low-fidelity. However, there are a few key insights that I developed through my research that can help unlock the potential of ubiquitous devices for modeling the full-body pose.

\subsection{Key Insights}

The first key insight is that user digitization is application dependent. Thus, if we understand the trade-off space between the applications and the estimation accuracy, we can get away with approximates rather than precise tracking. For example, if a user waves in VR, the digitization framework need not get radial velocity of the users hand perfectly, it just needs to show that the user is waving to correctly communicate their intentions. While the final goal of any system should be to maximize tracking accuracy and precision, understanding where estimates are useful is a great tool to unlocking pose based applications on ubiquitous devices.  

The second insight is that the human body, which are trying to model, is hierarchical. We know that our hands are connected to our elbows, and our elbows to our shoulder. There are physical limits to how far our joints can rotate, and many other spatial kinematic constraints that we can use as structural priors. Further, in addition to modeling a spatial kinematic skeletal tree, we can also account for temporal motion and behavior correspondences. 

Lastly, we can leverage the power of different modalities and device locations afforded by consumer devices to help estimate pose. In the subsequent sections, we will see what does leveraging all these insights look like in action. By the end of this chapter, I will showcase that full-body pose is not exclusive to the realm of specialized and external devices, but can now be enabled on mobile consumer devices. Albeit approximate full-body pose, its versatility becomes particularly evident in uncontrolled, real-world settings. This approach stands in stark contrast to methods bound by hardware limitations or external dependencies.

\subsection{Organization}

I start by showcasing the utility of modeling the pose of individual body parts and gradually build up to the complexity of full-body pose capture. First, I cover sensing approaches that model the head pose and eye-gaze of the user (Section \ref{eye-gaze}). I then shift my focus to another body part, namely the hands (Section \ref{hands}), which are our conduits to interaction with the world around us. I then look into full-body pose capture and augment these approaches to include the hands and head as well (Section \ref{whole-body}). Lastly, I cover approaches that extend beyond single person pose capture and extent to multi-user scenarios (Section \ref{multi-user}).

\section{Head and Eye-Gaze Estimation via Mobile Devices}
\label{eye-gaze}

\subsection{Introduction}
\label{sec:intro}

Computer interfaces with the ability to track a user's gaze location offer the potential for more accessible and powerful multimodal interactions, perhaps one day even supplanting the venerable cursor.
While useful for desktop computing, gaze also promises to be a powerful way to interact with phones, especially given the need to adapt to a variety of usage contexts (\textit{e.g.}, inability to use the touchscreen with encumbered hands). 
Specialized gaze tracking hardware -- either worn \cite{DBLP:conf/huc/KassnerPB14} or placed in the environment \cite{TobiiProLab} -- can track gaze with very high resolution \textit{i.e., }1.1~mm (0.45\degree) but, the need for specialized equipment is a significant barrier for consumer adoption. When relying on existing onboard hardware, research has primarily focused on user-facing RGB cameras. 
Unfortunately, gaze models utilizing this RGB data are too coarse for interactions with many user interface widgets, which are generally small on mobile devices.
To help close this gap, researchers have started assessing the value of depth cameras to improve performance \cite{DBLP:conf/etra/MoraMO14, DBLP:conf/icra/ZhouCLL17, DBLP:conf/aaai/LianZLHWLYG19}, but all research to date has focused on desktop-grade depth cameras (\textit{e.g.}, Microsoft Kinect V2 \cite{Kinect} , Intel Real Sense \cite{DBLP:conf/cvpr/KeselmanWGB17}). These sensors are much more capable than the depth cameras seen in smartphones, which must be very thin and comparatively lower powered.
Furthermore, much of this prior RGB+Depth (RGBD henceforth) gaze research had users maintain their head position in a highly constrained way (\textit{e.g.}, chin rest \cite{DBLP:conf/uist/SmithYFN13, DBLP:conf/cvpr/SuganoMS14}). This rigid requirement is at odds with the usual way a typical user interacts with a phone while walking, riding public transport, carrying handbags, \textit{etc}. Thus, it is important to build a gaze tracker that adapts to a user's changing context, uses existing hardware, and provides usable resolution. 

Here \cite{arakawa2022rgbdgaze}, I present a gaze tracker that uses an off-the-shelf phone's front-facing RGB and depth camera. I collected data from and implemented my system in recent Apple iPhones (X and above), which feature a 1080p user-facing camera and Apple's structured light TrueDepth camera (similar to the technology used in the Kinect V1 \cite{Kinect} and earlier PrimeSense models).
My mobile RGBD dataset of 50 participants is the first of its kind, offering RGBD data paired with user gaze location across a variety of use contexts.
I implemented a CNN model based on a spatial weights structure to efficiently fuse the RGB and depth modalities.
my model achieves 1.89~cm on-screen euclidean error on my dataset in a leave-one-participant-out evaluation, showing a significant improvement over existing gaze-tracking methods in mobile settings.
This result reaffirms the utility of fusing RGB and depth data, and offers the first benchmark for smartphone-based RGBD gaze tracking while a user is not simply sitting.

\begin{table*}

\caption{Comparison of my system with prior gaze tracking work. Grey-colored rows denote systems benchmarked using my dataset. Unconstrained studies are those where the distance and/or angle between the capturing device and user was not static.}
\label{tbl:rgb-dataset}
\begin{tabular}{r|c|c|c|c|c|c}
\toprule
\multirow{2}{*}{System}& \multicolumn{2}{c|}{Capture Modality} & 
\multirow{2}{*}{\begin{tabular}[c]{@{}c@{}}Mobile\\ Device\end{tabular}} &
\multirow{2}{*}{\begin{tabular}[c]{@{}c@{}}Unconstrained\\ Study\end{tabular}} &
\multirow{2}{*}{\begin{tabular}[c]{@{}c@{}}Calib\\ --Free\end{tabular}} &
\multirow{2}{*}{Gaze Error} \\ 
\cline{2-3} & RGB & Depth & & & \\ \hline

MPII Gaze \cite{DBLP:conf/cvpr/ZhangSFB15} & \checkmark & &  &\checkmark &\checkmark & 6.3\degree \\
EyeTab \cite{DBLP:conf/etra/WoodB14} & \checkmark & & \checkmark  & & \checkmark & 2.58~cm \\
EyeMU \cite{Kong2021EyeMU} & \checkmark &  & \checkmark  & \checkmark &  & 1.7~cm \\
iTracker \cite{DBLP:conf/cvpr/KrafkaKKKBMT16} & \checkmark &  & \checkmark  & \checkmark &  & 1.34~cm \\
iMon \cite{huynh2021imon} & \checkmark &  & \checkmark  & \checkmark & \checkmark & 1.57~cm \\
TabletGaze \cite{DBLP:journals/mva/HuangVS17} & \checkmark &  & \checkmark  & \checkmark & \checkmark & 3.17~cm \\
\rowcolor[rgb]{0.95, 0.95, 0.95} iTracker \cite{DBLP:conf/cvpr/KrafkaKKKBMT16} & \checkmark  & & \checkmark  & \checkmark  &  \checkmark & 2.77~cm  \\
\rowcolor[rgb]{0.95, 0.95, 0.95} Apple ARKit \cite{appleARKit} & \checkmark  & & \checkmark  & \checkmark  &  \checkmark & 6.38~cm  \\
\rowcolor[rgb]{0.95, 0.95, 0.95} \textbf{My System}  & \checkmark  & \checkmark  & \checkmark  & \checkmark & \checkmark  & \textbf{1.89 cm} \\ \bottomrule
\end{tabular}
\end{table*}

\subsection{RGB+Depth Dataset \& Collection}

I collected a first-of-its-kind dataset of RGBD on mobile devices. For data collection, I created an iOS application capable of recording and uploading gaze tracking data. The application runs on Apple iPhone X and above, as they all feature a high resolution front-facing RGB camera and a TrueDepth camera (640 $\times$ 480 depth map interpolated from a 170 $\times$ 170 IR dot pattern). For the data collection, I recruited 50 participants (mean age 25 years, 34 male, 16 female). Fourteen of them wore glasses during the data collection. Twenty of the participants were recruited through in-class recruitment and the remainder of the 30 were recruited using an online sign-up form posted on various social media sites. The app was delivered via TestFlight.

The custom iOS application asked participants to look at a target (red dot) that was moving on the screen. While the user gazed at the target, synchronized RGB and depth imagery was logged at approximately 8~Hz, along with the ARKit gaze prediction (which I capture as a state-of-the-art commercial benchmark). 
The speed of the dot movement was varied to add diversity.
The data collection was paused when the face was not detected using the Apple Vision Framework.
In a similar way to GazeCapture \cite{DBLP:conf/cvpr/KrafkaKKKBMT16}, I recorded device motion (9-axis IMU) sensor data synchronized to image data.
While I did not use this sensor data in my study, this could be a useful resource in future work.

\begin{figure*}
    \begin{center}
        \includegraphics[width=0.75\linewidth]{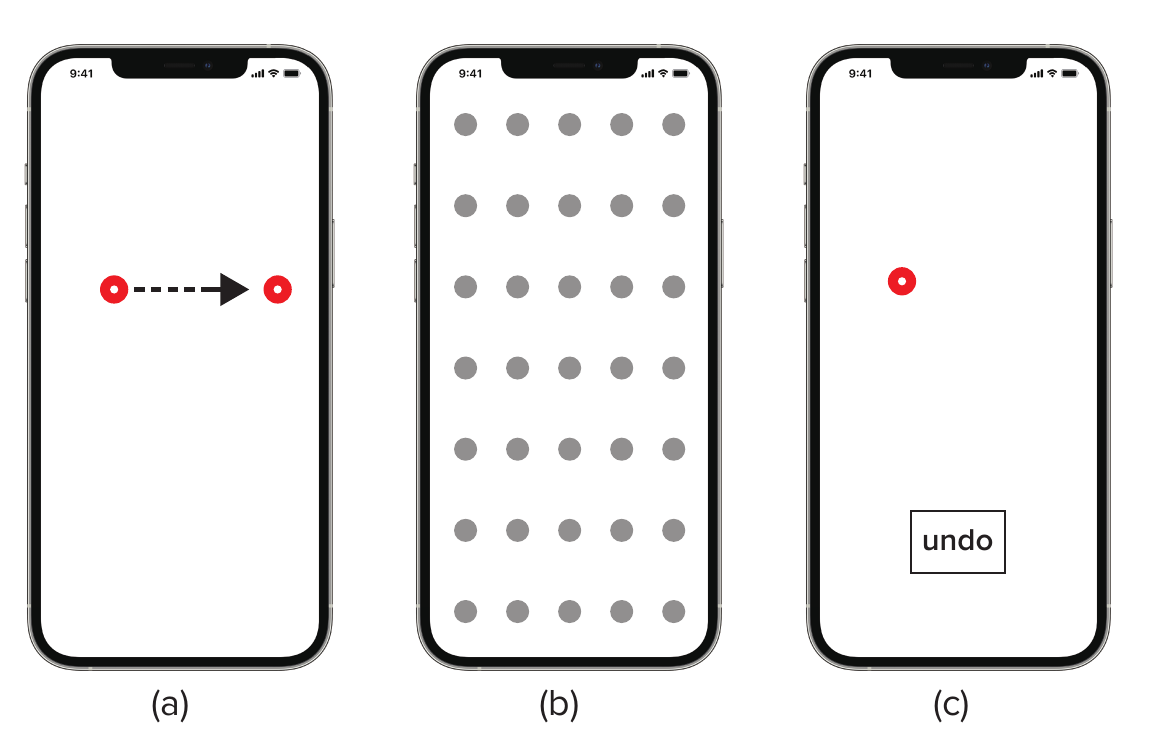}
    \end{center}
    \caption{My application for gaze data collection. (a) The target (red dot) moves around the screen. (b) The screen is divided by 35 fixed locations (illustrated here, but hidden from users) and the target moves from one to another. (c) Participants are able to stop by tapping the screen and ``undo'' a trial by pressing the button.}
    \label{fig:data-collection}
\end{figure*}

While using the app, the target dot was animated to cover various locations on the screen (\figref{fig:data-collection}(a)). Specifically (and similar to \cite{DBLP:conf/cvpr/KrafkaKKKBMT16, DBLP:journals/corr/XuEZFKX15, DBLP:journals/mva/HuangVS17}), I first pre-determined 5 $\times$ 7 = 35 fixed locations (\figref{fig:data-collection}(b)), and then the dot repeatedly moved linearly (vertically, horizontally, and diagonally) from one location to another in a random fashion such that it covered each location four times. 
In addition, I implemented an ``undo'' functionality that lets the participant jump back to the previous gaze location (in case they were not paying attention or did not follow the target). The ``undo'' button shown in \figref{fig:data-collection}(c) was hidden while the target was moving unless the participant tapped anywhere on the screen to stop its animation. 

To ensure data reliability, I followed an approach similar to Krafka et al. \cite{DBLP:conf/cvpr/KrafkaKKKBMT16}. First, during data recording, the app was kept in Airplane Mode to avoid any distractions via notifications. I further monitor user attention to ensure constant engagement with the application. This is done by introducing a color check mechanism during the task. Specifically, the color of the inner center of the moving dot changed to either white, green, or blue randomly during each animation (motion from one target gaze location to another). Users needed to perform a tapping action according to the color; tap nowhere if it was white, tap the right side of the screen if it was green, and tap the left side if it was blue. If they failed this color check, they were warned and the failed sequence was repeated.

The study consisted of four sessions spanning four distinct, yet common use contexts: standing, walking, sitting, and lying down. During each session, the participants were not given any specific instructions on how they should hold the device.
I note that they routinely changed hands and arm positions during the sessions. Each collection session lasted for approximately four minutes, and the order of the sessions was randomized for each participant. The study took roughly 20 minutes and participants were compensated with \$10~USD for their time. I also did not control for the environment, time of day or illumination during the data collection period. This led to high variability of data, critical in aiding the development of a robust, calibration-free gaze tracker.

I pruned the collected dataset by removing data points where the participants blinked. For this I employed an eye-aspect-ratio method \cite{cech2016real}. Roughly 2\% of my data consisted of blinks, which were dropped from analysis. In total, my final dataset consisted of 160,120 data points across 50 participants.

\begin{figure*}
    \begin{center}
        \includegraphics[width=0.9\textwidth]{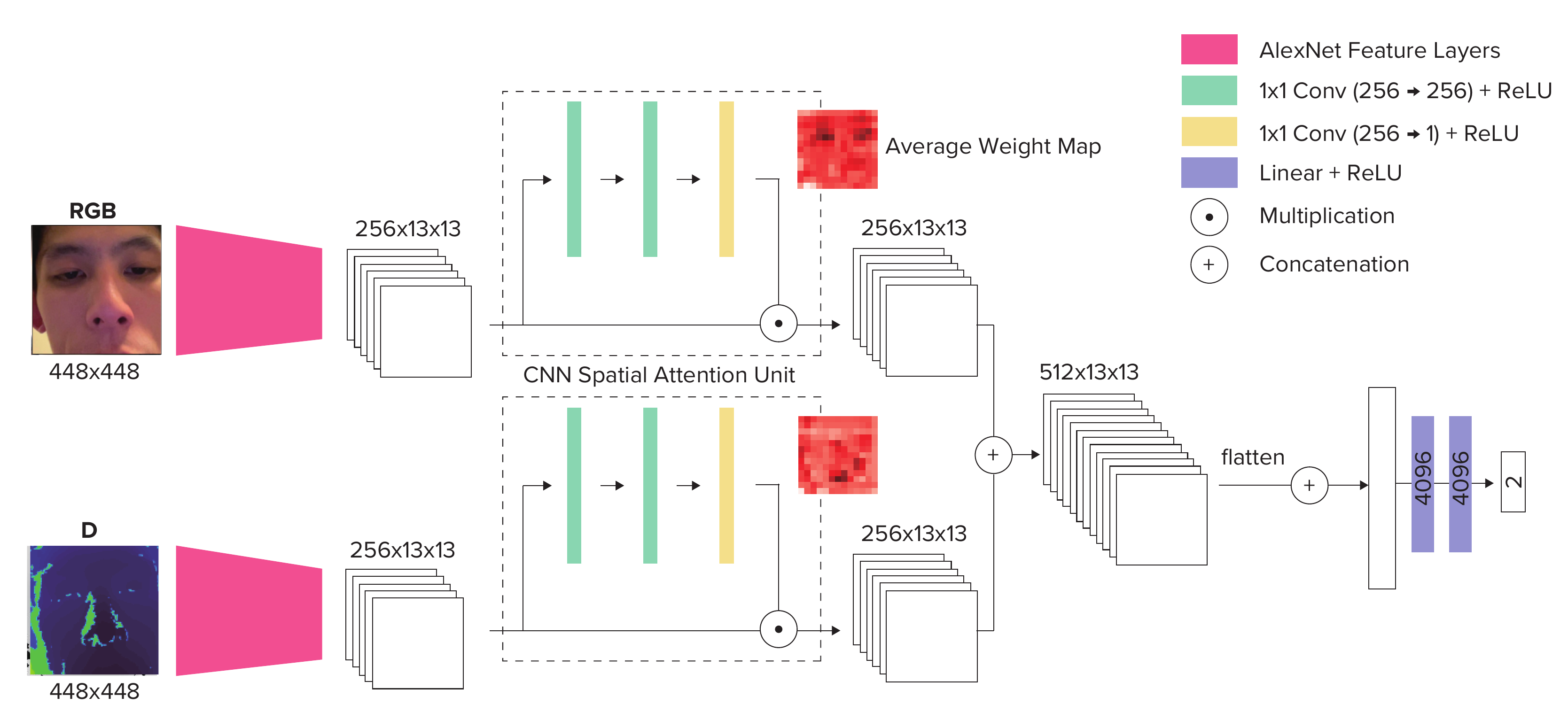}
    \end{center}
    \caption{Overview of my multimodal deep learning architecture. Input is the RGB and depth image of the user's face, with the 2D gaze location on the screen as output. The spatial attention maps of the corresponding RGB and depth maps are visualized in red heatmaps (darker color intensity denotes higher attention value).}
    \label{fig:nn}
\end{figure*}

\subsection{Implementation}

\subsubsection{Network Architecture}

I developed a multimodal learning-based method for estimating a user's gaze on a smartphone. I first crop the user's face using the Apple Vision Framework \cite{appleVision}.
The cropped face (448 $\times$ 448 pixels) in RGB and depth views serve as the input to my multi-input Convolutional Neural Network (CNN). The output is the predicted 2D gaze location in the screen coordinate frame of the smartphone. The overview of my CNN can be seen in \figref{fig:nn}. For my image-based feature extractor, I make use of spatial attention neural networks. This helps assign different information weights to different regions of the facial image. For example, it will automatically assign higher weights to the eyes or rather different parts of the eyes would be weighted differently based on their information entropy in the CNN. Prior work has found such approaches \cite{DBLP:conf/cvpr/ZhangSFB17} to be more accurate and computationally less intensive than those that crop different regions and feed them individually to a model \cite{DBLP:conf/cvpr/KrafkaKKKBMT16}. I warm-start my RGB and depth convolutional feature extractors (\figref{fig:nn} pink color) with AlexNet weights.
The embedding is then passed to three fully-connected layers with the ReLU activation, outputting the final two-dimensional gaze value.

\begin{figure*}
    \begin{center}
        \includegraphics[width=\textwidth]{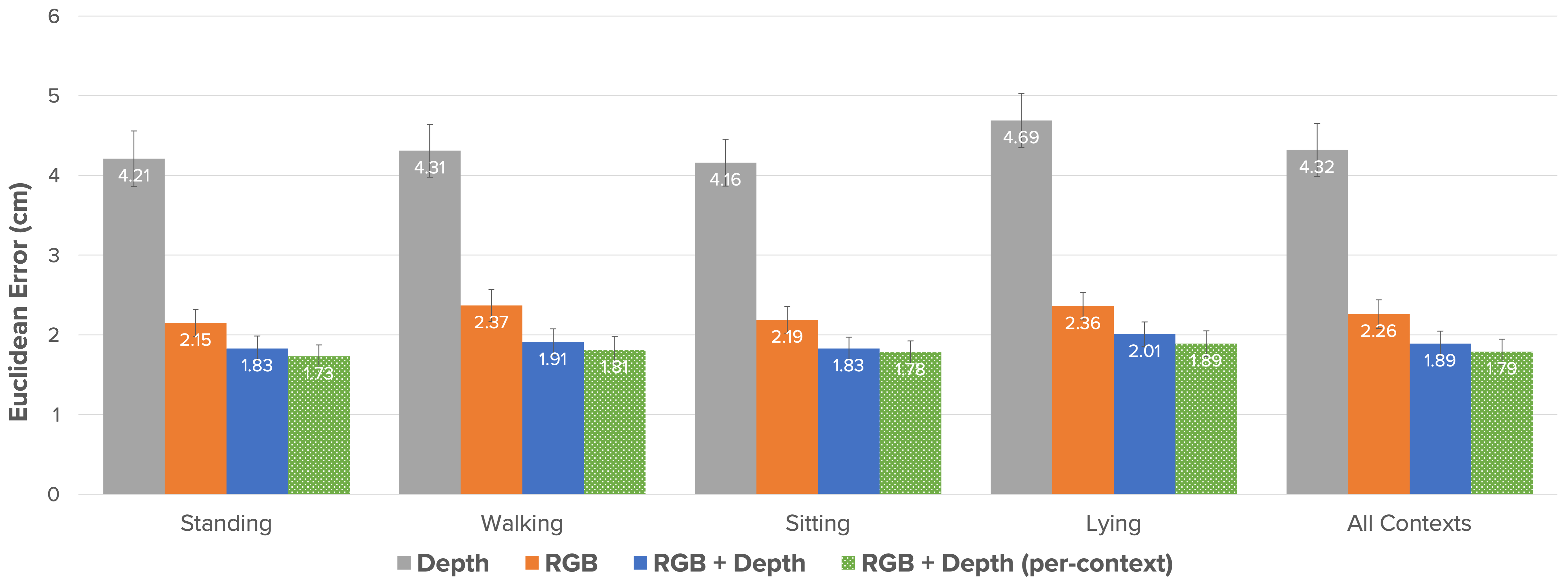}
    \end{center}
    \caption{Overall accuracy of my approach across different data input modalities (RGB and depth) and use contexts (sitting, standing, walking and lying). The error bars are standard error.}
    \label{fig:result-context}
\end{figure*}

\subsubsection{Training Protocol}
\label{sec:impl-protocol}

The model is implemented with PyTorch 1.9.1.
The RGB part of the model is first trained with the GazeCapture dataset while the depth part is initialized without any pretraining.
I use a batch size of 16 and update the model weights using the SGD optimizer with the initial learning rate 0.0005, the momentum 0.9, the weight decay 0.0001.
The learning rate is decayed by 0.1 for every five epochs.
I use the mean squared error between the ground truth and predicted gaze points as a loss function.
I train my model up to 20 epochs on an NVIDIA GeForce GTX 1080 Ti GPU, and it takes approximately 12 hours to train one fold in leave-one-participant-out procedure. 

\subsubsection{On-Device Model}
I deployed my model in a real-time iOS application. I first converted my trained PyTorch model to Core ML using Apple's CoreMLToolkit. A pair of synchronized RGB and depth frames are then pre-processed, which includes finding the face crop using Apple's Vision Framework \cite{appleVision} and normalizing the data between -1 and 1. This data is then sent to my CNN for prediction. On the iPhone 12 Pro Max, my model runs at 7~fps with an average latency of 121.3 ms (SD = 9.2 ms) from captured photo to gaze prediction. Using RGB alone, my system has a latency of 85.3 ms (SD = 7.4 ms) and runs at 10 fps. The tracking can run continuously for around 3.5 hours on the iPhone 12 Pro (battery capacity of 10.8 Wh). The model outputs a 2D gaze prediction, which is plotted on the screen. As a comparison point, Apple's Animoji feature, which digitizes people's faces and tracks their eyes, runs with a latency of around $\sim$110~ms on an iPhone 12 Pro \cite{ahuja2021pose}.
Similar to data collection, the predictions are paused when the face is not detected by the Apple Vision Framework \cite{appleVision}. Please refer to the Video Figure for a real-time demo.

\subsection{Results and Discussion}
In this section, I evaluate the efficacy of my multimodal model on my RGBD dataset. I first compute the performance metrics of my system across different input data modalities and use contexts, and then compare my system to RGB-based, state-of-the-art gaze tracking methods. 

\begin{figure}
    \begin{center}
        \includegraphics[width=\linewidth]{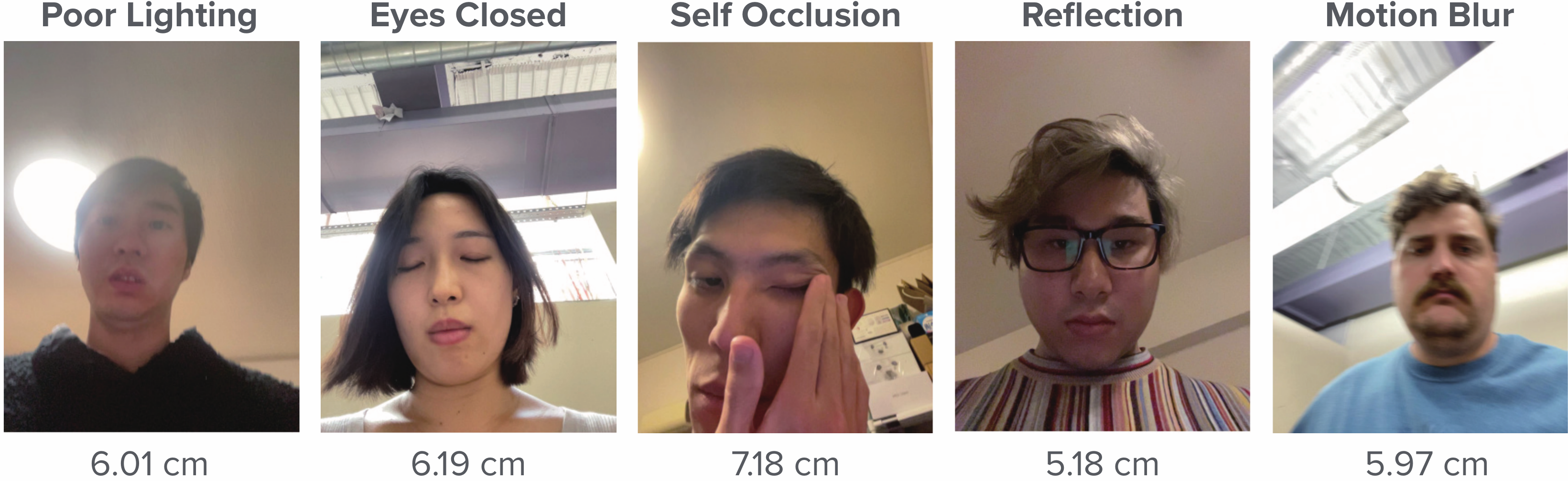}
    \end{center}
    \caption{Example images where my RGBD Gaze model had high error. Euclidean error is noted below each image.}
    \label{fig:error-analysis}
\end{figure}

\subsubsection{Overall Accuracy}

To evaluate the efficacy of my model, I follow a leave-one-participant-out protocol. The model is calibration-free, as no per-participant data is shared between the splits. Overall, my model achieves a euclidean gaze error of 1.89~cm (SD = 1.09~cm) when using RGBD data (\figref{fig:result-context} far-right chart).

Upon inspection, I find that my RGBD model has a very high error when the eyes are partially closed or occluded (by the user's hair, glasses frame, hands, \textit{etc}.). In these cases, it is impossible to resolve a full view of both eyes and the model falls back on head pose estimation for gaze. Other cases of error include poor lighting conditions, strong ocular reflections and motion blur. \figref{fig:error-analysis} showcases these sample error cases of the model.

To test the performance impact of each data modality, I trained the following model variants: depth-only, RGB-only, and RGBD. \figref{fig:result-context} summarizes this result. my RGB-only model has a euclidean error of 2.26~cm (SD = 1.27~cm), which falls to 1.89~cm (SD = 1.09~cm) with the addition of depth data. Anecdotally, upon visualizing the spatial attention maps of my model (\figref{fig:nn}), I find that the RGB attention model assigns a higher weight to the two eyes, thus focusing on eye gaze; while the depth model assigns a higher weight to the central region and edge of the face, thereby focusing on head pose.

\subsubsection{Effect of Use Context on Accuracy}
\label{sec:results-context}

Different use contexts result in different body postures \cite{ahuja2021pose} and face visibility \cite{DBLP:journals/mva/HuangVS17}, and also introduce artifacts such as motion blur. When I test my model across different use contexts (see \figref{fig:result-context}, blue bars) using the leave-one-participant-out protocol, I find that lying has the highest error (2.01~cm, SD = 1.18~cm) while sitting and standing have the lowest error (1.83~cm, SD = 1.10~cm and 1.83~cm, SD = 1.01~cm).
This is in line with prior work \cite{DBLP:journals/mva/HuangVS17} and can be attributed to the least facial visibility for the lying down context.
Compared to standing and sitting, walking has a relatively lower accuracy, that is, 1.91~cm (SD = 1.08~cm). 
This can be attributed to motion blur and head motion caused due to the movement of the smartphone in the walking scenario.

Prior work has successfully demonstrated the detection of different coarse body poses or activities (\textit{e.g.}, sitting, standing, running, walking, lying down) using smartphone motion data \cite{ahuja2021pose, casale2011human}. To quantify the performance effect of each use context, I tested the performance of a per-context calibrated model. Here I validate the model on data from only that particular context. Overall, this reduces error by 5.3\% (broken out by use context in \figref{fig:result-context}, green bars). 

\subsubsection{Comparison with Prior Work}
\label{sec:results-comparison}

A summary of comparative prior works can be found in Table \ref{tbl:rgb-dataset}. To the best of my knowledge, no prior work has made use of RGBD on smartphones for gaze estimation. I therefore benchmark my approach with two state-of-the-art RGB-based systems: Apple ARKit 4 ARFaceAnchor model \cite{appleARKit} and iTracker \cite{DBLP:conf/cvpr/KrafkaKKKBMT16}. Apple's ARKit provides an easy-to-integrate library for developers, and the iTracker model has been trained on a multitude of mobile devices (iPhone 4-6, iPad Pro, iPad Air). These two systems were run on the same data as my own model. ARKit and iTracker achieve a mean euclidean error of 6.38~cm and 2.77~cm, respectively. In contrast, my model, making use of RGBD, offered a much lower euclidean error of 1.89~cm. 

To test the efficacy of my RGB spatial attention model, I benchmark it on the GazeCapture test dataset utilized by iTracker \cite{DBLP:conf/cvpr/KrafkaKKKBMT16}. iTracker achieves an error of 2.04~cm (without any data augmentation) on their dataset. my RGB-only model achieves a similar error of 2.03~cm on the dataset (vs. 2.26~cm on my test dataset). This dip can be attributed to the varying use contexts and challenging capture scenarios of my dataset as well as the larger screen size of the devices used. I also find that my model achieved better accuracies compared to tablet-based gaze estimation works such as TabletGaze \cite{DBLP:journals/mva/HuangVS17} and EyeTab \cite{DBLP:conf/etra/WoodB14}, which reported an error of 3.17~cm and 2.58~cm respectively. Note that these comparisons are provided as a reference, as all these methods were tested on different datasets.

\subsection{Limitations \& Future Work}
While the accuracies of my system are promising, there are several key limitations that will need to be overcome before it is ready for commercial adoption. First is the accuracy of the system. Even with a calibration-free gaze error of under 2~cm, the accuracy falls short of the sub-millimeter accuracy afforded by dedicated eye trackers. In the future, this could be improved by collecting data across a wider array of mobile devices, scenes and users. The proliferation of depth cameras on smartphones and tablets (such as the Google Pixel 4 and the Apple iPad Pro) could help with training a more generalizable gaze tracking model.  

I also note that while the current contexts are encouraging and naturalistic, they can still be expanded. I can cover more situated contexts, for example, users interacting with the smartphone while driving, biking, or climbing stairs. Furthermore, rather than an experimenter conducting the study, I can increase the diversity of my dataset by crowd-sourcing the data collection (as done by Krafka et al. \cite{DBLP:conf/cvpr/KrafkaKKKBMT16} and Xu et al. \cite{DBLP:journals/corr/XuEZFKX15}). I believe that such a large-scale dataset consisting of RGBD modality can achieve high accuracy without per-user calibration, enabling practical gaze-powered mobile interactions.

\subsection{Conclusion}

This work explores the feasibility of head and eye gaze tracking on smartphones. I collected data from 50 participants across four use contexts and then trained a CNN model based on a spatial weights structure that can efficiently fuse my multimodal streams. Results demonstrate that my model offers improved accuracy, down to 1.89~cm euclidean error. While future work remains, this result suggests that RGB and depth information offers promise in enabling unconstrained mobile gaze tracking and could unlock a wealth of new and interesting end-user applications. 

\section{Hand Pose Estimation from Commodity Touchscreens}
\label{hands}


\subsection{Introduction}

Human hand interaction is a cornerstone of how we perceive and manipulate the world. Therefore, much effort in the research communities has gone into capturing and reconstructing hands during such manipulations to extract their spatial configurations with numerous applications in robotics, rehabilitation, Augmented and Virtual Reality.

In this work \cite{ahuja2021touchpose}, I focus on reconstructing hand poses during interaction with planar surfaces on the devices that we use on a daily basis: the screens of phones, tablets, laptops, and so on. I investigate the problem of recovering 3D joint positions from the spatial intensity map captured by the capacitive sensors that are integrated into such touchscreens. 

For this, I introduce \textit{TouchPose}, a deep learning model that estimates a fitting 3D hand pose based on a regressor that I devised and trained on the hand pose data captured from participants while they produced touch input on a surface with varying input postures.
I demonstrate that the hand regression model benefits from hard-parameter sharing multi-task learning by using a joint embedding space to concurrently estimate the corresponding depth image, the validity of the touch event, as well as the inferred hand pose in real time.

\begin{figure*}
\centering
    \centering
    \includegraphics[width=0.8\textwidth]{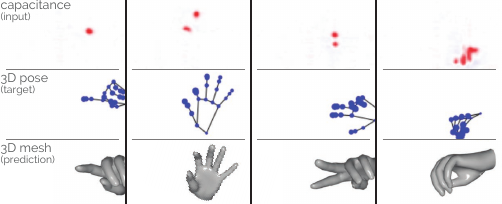}
    \caption[TouchPose sample predictions]{Top: Raw capacitive input images. Middle: Ground-truth hand poses for comparison. Bottom: Hand meshes rendered using skeletons that were estimated by TouchPose.}
    \label{fig:hand_mesh}
\end{figure*}

Unlike previous approaches that designed deep networks for \textit{individual} input parameters (e.g., touch classification~\cite{le2019investigating, chung2015quadratic}, finger angle~\cite{xiao2015estimating, mayer2017estimating}, inadvertent touch~\cite{schwarz2014probabilistic,hinckley2016pre}), TouchPose is a \textit{general-purpose model} that attempts to recover hand configurations independent of use-case (Figure~\ref{fig:hand_mesh}).
From each input frame, TouchPose produces 3D hand-pose estimates including finger classification, angle, and gesture, offering this information to UI developers for the purpose of interactive applications. Therefore, I see TouchPose as a use-case agnostic succession of existing work that trained custom networks with similar complexity on just individual use-cases.

Because TouchPose's training allows it to operate based on partial observation, i.e., only the parts of the hand that (almost) make contact, TouchPose can also infer 3D hand joints that lie \emph{outside} the touch-sensitive area as well as occluded points.
I show that TouchPose's reconstruction even generalizes to hand poses and orientations that the network has never seen before.

Collectively, TouchPose's contributions include:

\begin{itemize}
    \item a learning-based method to estimate 3D hand poses, depth maps and validity of touch events from capacitive images resulting from touch data on sensor surfaces.
    
    \item a dataset of capacitive touch images, aligned depth images, and annotated 3D hand poses, captured from 10 participants while touching the surface using various finger combinations, rotations, and angles.
    
    \item a series of interactive sample applications that are enabled by TouchPose's to support touch input and novel input techniques.
\end{itemize}

\subsection{Related Work}

TouchPose sits at the intersection of computer vision, graphics, and human-computer interaction. I start by reviewing the activities in the vision domain on 3D hand-pose estimation and then discuss how they influenced the research on interactive systems, particularly those building on optical and capacitive touch sensing.

\subsubsection{Hand poses from RGB and depth images}

Recovering hand pose configurations has been a long-standing problem in computer vision, often addressed using RGB and depth sensors.
Li et al.'s survey gives a comprehensive overview of methods and datasets in this context~\cite{li2019survey}.
Camera-based methods typically achieve higher accuracy compared to other sensing mechanisms given their high-resolution capture.
Prior approaches using depth sensors include model-based methods that use parameters to encapsulate physical constraints for the validity of results~\cite{oikonomidis2011efficient,xu2017lie, tang2015opening}.
In contrast, RGB-based hand pose estimation is more challenging due to depth ambiguity.
Yet, following the recent advances in deep learning, many recent projects have estimated 3D hand pose~\cite{mueller2018ganerated, spurr2018cross} and, more challenging even, 3D hand meshes \cite{ge20193d, zhou2020monocular} from single RGB images.

All vision-based efforts have in common that they start from higher-fidelity image data and reduce it to a 3D hand pose.
Analogous to frameworks that lift a pose from 2D to 3D~\cite{martinez2017simple}, wherein multiple plausible solutions exist for a given input, the problem I investigate is more under-constrained. TouchPose receives no sensory information from the whole hand and captures only subparts, which a considerably challenging and requires the method to extrapolate from sparse observations. 

\subsubsection{3D reconstruction from 2D touch imprints}

Numerous projects have investigated the problem of reconstructing the properties of objects above the touch surface from the contact they make.

Using the mutual-capacitance sensors in commodity touchscreen devices, researchers have leveraged the fact that they sense a small band of hover to predict finger poses.
Examples include Xiao et al.'s~\cite{xiao2015estimating} and Mayer et al.'s~\cite{mayer2017estimating} supervised methods to infer finger yaw and pitch angles that operate directly on capacitive images. 
Closely related to TouchPose, Chung et al. fit a hand model to the touch coordinates using a quadratic encoding~\cite{chung2015quadratic}, but rely solely on the location of touch points and do not make use of the fine-grained, higher dimensional data afforded by the capacitive imprints.
As a result, their method is affected by finger ambiguity when less than four fingers are in contact with the surface.

Through the use of self-capacitance sensing, prior work has been able to reconstruct input on as well as farther above the surface.
Rogers et al. devised a particle filter that simulated finger orientations based on the observations from a low-resolution sensor array to estimate finger motions and angles in 3D~\cite{rogers2011anglepose}.
PreTouch fuses such cues from surface and hover input to establish an anticipatory and retroactive hybrid touch model~\cite{hinckley2016pre}, showing a refined estimation of input intentions and locations.
Such hover sensing can also be used for detecting hand gestures above the surface~\cite{villar2018project} or to infer arm orientations and thus user position~\cite{zhang2019sensing}.

\subsection{TouchPose Dataset}

The TouchPose dataset consists of 65,374 samples collected across a total of 10 users for 14 different finger and whole-hand touch poses and gestures respectively. The data samples consisted of 72 $\times$ 41 capacitive image, along with corresponding depth map (captured via Kinect~\cite{microsoft_azure}) and hand pose (captured using Leap Motion stereo IR camera  - Ultraleap~\cite{leap_motion}) which were used as ground truth. The Leap returned a 3D hand skeleton with 21 joints: 4 for each of the 5 fingers and 1 for the wrist. As the Leap provides us only with hand skeletons, I collected depth data to aid in resolving the ambiguity of finger width (and, by extension, exact finger-touch points).
The capture of depth for the dataset was also motivated by potential future hand pose recognizers that could replace the skeletons from the Leap's low-resolution recording by analyzing the more accurate and higher-resolution depth frames.

\begin{figure*}
\centering
    \centering
    \includegraphics[width=\textwidth]{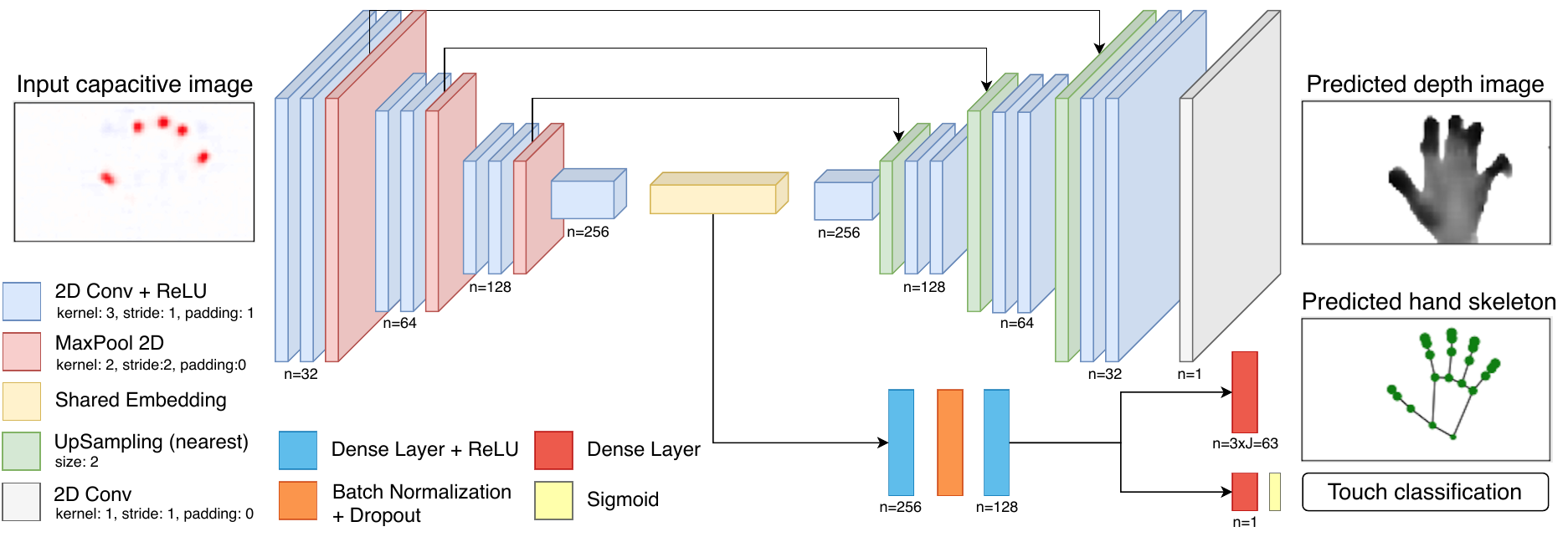}
    \caption[TouchPose multi-task CNN Architecture]{TouchPose CNN Architecture. It is a multitask learning framework that takes in a capacitive image as input and predicts the depth image, 3D hand pose, and touch classification for it.}
    \label{fig:arch}
\end{figure*}

\subsection{Multi-task Convolution Neural Network}

I propose a learning-based model that estimates the 3D pose of a hand $H$ touching a sensor surface based on the corresponding capacitive image $C$ as input. While it is intuitive that there exist multiple possible hand poses for a given input---especially in the case of few fingers touching the screen ---TouchPose  aims to recover the best fitting hand pose in a least-square sense.

TouchPose has two auxiliary tasks to improve the generalization of the model. The model is trained to not only predict the hand pose $H$ but also the corresponding depth image $D$ as well as the validity of the touch event. With these changes, the network now shares a common embedding for all three tasks, following an approach that is commonly known as hard-parameter sharing multi-task learning~\cite{ruder2017overview}. 

I model the capacitive frame $C$ and the depth image $D$ as real-valued tensors of dimensions $41\times72\times1$.
The capacitive image input frames are normalized between 0 and 1.
The hand pose $H$ consists of 21 joints, each represented by three Cartesian coordinates, and modeled as a real-valued tensor of size $21\times3$. The likelihood estimate $TC$ is a real-valued scalar, either 0 for fingertip events or 1 representing whole-hand events.

Figure~\ref{fig:arch} showcases the architecture of TouchPose. The multi-task CNN has 4,937,345 trainable parameters. During training, I use a batch size of 32 and update the weights using the Adam optimizer~\cite{kingma2014adam} with a learning rate of 0.001 and a decay of $5 \times 10^{-6}$. The network is implemented in Tensorflow and incurs an inference latency of 24\,ms per input.

As a final step, the 3D hand skeleton estimated by the multi-task CNN is fed to an inverse kinematic solver. This facilitates the rigging of a valid hand mesh (Figure~\ref{fig:hand_mesh}), reject outlier unnatural poses and allows TouchPose to continue animating the hands through sparse input periods. 

\subsection{Evaluation}

I outline three evaluation protocols to test the efficacy of TouchPose. 

\begin{itemize}

\item \textbf{Protocol 1:} I employ a 3-fold `leave-one-session-out' cross-validation wherein each fold consists of training data from two sessions and is tested on the remaining one. This is used to study the per-user effects (such as hand scale, geometry, etc.) on the accuracy.

\item \textbf{Protocol 2:} I use a 10-fold `leave-one-person-out' cross-validation wherein each fold consists of data from nine participants for training and the data from the remaining participant for testing. This introduces cross-subject variation in hand scales and pose styles.

\item \textbf{Protocol 3:} I use a 14-fold `leave-one-gesture-out' cross-validation wherein each fold consists of training data from 13 hand gestures and the remaining fold has a hand gesture that it has never seen before. This tests the generalizability of the model to unseen hand poses.

\end{itemize}

Unlike prior hand tracking methods that leverage global alignment, scaling or root-centric error calculations, I calculate all poses and errors in a touch-centric coordinate system.
Herein, the upper left corner of the touch panel is the origin.
This helps to account for global errors in orientations and different hand-scales without training custom models or calibrating for bone length. 

\subsection{Results}

\begin{table*}
\footnotesize
\centering
\begin{tabular}{l c c c c c c c}\hline
\textbf{EP} & \textbf{FingerID(\%)} & \textbf{Yaw Err ($^\circ$)} & \textbf{Pitch Err ($^\circ$)} & \textbf{EPE (mm)} & \textbf{EPE$_{v}$ (mm)} & \textbf{AUC} & \textbf{Depth Err (mm)} 
\\ \hline
1 & 91.1& 10.4  & 8.8 & 19.6  & 13.8   & 0.85& 20.7  \\
2 & 88.0 & 11.4 & 9.9& 21.8 & 15.8  & 0.83  & 22.2  \\
3 & 83.1  & 11.1  & 9.6 &  29.2  & 15.4  & 0.70  & 24.9 \\
\hline
\end{tabular}
\caption[TouchPose accuracy across evaluation metrics]{Quantitative results of different TouchPose variants across the evaluation metrics and protocols. Here, FingerID denotes the finger classification accuracies and EP stands for Evaluation Protocol.}
\label{tab:results}
\end{table*}

\subsubsection{Finger classification during touch}
\label{sec:fingerid}

I first evaluate the efficacy of TouchPose for classifying the finger that each fingertip interacting with the touchscreen belongs to. Under the cross-session cross-validation (Evaluation Protocol~1), TouchPose has the highest finger classification accuracy of 91.1\% (SD = 3.4\%), which falls to 88.0\% (SD = 10.4\%) for the cross-person scenario (evaluation protocol 2). Under Evaluation Protocol~3, where the model is tested on finger combinations it has not seen before, TouchPose has a mean accuracy of 83.1\% (SD = 13.3\%). 

\subsubsection{Depth Estimation from capacitive images}

While mutual-capacitance sensors are perceptive to hovering fingers up to 2--3\,mm, TouchPose substantially extends the possible depth range by two orders of magnitude to 255\,mm, albeit specific to fingers and hands by virtue of its learned model. The results of Evaluation Protocol~2 show that TouchPose's depth estimator achieves a mean error of 22.2\,mm across participants and all poses.
Of note, this error includes depth estimations that are \textasciitilde25\,cm from the surface and thus far beyond the sensitivity of any capacitive touch sensor. I further break down the analysis into hover bands above the surface for Evaluation Protocol~2. In the region of 10\,mm above the surface, TouchPose's depth estimation has an error of 4.2\,mm compared to the Kinect depth sensor (SD~=~0.7\,mm). In the 20\,mm band above the touch surface, this depth estimation error increases to 6.2\,mm (SD~=~0.8\,mm).

\subsubsection{Finger angle estimation during touch}

I evaluate the mean absolute error (MAE) in the yaw and pitch for fingers that are touching the surface. In summary, averaged across all evaluation protocols and fingers, TouchPose has a MAE of 11.0$^\circ$ for yaw and a MAE of 9.4$^\circ$ for pitch.

\subsubsection{3D hand pose estimation error}

To test the efficacy of the 3D pose pipeline, I make use of the following hand pose evaluation metrics:

\begin{itemize}
  \item \textbf{End-point-error (EPE):} This commonly used metric for 3D hand pose estimation is the mean Euclidean error between all the joints (= 21) of the Leap and predicted hand pose. 
  
  \item \textbf{End-point-error of visible joints (EPE$_{v}$):} Similar to EPE, but only computed for the finger joints touching the screen. For example, if only the index and ring finger are touching the screen, I will only compute the EPE for all the joints along those fingers. Hence, its calculated for fingertip events only.
  
  \item \textbf{AUC under PCK:} Area under the curve (AUC), which represents the percentage of correct 3D keypoints (PCK) of which the Euclidean error is below a threshold $t$, where $t$ ranges from 20\,mm to 50\,mm.
\end{itemize}

Similar to the previous finger classification and angular error-based results, cross-session cross-validation (EPE mean=19.6\,mm, SD=2.7\,mm) produces much lower errors than cross-person cross-validation (EPE mean=21.8\,mm, SD=7.1\,mm), showcasing that per-user data aids in performance. The higher standard deviation when validating cross-person can be attributed to participants' differing styles how they touched the surface, especially the pose of fingers that were not in contact.

The error of the visible joints (EPE$_{v}$) is more uniform across participants, with a mean $EPE_v$ of 15.8\,mm (SD=3.8\,mm). For the auxiliary tasks, TouchPose achieves a touch classification accuracy of 99.58\% and depth reconstruction mean absolute error of 22.2\,mm under Evaluation Protocol~2. Evaluation Protocol~3 proves to be the most challenging for TouchPose, with the model having its highest mean EPE of 29.2\,mm.

In all cases, as expected the EPE$_{v}$ is smaller than the EPE, showcasing fingers that touch the screen are correctly predicted and less ambiguous than the ones that do not. The EPE also decreases proportionally to the number of fingertips touching the screen, decreasing from 18.3\,mm for one finger to 12.0\,mm for four fingers touching the screen (reduction of 6.3\,mm).
This was expected---at the moment a user touches the surface, the degrees of freedom of the hand pose decrease (the touch locations limit the 3 degrees of freedom for the respective end effectors).
Therefore, as the fingers make contact with the surface, the more constrained the search for the whole hand. 

\subsection{Demonstration Applications}

\begin{figure*}
\centering
    \centering
    \includegraphics[width=\textwidth]{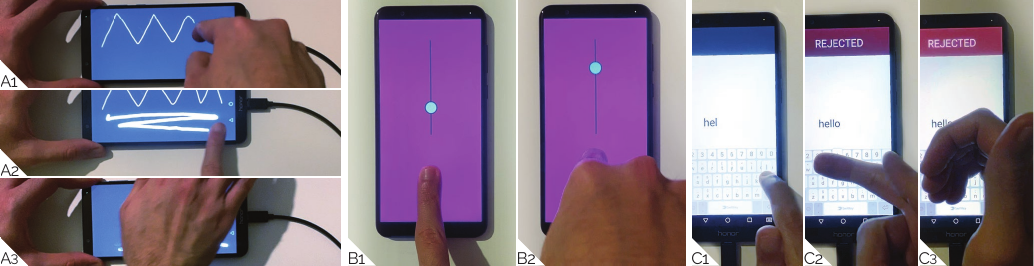}
    \caption[TouchPose demonstrative applications]{TouchPose demonstrative applications. (A)~This finger- and hand part-specific drawing app (A1--2)~infers stroke width from attack angle or (A3)~smudging. (B)~This 3D joystick is quick to operate through finger angle. (C)~Inadvertent touch rejection allows (C1)~regular typing but rejects (C2)~fingers at wrong orientations or (C3)~other hand parts.}
    \label{fig:apps}
\end{figure*}

TouchPose can be used as a general-purpose model that performs independent of use-case, providing hand poses (and by extension finger ID, angle, gesture) to UI developers. I showcase a few of the many possible interactive applications TouchPose can enable as shown in Fig.~\ref{fig:apps}. 

Using the capacitive touchscreen on an Android phone (Honor~7X, 15\,cm display, capacitive input offloaded and processed following prior approaches~\cite{le2018infinitouch}), I illustrate examples that make use of TouchPose's unique capability to recover hand and finger poses as well as implicitly identify the part of the hand that is in contact with the screen.
Fig.~\ref{fig:apps}A shows a drawing app using this to select the input tool and parameterize the stroke width according to the finger angle in pitch and yaw, all with a single input with feedback displayed in real-time.
TouchPose can also power applications to afford quicker input to UI controls, such as the slider in Fig.~\ref{fig:apps}B that adjusts based on finger pitch, making TouchPose a suitable processing layer for modeling applications on mobile devices where surface area is scarce.

A final app, which could also prove useful most immediately, processes input events into parts of the hand and finger to assess their suitability for the given input operation.
Unlike current text editing apps on touch devices, the app shown in Fig.~\ref{fig:apps}C incorporates TouchPose for transparent processing of input events, letting (C1)~valid type events triggered by fingers pass while rejecting inadvertent input caused by (C2)~invalid parts of the finger or palms and (C3)~sides of the hand.
Inadvertent touch detection is particularly enticing as touchscreens lack the capability of distinguishing between touch types.
This problem may be even more severe on larger-screen devices such as tablets or tables, where users tend to rest their palms or wrists while providing input or while writing with a stylus to support accurate input.

\subsection{Limitations}

While TouchPose demonstrates feasibility, there are several limitations that will benefit from further refinement.
In its current form, it cannot handle multiple hands interacting simultaneously. I also acknowledge that the current implementation cannot disambiguate finger classes in single-touch events.
TouchPose recovers ``partial'' hand poses, as there is inherent ambiguity in missing contacts. However, I note that as more fingers make contact, ambiguity drops and finger identification confidence improves. In future, higher-resolution touchscreens in the future and improved touch-sensing ranges~\cite{hinckley2016pre} will provide more 3D cues on the shape of the hands. This will help alleviate single-touch ambiguities and resolve hovering fingers. Lastly, TouchPose is enabled by a multi-task convolutional neural network-based architecture. While it could directly run on smartphones (e.g., using Tensorflow Lite), it produces computational overhead compared to standard touch event detection pipelines that are designed for low latency.

\subsection{Conclusion}

I have presented the first general-purpose estimator for hand pose recovery using capacitive images from a touch surface alone.
TouchPose performs this task by implementing a multi-task architecture, predicting depth maps, hand poses and validity of touch events.
For data acquisition, I devised a capture rig that integrates a mutual-capacitance touch sensor, a short-range depth camera below, and a stereo IR camera above for hand pose labeling.
10 participants provided touch events using various hand poses and combinations of fingers, touching, sliding, and gesturing on the touch surface while the apparatus captures synchronized frames from all sensors.
The final dataset comprises 65,374 pairs of capacitive images, depth maps and annotated 3D hand poses and served for training and testing the TouchPose model.
TouchPose estimates 3D hand poses from just the intensities and imprints left on the surface, reaching an average end point error of 21.8\,mm across the hand joints, including those that lie outside the touch area.
In addition, TouchPose generalizes to hand gestures it has never seen before. Taken together, I believe that TouchPose will help future developers to effortlessly attach interactive behavior to input \textit{semantics}, advancing from today's na\"ive notion of touch input as 2D coordinates and interpreting input in the context of dexterous hand pose.

\label{TouchPoseChapter}

\section{Holistic Full-body Digitization using Smartphones}
\label{whole-body}


\subsection{Introduction}

While prior works only focus on reconstructing particular body parts (such as hands or head), the human body is far more expressive and offers many degrees of freedom for human-computer input purposes. Technologies able to digitize a user’s full-body could enable new interactive experiences beyond the touch-centric (and occasionally IMU-driven) input that I see on contemporary mobile devices.

Today, full-body motion capture is most closely associated with computer-generated imagery in blockbuster films, using expensive multi-camera rigs and special suits with markers. However, as technologies have improved, consumer-oriented uses have become possible. For instance, there are now several companies offering small sensors, worn on the body, that digitize the wearer’s pose for use in more immersive VR experiences \cite{Antilatency, Vive}. Of course, the bar for consumer acceptance is high, and this highly-instrumented approach seems unlikely to go mainstream in the near future. A decade ago, Microsoft took a different approach with its XBox Kinect sensor \cite{microsoft_azure}, a \$150 accessory depth camera that could capture users’ pose without any worn instrumentation. A variety of interactive, pose-enabled games proliferated, crossing genres including sports, dance, and role-playing games.

Regardless of whether the sensors are worn or external, the necessity for extra devices, plus the added cost of that hardware, dampens the likelihood of mass adoption. More importantly, both approaches preclude many interesting uses of body digitization when people mobile and outside of controlled settings. In response, I set out to develop a full-body pose estimation system that could run entirely self-contained on a smartphone held normally in one’s hand. My system can work on the go, offering new avenues of interactivity \textit{anywhere} and without prior setup. For this reason, I call the system \textit{Pose-on-the-Go}. 

Achieving this vision required leveraging almost every sensor at our disposal in modern smartphones, including the front and rear cameras, user-facing depth camera, capacitive touchscreen, and IMU. I fuse data from these disparate sensors to rig a real-time, animated skeleton of the user as they operate their phone. Pose-on-the-Go is the first system to demonstrate full-body pose estimation using an unmodified smartphone held in the hand. This affords Pose-on-the-Go wide applicability and superior practicality over other methods, which almost all require special instrumentation. An additional contribution is a rigorous study, benchmarking against a true gold standard - a professional-grade Vicon optical tracking system.

I believe exposing live user pose (even a coarse approximation as I demonstrate) as an API on mobile devices could enable some very creative and novel interactive experiences. For example, one of the example applications is a 3/4 perspective space shooter where the user’s virtual on-screen character matches their live body pose, offering a unique level of embodiment not previously seen in smartphone gaming. Indeed, a significant benefit of being software-only is that many recent smartphone models could be enabled via an over-the-air update, and the software could run as a background service on top of which developers could build pose-enabled apps.

\subsection{Implementation}

\begin{figure}
  \includegraphics[width=0.9\columnwidth]{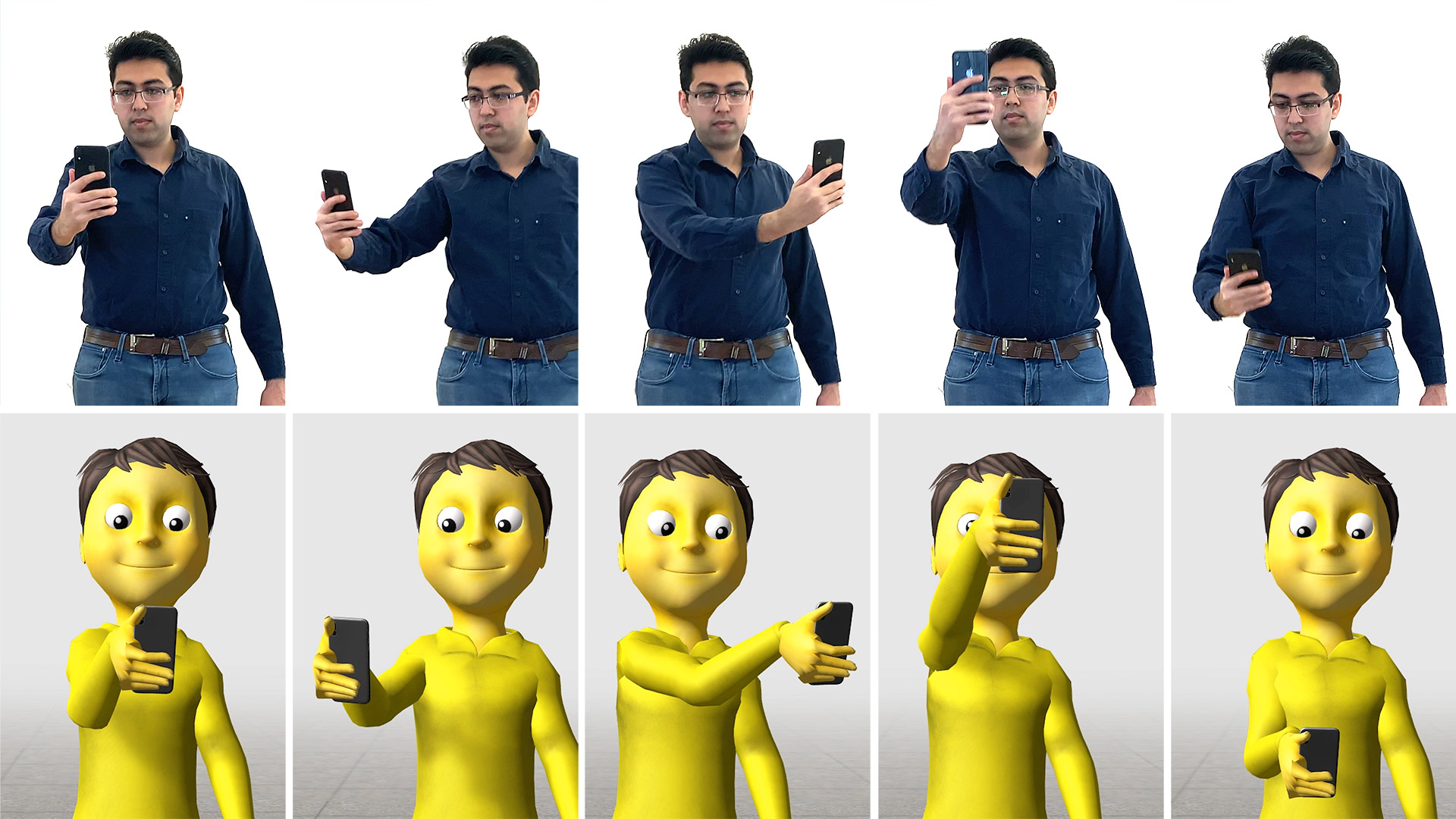}
  \caption{Pose-on-the-Go tracking a user’s arm pose and eye gaze.}
  \label{fig:fig_arm}
\end{figure}

I built the proof-of-concept implementation on iOS, which has an API that allows both the front and rear cameras to be streamed simultaneously (a feature now available on some Android smartphones, and likely to be mainstream soon). iOS also offered a robust set of APIs across all of the features I wished to implement (face tracking, pedometer, locomotion mode prediction, 6-DOF absolute spatial tracking, etc.), and many of the heavier-weight APIs are optimized to take advantage of hardware acceleration, offering us some computational headroom. As a development device, I selected an iPhone XR, which is Apple's mid-tier offering. 

\subsubsection{Inverse Kinematic Model \& 3D Engine}

There are many inverse kinematic (IK) SDKs available, both open source and commercial. I chose to use Root-Motion’s VRIK library~\cite{root_motion} running on the popular Unity engine~\cite{Unity}, which has native iOS support and an extensive set of tools and assets for creating example apps. As with the hardware, Pose-on-the-Go does not have any strong dependencies and should work with practically all human IK packages. To facilitate rapid prototyping, I wrote a thin smartphone app responsible for interfacing with all sensors and streaming data to a MacBook Pro laptop (3.1 GHz Dual-Core Intel i5, 16 GB RAM) over WiFi, where the Unity-based implementation, along with the IK solver ran. With additional engineering, it should be possible to run the entire process on the smartphone.

\subsubsection{Head Position \& Orientation}

The first step of the process is to establish the position and orientation of the head relative to the phone. For this, I use the user-facing camera and ARKit’s ARFaceAnchor API~\cite{ARFaceAnchor}, which offers 6-DOF head tracking. Note this does not use the iPhone’s user-facing depth camera.

\subsubsection{Eye Gaze}

The same ARFaceAnchor API \cite{ARFaceAnchor} also provides an estimate of the gaze vectors for each eye. Although too inaccurate to enable gaze targeting on the phone’s screen, it is more than sufficient to realistically animate the eyes of the avatar, providing another dimension of expressivity (see Figure \ref{fig:fig_arm}).

\subsubsection{Torso Orientation}

While ARKit comes with advanced, built-in functionality to track heads, it offers no facilities to track the user’s lower body. Critically, body orientation needs to be established as it moves independently from the head. To capture chest \textit{yaw}, I use the iPhone XR's user-facing depth camera to exact two bilaterally-symmetric patches on the user’s torso and compute a chest yaw and pitch vector. 

\subsubsection{Phone Orientation}

Using its IMU, the phone tracks its absolute 3-DOF orientation, establishing both north and down (i.e., gravity vector). I use this data to animate the avatar's hand holding the phone. More importantly, by combining this 3-DOF data with the aforementioned 6-DOF tracking of the user's head, I can now correctly orient the head with respect to the world, from which I can ``hang'' the rest of the user’s body.

\subsubsection{Arm \& Hand (Holding Phone) Pose}

At this stage, I know the relative spatial arrangement between the phone, head and torso. I assume the phone is held in one hand, and that an arm links the two (Figure \ref{fig:fig_arm}). As the elbow and wrist joints are rarely seen in the user-facing camera view, I instead use the IK solver to generate a likely arm pose, articulating the avatar’s elbow to connect the two points (i.e., shoulder to phone). As briefly noted in the last section, I also take into account phone orientation to articulate the wrist joint. Finally, I assume the phone is held in a standard grasp, and so I pose the avatar’s fingers to match.

\subsubsection{Arm \& Hand (Not Holding Phone) Pose}

For the arm not holding the phone, there is very little data to operate on. Although the shoulder is often visible in the user-facing RGB and depth cameras, once can almost never see the elbow, wrist, or hand. Fortunately, it is not uncommon for users to employ their other hand for input on the touchscreen. In such cases, I can use the cartesian touch screen location of the finger, in combination with the 6-DOF location of the phone relative to the torso, to pose a plausible arm. Furthermore, TouchPose (see Chapter~\ref{TouchPoseChapter}) can then be employed on the capacitive image to estimate the full-hand pose.

\subsubsection{Absolute ``World'' Position}

So far, I have only discussed the relative spatial arrangement between body parts and orienting the body with respect to the gravity vector. Such data alone would allow Pose-on-the-Go to provide an upper body pose, locked to a frontal view (e.g., fixed to the chest normal, allowing the head to look around). With the addition of absolute 6-DOF ``world'' tracking, several more expressive dimensions can be enabled.

One of the main reasons I selected iOS as the development platform was ARKit’s best-in-class absolute 6-DOF tracking. Apple’s hardware-accelerated, inside-out sensing implementation combines both visual odometry using the rear camera and data from the iPhone’s IMU. This allows the phone to track its movement and position in 3D space. Since Pose-on-the-Go positions the head relative to the phone, the torso relative to the head, the arms relative to the torso (and so on), I can use the phone's spatial data to not only translate the avatar accurately in space, but also rotate the avatar to match the direction the user is facing.

\subsubsection{Locomotion Mode \& Leg Animation}

The 6-DOF location of a user's body allows to animate avatar locomotion, despite having no direct sensor data for the legs. To achieve this, I translate the avatar's upper body and run VRIK's locomotion solver, which contains animation support for bipedal locomotion. Stride length is a static parameter, and so to solve for different movement speeds, the leg animation is simply sped up or down, though this can produce unrealistic results. Additionally, small shifts in the user's posture and occasional position tracking errors even when the user is standing still can cause the IK solver to take errant steps. 

To improve pose and animation quality, I leverage iOS's CMMotionActivity API \cite{CoreMotion}, which provides the following locomotion mode predictions: stationary, walking, running, cycling, automotive and unknown, along with a confidence score. I found this prediction to a reliable filter, verses relying on motion alone to animate the legs. When the phone reports that a user is stationary, I do not animate the legs, except when turning the body. When walking, I set the stride length to that of a typical walk, while running requires a longer gait. By setting these parameters appropriately to a user's anthropometrics, a more realistic animation is achieved.

Although this process offers only a very coarse approximation of leg movements, when taken together with the full-body avatar, the output is reasonably convincing. Anecdotally, I found users to be quite forgiving – it may be that people do not attend to their absolute leg position (e.g., while walking) at the same level of scrutiny as their arms or head, where spatial errors are immediately commented upon. 

\begin{figure}
  \centering
  \includegraphics[width=0.9\columnwidth]{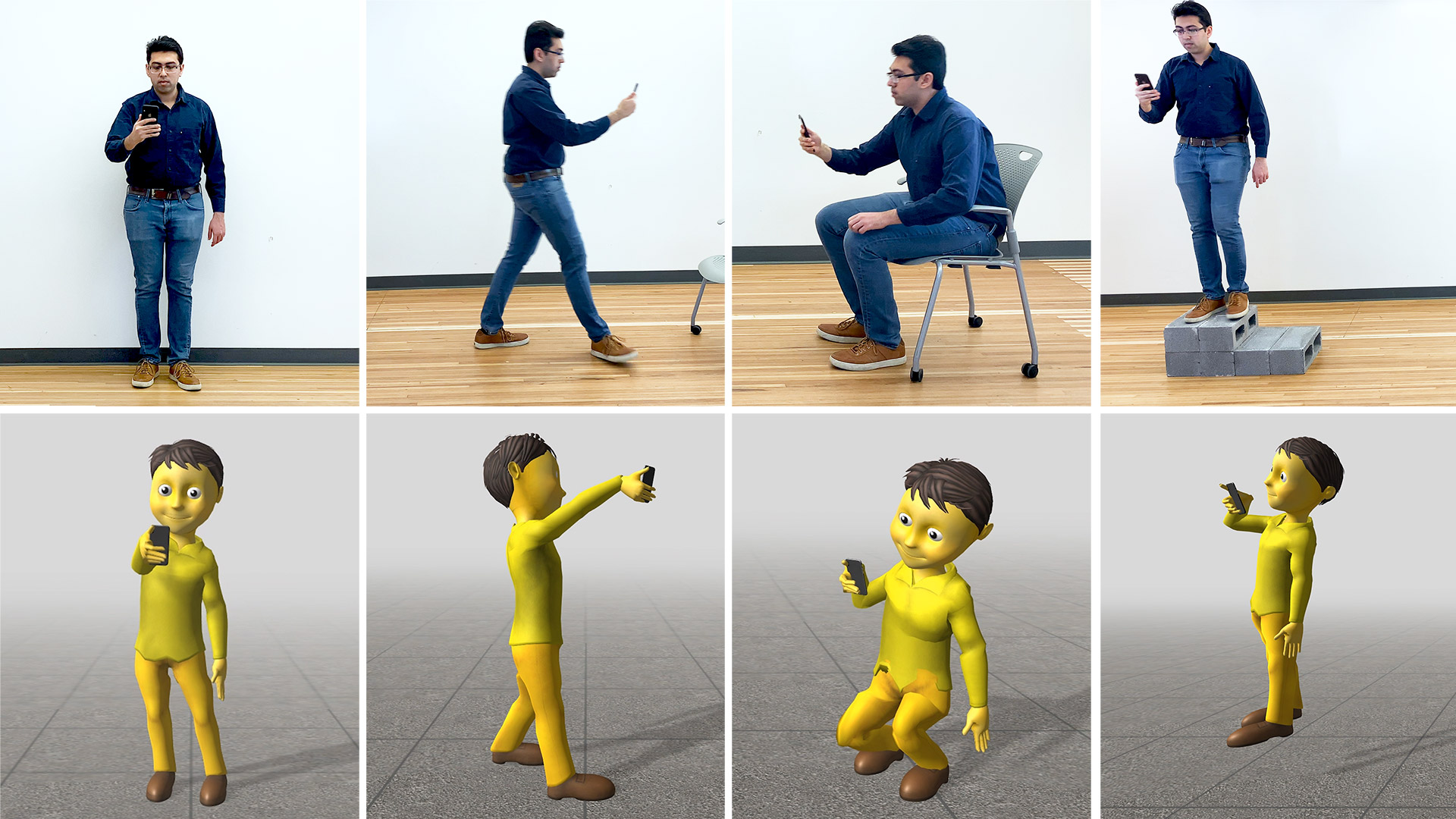}
  \caption[Pose-on-the-Go tracking a user’s leg pose and locomotion]{A user standing (far left), walking, (center left), sitting (center right) and standing on a box (far right) is mirrored by the avatar.}
  \label{fig:fig_movement}
\end{figure}

\subsubsection{Sittings/Standing \& Body Height}

Locomotion animation primarily leverages a user's X/Y translation in an environment. However, the Z-axis (e.g., elevation) is useful in detecting and representing other poses, such as sitting and stepping onto objects (Figure \ref{fig:fig_movement}). As I have absolute 6-DOF position tracking, I  simply calculate if the head is above or below its starting height, and correspondingly rig the avatar with this constraint. In cases where the user is now higher than standing, I can infer the user has stepped up onto an object and show the legs leaving the floor (Figure \ref{fig:fig_movement} far right). If the user is lower, the IK solver bends the legs constrained by the floor plane (Figure \ref{fig:fig_movement} center right). In general, this process is pretty crude as there are many ways to e.g., sit, squat, and kneel. 

\subsubsection{Data Synchronization}

The system implementation is highly asynchronous, with a wide variety of sensors and processes running at different framerates. Most notably, the head tracking pipeline using the front camera runs at 15.5FPS, the depth camera-driven torso normal extraction runs at 11.8FPS, and the phone's 6-DOF spatial tracking runs at 19.9FPS. I set other low priority processes, such as reading the finger touchscreen position, to 5FPS. All of this data is passed to the IK solver running in Unity, which is set to run at 60FPS, matching the screen refresh rate. 

Fusing many data streams of varying framerates requires some strategy for effective alignment. For the prototype implementation, a pool of threads asynchronously receive and store the most recently received frame of data for each sensor. The main process loop, which runs at best effort given available CPU and GPU resources, simply uses the most recent data reported by all sensors. Although simple, I found this approach to be sufficient for the proof-of-concept system. An approach taking into account time-since-data-received could potentially extrapolate better live estimates of sensor values, and yield higher quality and lower-latency pose output.

\subsubsection{Framerate, Latency and Power Draw}

As noted above, Unity (which runs IK, animates and renders the avatar) runs at 60 FPS, though the underlying sensor data arrives at a variety of frequencies. The best measure of end-to-end latency is ``motion-to-photon", which is the time taken between a user's motion and when the device's pixels reflect that change. For this, I used a 240 FPS camera to record the user and screen in the same scene. I found an average motion-to-photon latency of 358ms (SD=63ms). The implementation is very much a proof of concept, with minimal optimization, and thus these numbers should be viewed as a performance upper bound. For reference, Apple's Animoji feature \cite{Animoji}, which digitizes people’s faces as characters, runs with a latency of around \textasciitilde $150$ms on an iPhone XR. I believe comparable performance should be achievable, certainly on future smartphones, which continue to make significant strides in compute power.

Pose-on-the-Go requires many sensors to be running, including three different cameras, which is energy-intensive. To quantify this, I measured the power draw with the iPhone XR on its home screen and also running the Pose-on-the-Go daemon. I found a power consumption delta of \textasciitilde $5.3$W, which includes running all sensors and system processes like ARKit and CoreMotion. In practice, I found the iPhone XR could run continuously for \textasciitilde $2$h, which closely matches what one would expect from the iPhone XR’s 11.12Wh battery rating. Note that this power consumption number does not include running IK (not particularly intensive) or application graphics (highly variable depending on the app), which will add further burden.

\subsection{Evaluation}

I performed a series of studies, benchmarking Pose-on-the-Go against a professional-grade optical tracking system, which serves as a gold standard ground truth. Specifically, I used a Vicon system~\cite{Vicon}, with twelve MX40 cameras and four T160 cameras capturing at 120FPS. Vicon’s software and Pose-on-the-Go produce equivalent 3D joint data (14 joints: head, torso, left shoulder, right shoulder, left elbow, right elbow, left wrist, right wrist, left hip, right hip, left knee, right knee, left foot and right foot), but in different coordinate systems. For analysis, I perform an extra post processing alignment step, designating the torso of both pose outputs as the body origin. I recruited 8 participants for the user studies (2 female) with ages ranging from 18 to 39 years (M=27.6 years); all were right-handed.

\subsubsection{Head Orientation}

I asked participants to face a wall 1.5m away with four markers arranged in a rectangle. I asked participants to hold the phone in front of them in a natural position and perform the following head movements five times in a row, using the markers as a positioning guide.

\begin{itemize}
    \item Pan left and then right across the middle of the markers.
    \item Pan up and then down across the middle of the markers.
    \item Pan clockwise around the perimeter of the markers.
    \item Pan counterclockwise around the perimeter of the markers.
    \item Pan across the markers in a “figure eight” pattern.
    \item Roll the head to comfortable extremes. 
\end{itemize}

These motions allowed to evaluate the performance of Pose-on-the-Go's estimation of head yaw, pitch and roll. Across all participants, Pose-on-the-Go had a mean yaw, pitch and roll angular error of $6.4^{\circ}$ ($\textit{SD}=3.0^{\circ}$), $5.4^{\circ}$ ($\textit{SD}=1.6^{\circ}$) and $10.7^{\circ}$ ($\textit{SD}=4.0^{\circ}$) respectively.

\subsubsection{Torso Orientation}

In this task, I asked participants to rotate their chest left-to-right and back again (i.e., yaw), while trying their best to keep the phone held in the same position. This motion was completed five times per participant. From this data, I found Pose-on-the-Go deviated from the Vicon ground truth by a mean angular error of $15.9^{\circ}$ ($\textit{SD}=10.1^{\circ}$). 

\subsubsection{Arm Pose (Hand Holding Phone)}

To gauge the accuracy of arm joint tracking, I asked the participants to perform the following motions with the smartphone five times each:

\begin{itemize}
    \item Move the phone away from the body, and then towards.
    \item Move the phone left-to-right and then right-to-left.
    \item Move the phone up and then down.
    \item Move the phone clockwise in a \textasciitilde $50$cm circle.
    \item Move the phone counterclockwise in a \textasciitilde $50$cm circle.
    \item Move the phone in a “figure eight” \textasciitilde $50$cm tall.
\end{itemize}

With the data from this task, I computed the spatial error for participants’ wrists, elbows and shoulders. I found that Pose-on-the-Go had a mean 3D euclidean error of 18.0cm (SD=3.0cm) across all joints. Broken out by joint, wrists had the greatest error of 27.4cm (SD=4.7cm), followed by elbows (M=17.0cm, SD=3.9cm), and shoulders (M=9.7cm, SD=2.1cm). Unsurprisingly, error increases as the IK solver tries to estimate joints farther along bone linkages (here, the torso is the body origin). 

\subsubsection{Hand Orientation (Hand Holding Phone)}

I asked the participants to hold the phone in their dominant hand and rotate it left-to-right (yaw), up-and-down (pitch), and twist it while keeping the screen facing them (roll), five time each using only their wrist (as best possible). From this data, I calculated a wrist angular joint error yaw, pitch and roll of $11.5^{\circ}$ ($\textit{SD}=2.3^{\circ}$), $9.1^{\circ}$ ($\textit{SD}=3.3^{\circ}$), and $8.9^{\circ}$ ($\textit{SD}=4.9^{\circ}$), respectively.

\subsubsection{Arm Pose (Hand Not Holding Phone)}

I asked the participants to use their non-dominant hand to touch the screen five times in a row, simulating typing, and then drop their arm to a resting position. This process was repeated five times, for a total of 25 screen taps. On this data, Pose-on-the-Go achieved an average 3D euclidean error of 20.9cm (SD=5.1cm), 13.4cm (SD=4.4cm), and 6.6cm (SD=1.2cm) for the wrist, elbow and shoulder joints respectively.

\subsubsection{Leg Pose (Sitting/Standing)}

In this task, I placed a chair and 25cm tall pedestal into the Vicon tracking area, 2.5m apart. I asked participants to sit in the chair for a few seconds, then stand up and walk over to the pedestal, and finally to step onto the pedestal. This procedure was repeated five times and provided a wide variety of poses to evaluate the quality of leg posing.

There was no significant difference between left and right leg accuracy, and so the results are combined. I found a mean 3D euclidean error of 8.9cm (SD=1.6cm), 14.2cm (SD=2.3cm), and 23.7cm (SD=3.1cm) for the hips, knees and feet respectively. This follows a similar trend as the arms, where the IK solver gets increasingly errorful father along bone linkages.

\subsubsection{World Tracking}

All of the analyses discussed so far have reported results in body coordinates (torso as the origin). However, Pose-on-the-Go also tracks and animates avatars moving around the world, and so I can evaluate the fidelity of absolute “world” position tracking. For data collection, I asked participants to walk the following paths continuously for 20 seconds each (with the aid of floor markers).

\begin{itemize}
    \item Walk forwards 2m, then backwards 2m (repeatedly).
    \item Sidestep left 2m and then right 2m (repeatedly).
    \item Walk in a circle (2m diameter).
    \item Walk the perimeter a $2\times2$ meter square.
    \item Walk in “figure eight” pattern in a $4\times2$m area.
\end{itemize}

I use the avatar’s torso point for analysis. Across all walked paths and participants, I found a mean 3D world euclidean error of 11.2cm (SD=5.7cm). Across X, Y and Z axes individually (Z is up), the mean euclidean errors are 11.6cm (SD=6.0cm), 15.2cm (SD=9.5cm) and 6.9cm (SD=3.9cm) respectively. 

\subsubsection{Head Height Tracking (Z)}

The “world” tracking procedure described in the previous section did not vary the height of the user (other than natural variation during locomotion). For this reason, the height of the head varied by \textasciitilde $20$cm over the entire data collection period. To more fully evaluate head height, I instead use the data collected in the previous Leg Pose (Sitting/Standing) study, where users were asked to sit and step onto a pedestal. Using this data, but now with absolute tracking, I found a mean head euclidean Z-axis error of 9.6cm (SD=4.6cm). 

\subsubsection{Locomotion}

I use the data from the body “world” tracking study to estimate the fidelity of Pose-on-the-Go’s locomotion animation. Rather than spatially compare every step taken, which I previously noted can be 180 degrees out of phase, I instead evaluate the movement sequence holistically to see if the avatar is locomoting in a faithful way. For this, I compare the number of steps the user took in reality vs. number of steps taken by the Pose-on-the-Go avatar. Across all participants, the average step count error is 6.3\% (SD=2.74\%). In other words, if a user takes 100 steps, their avatar will have been animated taking 6.3\% more of fewer steps on average.

\subsubsection{Obstacle Course}

All of the prior studies focused on different sets of joints in purposely designed tasks. To test the efficacy of the full-body motion capture system, I conducted a less constrained study where the participants were asked to walk in a $4\times2$m area in a “figure eight” repeatedly. During this I asked them to look at different targets randomly and occasionally asked them to sit down, stand on the pedestal and touch their screens (akin to an ``obstacle course" methodology, see e.g., \cite{bao2004activity, bao_obstacle}. I then calculated participants' full body 3D euclidean error and also their world position error. 

In this task, the average Euclidean error is 20.9cm (SD=2.6cm) across all participants and all joints. I found that the legs were out of phase 39.8\% of the time (which I corrected for when computing leg joint error). The world tracking accuracy broken out by X, Y and Z axes is 30.5cm (SD=17.7cm), 32.8cm (SD=19.0cm) and 10cm (SD=3.3cm) respectively. 

\subsection{Example Uses}

To illustrate the utility and feasibility of Pose-on-the-Go, I selected three use domains – gaming, social apps, and health – and created two proof-of-concept demo apps for each category. 

\begin{figure}
  \includegraphics[width=\columnwidth]{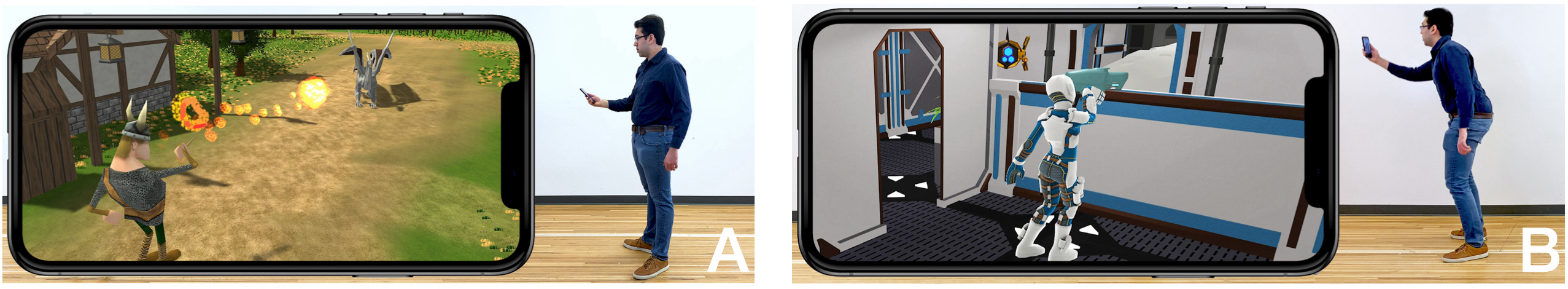}
  \caption[Pose-on-the-Go Example Application: Gaming]{A) A fantasy game, where users can fight with swords and cast spells using arm movements. B) A futuristic third-person shooter, where a user can physically run and duck, while using their phone to aim and shoot a laser weapon.}
  \label{fig:game}
\end{figure}

\subsubsection{Gaming}

The first demo application I created was a third-person fantasy game (Figure \ref{fig:game} A). The user can select among different weapons by tapping on the screen, which are controlled through arm motion. For example, a sword can be swung in the air, using the phone as a proxy for the hilt. I also included a magic wand, where various spells can be cast by tracing different paths in the air (e.g., a circle to cast a fireball). In this example app, walking forwards/backwards and turning left/right could be controlled on the phone using up/down and left/right swipes on the phone’s screen, allowing the user to stand or sit without moving their body. 

For the second game –- a third-person futuristic shooter –- I utilize the user’s actual absolute body position and pose (Figure \ref{fig:game} B), allowing them to physically walk around the virtual environment, and also duck behind cover to avoid enemy fire. In this game, the phone is a proxy for a handheld laser, allowing the user to both aim and shoot with their arm.

\subsubsection{Social Apps}

There are many smartphone apps that capture a user’s face and digitally transform or augment it for social purposes, such as Apple’s Animoji \cite{Animoji} and Snapchat's Lenses \cite{SnapLenses}. The latter software is also an example of full-body AR augmentation, through this app augments other users (i.e., captured through the rear facing camera) rather than the holder of the phone. Such AR augmentation and ``avatarization'' has value in both entertainment and professional collaborative contexts \cite{benford1995user, brown2004cscw, piumsomboon2018mini}. With Pose-on-the-Go, I can immediately extend Animoji-style face capture with full-body versions. As an example, I created a bear avatar that can wave, hug, jump, and make a wide variety of other expressive, full-body motions (Figure \ref{fig:animoji} A). 

\subsubsection{Health}

\begin{figure}
   \includegraphics[width=\columnwidth]{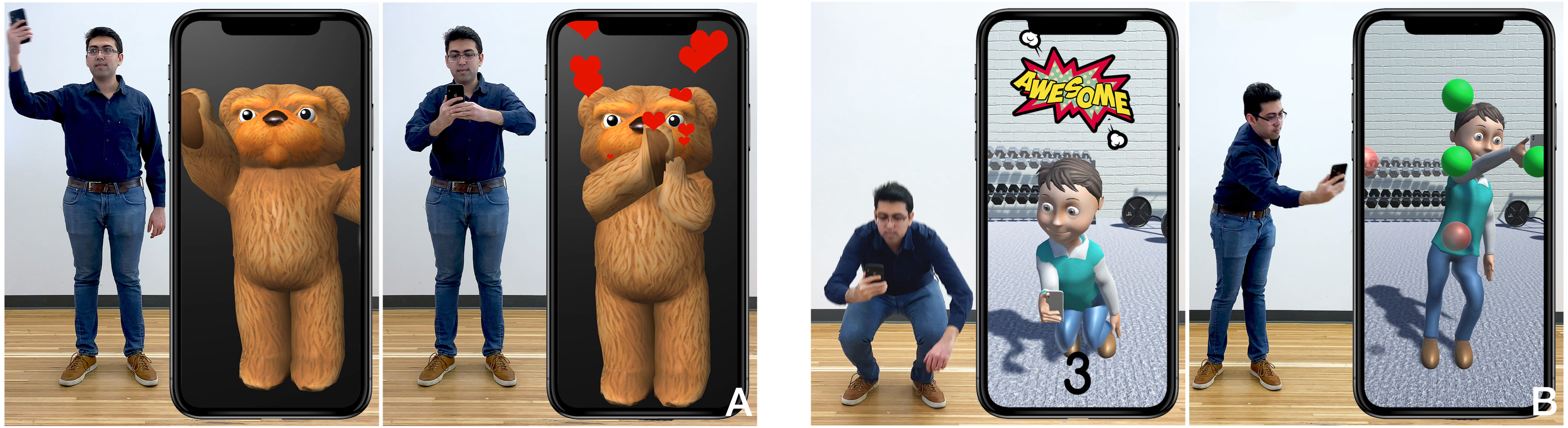}
   \caption[Pose-on-the-Go Example Application: Social Avatars and Health]{A) Full-body “Animoji” enabled by Pose-on-the-Go, where users can wave (left) and hug (right). B) A tool for health sensing such as counting squats (left) and reaching targets for physical rehabilitation (right).}
   \label{fig:animoji}
\end{figure}

Finally, there are many examples in fitness and rehabilitation where capturing a user's pose could be especially valuable. As a proof of concept, I created two applications. The first is a fitness app (Figure \ref{fig:animoji} B), where exercises like squats and lunges could be counted or timed, as well as evaluated for quality (e.g., squat height, lunge distance). As a second example, I created an app that presents a series of targets that the user must reach with their hands and maintain balance (Figure \ref{fig:animoji} B), which could be part of a larger physical therapy regiment. 

\subsection{Limitations}

It is important to note that Pose-on-the-Go is only an estimation of full-body pose, and it not meant to compete with high-accuracy tracking systems (e.g. Vicon), which are used in film post-production and similar uses requiring high fidelity. Indeed, this is the inherent tradeoff that comes with building a system that requires no new, special or external sensors, and instead attempts to maximize use of the data already available to it. Nonetheless, Pose-on-the-Go's accuracy should be sufficient for a wide range of casual gaming and social apps. That said, there are several areas where the system falls short, which are worth reviewing, pointing to potential future work.

First is incomplete data about the body, most notably the arm not holding the phone, which is almost never digitized except when the user touches the screen. This potentially could be remedied with the advent of wider field-of-view, user-facing cameras, which might better capture the elbow, or at least the upper arm with respect to the shoulder joint. The hand also might occasionally appear. The other body location I can only loosely approximate are the legs, animated based on absolute body motion and fixed stride lengths derived from a locomotion mode prediction. This is sufficient to animate the avatar, but is not true body capture. For this reason, walking animations can be totally out of leg phase from reality. Other joints, such as the hips and knees are entirely estimated by the IK solver, and are essentially an interpolation between other known pose data. 

Latency is another limitation of the current implement. At \textasciitilde $350$ms, it starts to degrade the realism of the full-body tracking. Experiences would have to be designed to accommodate such latency, an issue also faced in many Xbox Kinect titles. In particular, the arm not holding the phone can only start animating once a finger touches the screen and has lag closer to one second (when animation is applied) with no immediate avenue for improvement. Of course, even commercial systems have some lag – Apple's Animoji feature, which also fuses data from RGB and depth user-facing cameras, has a latency of \textasciitilde $150$ms. 

Another limitation is computational complexity and power demand. This will require significant optimization efforts, but I do believe it is possible as demonstrated by Animoji-type features found on many smartphones today, as well as pose tracking of other users (via the rear camera) in e.g., PoseNet~\cite{Papandreou:2018}, SnapChat~\cite{SnapLenses}, and ARKit~\cite{ARFaceAnchor}. These processes are also heavyweight, but have been highly engineered and take advantage of hardware acceleration in order to run at interactive speeds. Of course, Pose-on-the-Go has to contend with more sensors, including all three iPhone cameras, which impacts battery life when running applications utilizing full-body pose features. 

\subsection{Conclusion}

I have presented Pose-on-the-Go, a sensor fusion approach that allows smartphones to estimate their owners' full-body pose using only internal sensors. I benchmark the pose tracking output against a ``hollywood-grade'' optical tracking system requiring retroreflective markers. While only a coarse estimation of body pose, it nonetheless opens up new and interesting whole-body applications, ranging from mobile AR games to more expressive social and collaborative interactions.

\label{chapterPOSEOTG}

\section{Towards Multi-User Digitization in Mobile XR Experiences}
\label{multi-user}


\subsection{Introduction}

\begin{figure*}
  \centering
  \includegraphics[width=\textwidth]{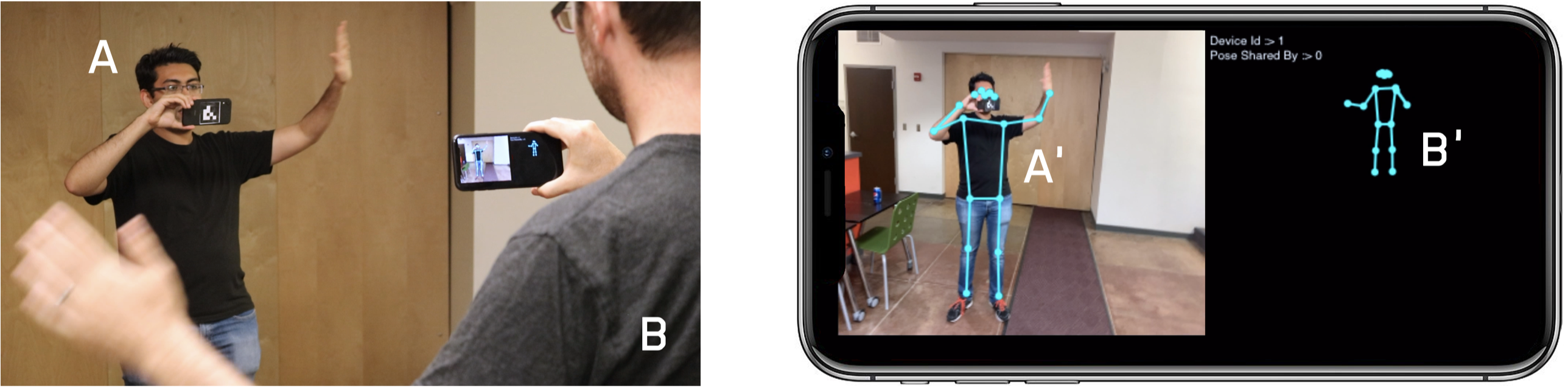}
  \caption[BodySLAM running on smartphones]{In this example scene (left image), two users (A and B) face one another, perhaps playing a mobile AR game where full-body tracking could be valuable for expressive input. Unfortunately, neither phone is able to see – and thus digitize – its owner’s body. However, User B's phone (right image) can see User A (and vice versa) through its rear facing camera. BodySLAM uses this view to capture and digitize the body, hands and mouth of User A and shares the data (visualized as A’). User A does the same for User B, providing ad hoc full-body tracking (B’) without having to instrument either the user or environment. }
  \label{fig:bodyslam_teaser}
\end{figure*}

My previous pose capture approaches such as Pose-on-the-Go focused on the digitization of a single user. On the other hand, many prior approaches, such as GymCam in Section \ref{chaptergymcam} make use of an external camera for tracking multiple users. Here I will explore the concept of multi-user digitization, but without the need for an external sensor. To this extent, I present BodySLAM - a system takes an orthogonal approach and tackles making multi-user digitization (Figure \ref{fig:bodyslam_teaser}) more practical. To do so, it takes advantage of an emerging use case: co-located, multi-user AR/VR experiences. Although nascent, the application space is already diverse, ranging from co-located 3D modeling \cite{Benko:2004, Smisek:2013} and AR-augmented collaborative spaces \cite{facebookVR, Schmalstieg:2002}, to tele-medicine and multi-player games \cite{Alexiadis:2011, Billinghurst:2000}. 

BodySLAM makes it possible to have shared and ad-hoc extended reality experiences in dynamic environments without the need for any additional hardware or setup. In such contexts, participants are often able to see each other’s bodies, hands, mouths, apparel, and other visual facets, even though they generally cannot see their own bodies. Using the existing outwards-facing cameras on smartphones and AR/VR headsets (e.g., Microsoft HoloLens, Oculus Quest), these visual dimensions can be opportunistically captured and digitized, and then relayed back to their respective users in real time (enabling e.g., full-body motion capture and hand gesture recognition) without any special instrumentation. This is the key insight that motivated the work on BodySLAM. 

\begin{figure*}
  \includegraphics[width=\textwidth]{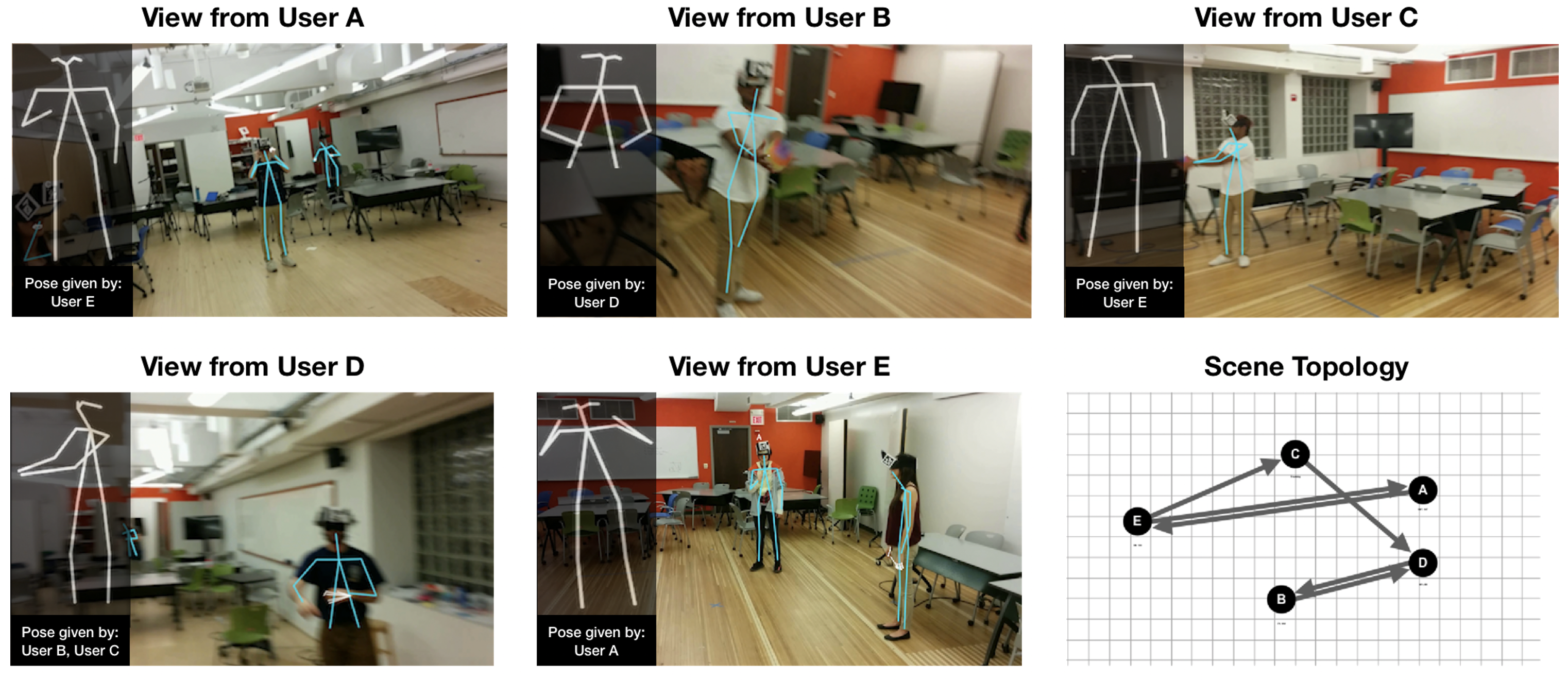}
  \caption[BodySLAM system overview]{Example camera views (A-E) from co-located users wearing AR headsets, which BodySLAM uses to digitize the body pose, hand gesture, mouth state and apparel of participants. This data is then relayed back to users (shaded insets), offering advanced functionality without the need for worn instrumentation. BodySLAM also generates a scene topology (bottom right).}
  \label{fig:system}
\end{figure*}

The system name was inspired by SLAM (simultaneous localization and mapping) techniques for mapping unknown environments \cite{Cadena:2016, Dissanayake:2001}. In these systems, many viewpoints are used to reconstruct the geometry of the environment. In a similar vein, BodySLAM uses disparate camera views from many participants (Figure \ref{fig:system}, User views A-E) to map the geometric arrangement of other users in the environment (Figure \ref{fig:system}, bottom right), as well localize the capturing user's position in the scene. the system also captures fine-grained details, such as hand and mouth pose, as well as visual attributes such as apparel. When a person is seen by two or more users (Figure \ref{fig:system}, views of User D from Users B and C), I also estimate 3D pose data.

\subsection{Related Work}

There has been significant prior work in motion capture and reconstruction of scenes. I have already covered prior works in body digitization in Section \ref{sec_pose_rel_work}. Thus, I will now cover prior works on multiview geometry and similar reconstruction techniques, concluding with efforts in the domain of creating shared AR/VR experiences.

\subsubsection{Scene Reconstruction}

There has been a plethora of research in the robotics and computer vision community on scene and multiview reconstruction. These include the aforementioned simultaneous localization and mapping (SLAM) based techniques that help place objects in an unknown environment \cite{Dissanayake:2001}. Visual SLAM approaches can rely on multiview stereo \cite{Cadena:2016}, visual odometry \cite{Nister:2004} and structure from motion \cite{Koenderink:1991}. Similarly, BodySLAM uses multiple views to reconstruct user topology, and multi-view stereo to estimate 3D pose information.

\subsubsection{Collocated AR/VR}

The HCI community has lightly explored collocated AR/VR experiences through different enabling technologies. For instance, Side-by-Side \cite{Willis:2011} used handheld projectors (emitting both visible and infrared light) to create multi-user experiences, such as gaming. SynchronizAR \cite{Huo:2018} demonstrated collaborative gaming through a smartphone AR app with indoor localization. HoloRoyale \cite{Rompapas:2019} provides a comprehensive design space exploration of large-scale high-fidelity collaborative AR experiences. Commercially, Apple’s ARKit allows for shared AR experiences, where many users can join using their own smartphones. Finally, Imaginary Reality Games \cite{Baudisch:2013} creates virtual playgrounds for participants to play sports. 

\subsection{System Architecture}

I now describe the three high-level components that form the basis of BodySLAM. 

\subsubsection{Exemplary AR/VR Prototypes}

I created three exemplary prototypes that cover different AR/VR modalities. These include an iPhone XR smartphone for mobile (i.e., ``passthrough") AR, a HoloKit headset \cite{HoloKit} for AR (center), and a Google Cardboard \cite{Cardboard} for VR (right). I fitted each of these devices with printed 7x7 cm ArUco tag \cite{Munoz:2012} for spatial tracking. For the two headsets, unique tags are placed on all four sides for user identification from any viewpoint. For the mobile AR prototype, a single tag is placed on the back of the phone.

\subsubsection{Client Software}
\label{clientsubsec}

BodySLAM employs a client-server paradigm. I created two client implementations, with different approaches for computation. First is a smartphone app that streams camera data using RTSP over WiFi to our backend server for computer vision processing. Although this architecture can leverage greater CPU and GPU resources, it incurs a latency penalty. On average, it takes 110 ms for video to reach the server, followed by 154 ms of processing, and a further 10 ms for processed data to be returned to the client. In total, our system latency is roughly 300 ms at a frame rate of 9 Hz.

The second client implementation runs all computer vision on the smartphone. For this, I use PoseNets \cite{Papandreou:2018}, a lightweight pose estimator that reports body keypoints. Although missing hand and face tracking capabilities, it nonetheless offers a useful proof-of-concept for a more self-sufficient system. ArUco detection and tracking also runs locally, using JSArUco \cite{JSArUco}. Processed data is then sent to the server for final processing (e.g., 3D pose via stereo correspondence) and distribution to other clients. End-to-end latency is around 150ms, and the full stack runs at ~12Hz on an iPhone XR (see demo in Video Figure). Given the tremendous strides phone manufacturers are making with hardware accelerated deep learning, a fully featured and full framerate pose tracking pipeline should be possible in the near future. 

For the VR client software and the two AR clients, I sync the overlaid pose of in-view people with the gyroscope of the smartphone, so that the latency of the real-time view (limited by camera FPS in the VR and Mobile AR implementations) is decoupled from the FPS of pose tracking. For demonstration, the interface is chiefly a passthrough view, on top of which I superimpose the wearer’s body, hand, face, skin and apparel if provided by others.

\subsubsection{Backend}

the backend (i.e., cloud) server is a 12-core Intel Core i7 with three Nvidia 1080 Ti GPUs. The server software is a multi-threaded listener, which listens for connections and data from clients. The server maintains global state and multicasts it back to registered clients. For the AR and VR headset prototypes, the client smartphones stream camera frames over WiFi to the backend for computer vision processing. For the mobile VR prototype, all computer vision processing happens on the smartphone and only pose data is transmitted to the server.

\subsection{Processing Pipeline}

I now describe the core processes that form the main computer vision and machine learning pipeline.

\subsubsection{Identifying and Localizing Users}

The first stage of the computer vision pipeline is to find and track participants in each camera view. As noted in Apparatus, users are assigned unique ArUco tags. For the headsets, four tags are used, which identify the user and the side of the head. The front-facing ArUco tag is considered the user’s origin, and the left, right, and rear tags have known rotational and spatial offsets to this origin. If multiple tags from one user are visible, I use the tag most-frontal to the viewer, as this provides the most stable tracking result. In the self-contained mobile AR client I use JSArUco for tracking; for the backend computer vision pipeline I use OpenCVs ArUco marker implementation \cite{OpenCV}. In both cases, I use a time-to-live of one second to provide stability against momentary losses in tracking (e.g., occlusion, motion blur). The result of this process is the relative 3D position and orientation of every user that is seen by others, with wearers as the origin.

\subsubsection{Scene Topology}

As users cannot see themselves, it means everyone is given their position and pose data by someone else. This pairwise information is used to create a directional graph, with distance and angle information. Note that users that see no one can still be added to the graph if at least one person sees them. Also, a user that is not seen by anyone can still be added to the graph if they see at least one person, as show-cased in Figure \ref{fig:system}.

To create a global scene topology, the software finds the most complete graph with a depth-first search. Although links are directional in reality (i.e., one person sees another), the origins can be inverted to make them functionally bi-directional. I move from one person to another, building a unified 3D coordinate system by multiplying individual user’s transformation matrices as they are added. This way the whole operation is limited to the order of O(V+E) where V is the number of people and E is the number of people seen by each person.

\subsubsection{Body, Hand and Mouth Pose}

Once participants are localized in a common coordinate system, the next step is to extract fine-grained body details. For this, I use PoseNets \cite{Papandreou:2018} on the mobile AR client and OpenPose \cite{Cao:2017} running on the backend server for the worn VR and AR clients. Both packages provide body keypoints, however only OpenPose provides hand and face keypoints.

\subsubsection{2D Pose via Multi-Viewpoint Selection}

If a user is seen by exactly one person, I must make best use of this limited keypoint data. Unfortunately, this view is rarely frontal, and so I attempt to transform it to a frontal view using the algorithm in Kostrikov et al. \cite{Kostrikov:2014}, which estimates 3D keypoints given 2D keypoints. With this estimated 3D output, I can rotate to a synthetic frontal view. 

In multiuser settings, where a user might be seen by several people, the software selects the most frontal of the views for this transformation. Although other views might contain more tracked keypoints, I found in practice that the more frontal the view, the truer the pose following rotation.

\subsubsection{3D Pose via Stereo Correspondence}

In cases where users are seen by two or more people, I can estimate 3D body pose. For this, I take the pair that minimizes the reprojection error of the 3D pose back onto the image. For each selected pair, I move all the cameras to a homogeneous coordinate frame using the 6 DOF data from users’ ArUco tags. I then run a 3D point triangulation \cite{Hartley:1997} to estimate the 3D position for each body keypoint. As before, I rotate this view to be frontal when presenting the data to the wearer in the client app.

\subsubsection{Skin and Apparel Color Estimation}

For skin color, I extract patches from the neck, hands and lower face (which are the least likely to be occluded by clothing) and compute the median color. For shirts, I take an image patch between the hips and torso keypoints, and similarly compute the median the color. For pants, I extract a patch from above the left and right knees of a user.

\subsubsection{Hand Gesture Recognition}

As a proof-of-concept of a downstream use of pose data, the system performs real-time hand gesture recognition. I support five gestures: okay, thumbs up, high five, fist and peace sign. For machine learning features, I calculate unit direction vectors from all hand keypoints to the wrist. These converts the data from a higher-dimensional pixel-space to a scale invariant feature space. These are fed to a standard multiclass classifier (MLP; sklearn, default parameters).

\subsubsection{Mouth State Recognition}

I also perform mouth state recognition, supporting neutral, smile and mouth open gestures. Similar to the hand gesture process, I compute unit direction vectors for all mouth keypoints using the left mouth corner as the origin. As before, I pass this feature vector to a MLP classifier (sklearn, default parameters) for prediction.

\subsection{User Study}

To better evaluate the performance of BodySLAM in different use contexts, the study procedure varied the activity (static vs. dynamic), distance between participants (2 or 4 meters) and group size (one-on-one vs. small group). As a proof-of-concept apparatus I used the Holokit AR prototype, so that participants could see each other and BodySLAM data. the backend software saved processed camera frames for later evaluation. All classifiers were pre-trained on five independent users and ran live during data collection.

\subsubsection{Static Activity}

I chose a meeting context as an exemplary static activity. In the one-on-one condition, this was equivalent to a standing, face-to-face discussion. For this, I recruited three pairs of participants, and had them repeat the study procedure at 2 and 4 meters apart. At each distance, I asked both participants to perform the following actions sequentially, in a random order:

\begin{itemize}
    \item \textit{10 body gestures}: hands by side, right hand raised, left hand raised, arms crossed, hands behind head, arms stretched horizontally, hands on hips, sitting down with arms at rest, sitting down with arms crossed, and sitting down with chin resting on hands.
    \item \textit{5 hand gestures}: okay, thumbs up, high five, fist and peace sign.
    \item \textit{3 mouth states}: neutral, smile and mouth open.
\end{itemize}

Participants saw their live pose and gesture classification results in the AR overlay, which they judged and verbal stated to be correct or incorrect, which was recorded by the experimenter. Participants pairs repeated the two distance conditions three times each, for a total of six collection sessions. 
For the small group condition, I mimicked a conference room setting. I recruited two groups of 5 participants, who were seated around a table. Participants completed the same 10 body gestures, 5 hand gestures and 3 face gestures using the same procedure above.

During the study, BodySLAM also captured shirt, pant and skin color. After completing all sessions, I had users select their own skin color on a printed Fitzpatrick scale \cite{Sachdeva:2009}, which I later compared to BodySLAM’s estimate. To assess the quality of shirt and pant color estimation, the extracted colors were shown to participants, who judged it as either accurate or inaccurate.

\subsubsection{Dynamic Activity}

As an exemplary dynamic activity, I used a ball game (Figure \ref{fig:system}), where participants were told to achieve the highest number of consecutive passes without dropping the ball. After fitting participants with the Holokit AR prototype, I let them play for five minutes, during which BodySLAM ran continuously on all headsets. Distance between participants varied as they moved around the space, as did their head direction and pose, especially when having to pick up dropped balls. For the one-on-one condition, I recruited three new pairs of participants to play the game. For groups, I recruited two sets of five players.

\subsection{Results}

BodySLAM tracks many user dimensions, and as such, I break the discussion of the results into four parts: user registration in the scene topology, keypoint tracking, gesture recognition, and skin/apparel extraction. As noted previously, all classifiers were trained before the study, and thus all accuracy numbers reported here are cross-user results.

\subsubsection{User Identification and Tracking}

Across all conditions, the average percentage of participants captured and registered in the scene topology was 95.2\%. In the one-on-one conditions, both static and dynamic activities, registration was 100\%. Even when one user had to bend down to pick up a dropped ball, the other user was able to capture them and add them to the global scene topology. Surprisingly, this performed better than the static small group condition (89\%). In the latter setting, users were sitting fairly close, and often only captured one or two other participants in the camera’s field of view. Depending on the group foci, this sometimes led to disjoint graphs (e.g., two participants looking at each other, but not seen by anyone else). The group ball game had the worst registration performance (87\%), which I found was chiefly due to high motion blur causing ArUco tracking to fail.

\subsubsection{Body, Hand and Mouth Keypoints}
\label{keyptsubsec}

I considered capturing 3D ground truth keypoints for users' bodies, hands and mouths using a professional optical tracking system. This required dozens of markers to be worn by the participants, which was cumbersome to setup and prone to breakage in the ball game condition. I also found that the markers interfered with the computer vision pipeline, especially on the hands and mouth. I instead decided to avoid instrumenting the participants and use human annotators to post hoc code the live tracking output from the system.

For the one-on-one conditions, I extracted one frame every second (image with keypoints overlaid), yielding 1184 frames. For the small group conditions, I randomly sampled 1000 frames. These frames were split equally between two annotators, who coded each frame as correctly or incorrectly registered to all users visible in the view. A correct registration required 1) more than 80\% of visible keypoints in the frame to be detected and 2) that all keypoint centers intersected with their respective body joint. Overall registration accuracy was 95\%, 93\% and 83\% accuracy for body, hand and mouth keypoint registration. 

\subsubsection{Self-Assessed Body Pose Quality}

As described previously, participants were shown their body pose (provided by other participants via BodySLAM) in their AR headsets. In the static activity conditions, participants were asked to perform one of 10 possible body poses. The experimenter then verbally asked, “does the pose you see on your screen match the pose you are holding.” In 94\% of instances, participants agreed. Failure cases were usually due to gross keypoint registration errors.

\subsubsection{Hand and Mouth Gesture Recognition}

BodySLAM uses hand and mouth keypoints for multiclass gesture classification. In the static activity conditions, I found a mean hand classification accuracy of 88\% and a mean mouth state recognition of 91\%. Unlike body pose, which benefits from greater distance between participants (such that the whole body is visible), hand and mouth classification benefit from being closer in order to provide sufficient camera resolution to resolve fine details.

\subsubsection{Skin Color and Apparel Detection}

As described above, participants selected their own skin color from a printed Fitzpatrick scale \cite{Sachdeva:2009} during the study. I compared this number to BodySLAM’s skin color estimate and found a mean error of 0.7 (SD=1.0). For pant and shirt color, all of the participants rated BodySLAM’s estimated colors as accurate.

\subsection{Simulation Study}

\begin{figure*}
  \includegraphics[width=\textwidth]{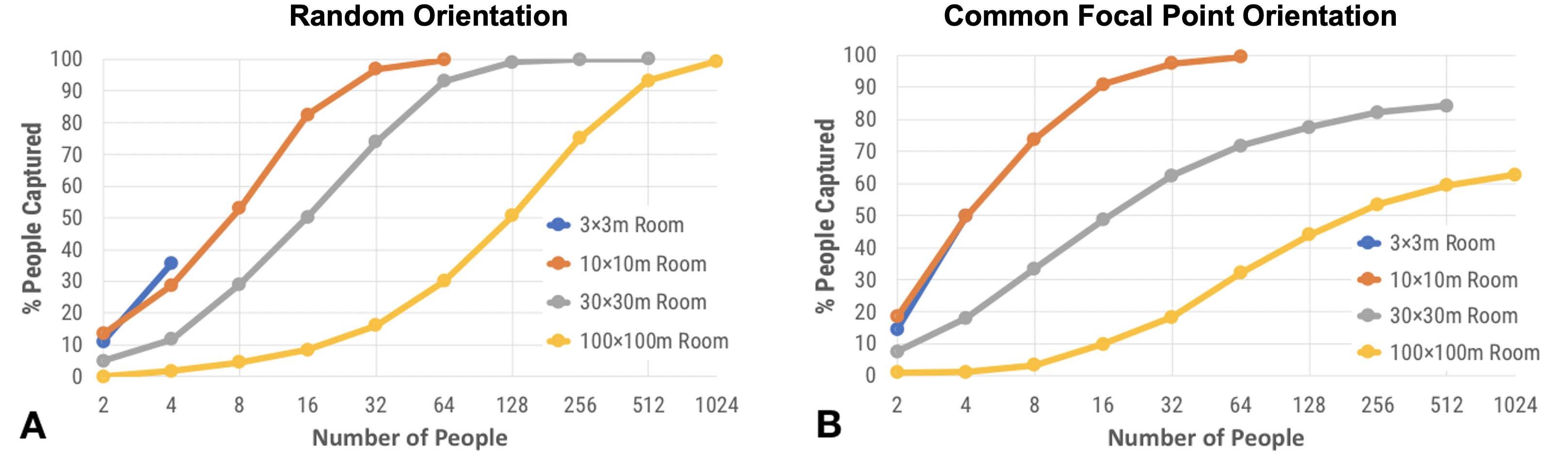}
  \caption[BodySLAM Simulation Study Accuracy]{Percentage of bodies captured for random (A) and common focus body orientations (B) in simulated rooms of different size and occupant count.}
  \label{fig:sim_acc}
\end{figure*}

To explore how BodySLAM might scale to larger spaces and numbers of people (e.g., conferences, stadiums), I ran software simulations in virtual rooms. I tested different virtual room sizes: 3x3, 10x10, 30x30 and 100x100 meters). I also varied the number of people in the room, as well as their orientation behavior (random body orientations or common focus body orientations). 

I used a 90° virtual camera field of view, matching an average mid-tier smartphone camera. Maximum body registration range was set to 14 meters, which I found to be the practical limit of ArUco tag detection at HD camera resolution. I model users as 50 cm circles that cannot be closer than 50 cm to one another, though I note that I do not model occlusion. Each combination of parameters was simulated 100 times, with detection statistics averaged. Results can be seen in Figure \ref{fig:sim_acc} broken out by random (A) and common (B) focus body orientations.

\subsection{Limitations}

The most immediate limitation of the approach is its heavy computational requirement. Running computer vision (pose modes and ArUco detection) on mobile hardware is taxing. the process runs at around 12 FPS on an iPhone XR, much slower than native camera frame rate, and more critically, has a noticeable lag for users (~150ms; see Section \ref{clientsubsec}). Note that this lag applies to BodySLAM data only, and not the AR/VR graphics, which can run at full frame rate. Nonetheless, a mismatch between e.g., a user seen in AR and their body data could induce discomfort. Processing latency can also produce incorrect 3D pose estimations due to out of sync stereo correspondences. Fortunately, smartphone manufacturers are increasingly including hardware accelerated deep learning capabilities, which should improve performance in the coming years. For example, BlazePose \cite{Bazarevsky:2020} makes use of such optimizations and can run at above 30 Hz on modern smartphones. 

I also note that BodySLAM is limited by the field of view of the rear-facing camera and occlusion in the environment. A user can lose pose tracking if they are not in the field of view of any other user. The likelihood of this occurring can be decreased by the use of ultra-wide-angle cameras, which are becoming more common in the market (e.g., Samsung S10 has 123° FOV, iPhone 11 has 120°).

Another issue with the current design is that the apparel color estimation can only work for apparels with single colors. Thus large printed designs and multi-colored apparels will lead to decrease in accuracy. In such cases a texture based approach might work better than color estimation. Furthermore, since the color of the apparel is extracted based on joint locations, it is more likely to work on garments that fully cover the wearers' body (and fail on e.g., short shorts and crop tops). 

As noted in Section \ref{keyptsubsec}, I could not not use an optical tracking system for ground truth as it interfered with the computer vision pipeline. Instead, I had human annotators subjectively rate each keypoint as correctly placed or not (i.e., a binary rating), as opposed to a continuous spatial accuracy metric such as euclidean error. This experimental compromise permitted us to run a user study, but at the expense of reporting precision.  For some insight into per-joint error, accuracy benchmarks on public datasets can be found in \cite{Cao:2017, Papandreou:2018}.

I also acknowledge that the use of ArUco for person detection and relative 3D spatial localization is inelegant, requiring instrumentation of headsets or smartphones. Additionally, although I have a tracker and time-to-live associated with each detected ArUco marker, in cases of extreme occlusion and motion blur, this process can fail in registering users. In such events markerless based ID approaches could be used. For the mobile AR use case, face recognition could be used to dispense with the need for an ArUco tag. In the two headset form factors, the face is mostly occluded, and so matching would have to occur based on body biometrics, apparel and other person attributes as seen in deep learning based person re-identification \cite{Lin:2019, Zheng:2019}. 

Finally, the current implementation relies on the cloud to collect and disseminate pose data, and also run advanced features such as 3D pose estimation using views from several users. This requires internet connectivity and contributes almost half of the system’s total latency. In the future, it could be that proximate devices create ad hoc wireless networks to share pose information more directly with others.

\subsection{Conclusion}

I have described the work on BodySLAM, which shows that it is possible to use the existing cameras in AR/VR experiences to opportunistically capture the bodies, hands, mouths and appearance of participants in multi-user settings. This offers functionality that would otherwise have to be achieved with additional special-purpose sensors, either worn or installed in the environment. I combine user studies with software simulations to evaluate how well the system scales across rooms and group sizes. While there are innate limitations of the ad hoc approach, its software-only nature makes it unique in the literature. In future iterations, mobile versions of BodySLAM can be combined with sensor fusion techniques as those presented in Pose-on-the-Go for an even higher fidelity pose capture of the users, further decreasing its reliability on opportunistic capture.

\label{chapterBODYSLAM}

\part{Improving User Practicality while maintaining Digitization Richness}
\label{part:improving_practicality}

\chapter{Overview}

Computing devices such as microphones and cameras are becoming increasingly prevalent in our daily lives, and in the consumer devices around us. Much of my prior, research detailed in Part \ref{part:improving_richness} has made use of these sensors. They showcase that full-body pose capture from ubiquitous devices like smartphones are possible and useful for certain applications. While these approaches present new avenues for increasing user digitization richness, their practicality can further be enhanced by tackling two key challenges.  

\section{Privacy and Invasiveness}

The first pertains to tackling the the privacy and invasiveness challenges presented by cameras and microphones as a data stream. The high fidelity inherent in these sensors is a double-edged sword. In general, as the richness of the sensing modality increases, so does the range of states and events that it can sense. Unfortunately, privacy implications tend to also grow in lockstep, to the point where many people do not want to use such sensors. For example, a high-fidelity sensor like a microphone can tell what activity you are doing (\ref{ubicoustics}), which device you're addressing (\ref{fig:dov}) and many other valuable facets. But at the same time, it can identify you through your voice, log conversations, etc. - dimensions you might not want e.g., smart refrigerator to know. On the other hand, low-fidelity sensors are generally much more specific in the types of events they can sense. For instance, a smart smoke alarm can detect smoke but does not know who you are or what you are doing. Ideally, we want approaches that have a low potential for privacy invasiveness but retain high fidelity, to power rich, end-user applications. As I will describe in greater detail in Chapter \ref{chaptervid2doppler}, I have explored an approach that advocates for inherently privacy-aware sensing systems.

Such a sensing modality would enable the digitization of the user analogous to that of a camera and microphone, but with the low invasiveness of a sensor such as a barometer. It is to be noted that even though we move towards more lower-fidelity signals that do not encode the representation of the user in ``human understandable" format while looking at the signal, computers can still decipher representations of the user from it. Thus, such inherently privacy-aware sensing systems are another tool in the arsenal to other approaches already present. They are not a replacement for policy based methods, or other techniques such as featurization and computing on the edge, differentiable privacy or data encryption techniques. Rather, they are to be used in conjunction with them to enable even higher security. We will cover this in greater detail in the following Chapter and particularly in Section \ref{discussion_privacy}.

\section{Passive and Longitudinal Sensing}

The second challenge is the opportunistic nature of the approaches discussed thus far. The prior works described in this thesis that capture the whole-body such as Pose-on-the-Go \ref{chapterPOSEOTG} and BodySLAM \ref{chapterBODYSLAM} rely on the user always holding the phone in their hands or carrying them on their headsets to enable pose capture. This limits their ease-of-use and applicability in-the-wild. Ideally, we want a solution that enables accurate and holistic full-body pose but without the constraint of the phone always being actively in the user’s hand. Further, it should still be minimally invasive and should not be limited to a particular device configuration (e.g. only smartphone in the hand). It should be lightweight enough to run on the edge, adaptive to the different consumer devices the user carries on with them and should ideally enable pose observations from low-fidelity data steams that are privacy-aware. Such properties would make it a practical and powerful tool for long-term behavior modeling. Chapter \ref{chapterIMUPoser} covers such a proposed approach in greater detail.  

\chapter{Privacy-Aware User Digitization}

\section{Introduction}

Future smart homes and offices will need to be able to digitize their users in order to intelligently adapt to the environment and respond to their users’ needs. In addition to IMU, capacitive, camera and audio based approaches I have discussed, we also need to bring new sensors to bear, to inform the next generation of computing platforms. These can either be mobile or even be ambient sensors, to track users no matter what devices they carry on their person. If these devices are going into peoples homes, offices and settings of daily use, we need to tackle several practicality and deployability considerations, one important aspect of which is privacy and invasiveness. 

An incredible variety of technical approaches for digitizing the user's have been considered over many decades of research. While tagging every object in the environment with sensors can be used to infer activity \cite{tapia2004activity, tapia2007portable}, this approach is expensive, hard to maintain, and often visually obtrusive. Therefore, the trend has been towards centralized sensing, either a worn device (\textit{e.g.}, smartwatches) or with \textit{e.g.}, microphones or cameras operating in an environment \cite{khurana2018gymcam, ubicoustics}. While more practical, these high-fidelity sensors also raise significant privacy concerns. Indeed, many users are wary of microphones and cameras recording them in their homes, especially after recent data leaks \cite{google_data_leak}. For this reason, there is renewed interest in identifying and exploring sensing modalities that are inherently more privacy preserving, yet sufficiently rich to enable fine-grained activity recognition.

In this work \cite{ahuja2021vid2doppler}, I explore one such sensor: the millimeter wave (mmWave) Doppler radar. Owing to their extensive use in security and automobile applications, the price of these sensors has fallen dramatically, to even just a few dollars for basic units (\textit{e.g.}, RCWL-0516, HB100 and LV002 Doppler sensors). More sophisticated frequency-modulated continuous wave (FMCW) sensors cost around \$30~USD \cite{TiDigikey}. Both types of radar sensors are solid state and small enough to be integrated into consumer devices, such as smart speakers and smartphones \cite{lien2016soli}. These radar sensors emit a known RF signal, and any motion in the scene (either from users or objects) causes reflected signals to be Doppler-shifted, which can then be used to create a 1-D Doppler plot. In the case of FMCW sensors, a 2-D plot of range \textit{vs}. the Doppler shift of signals can be produced. Although some biomechanical attributes are expressed in the Doppler signal (\textit{e.g.}, limb gait while walking), this has only been shown to recognize people from a small set of users \cite{raj2012ultrasonic}, and not from the population at large. Indeed, it would seem hard to be embarrassed by leaked Doppler data, in contrast to a video or audio recording that can easily reveal identity and capture sensitive content ~\cite{caine2006benefits, phelps2000privacy}. 

That said, Doppler radar faces a significant challenge: bootstrapping machine learning classifiers. Unlike audio and computer vision approaches that can draw from huge libraries of videos to train machine learning models, Doppler radar has no existing large datasets. All prior Doppler sensing work I could find in the literature had to collect their own bespoke training data for recognition tasks. The scale of data appears to be so limited that the full potential of techniques like deep learning techniques remains to be seen.
    
In this work, I propose a unique software pipeline that allows unstructured videos to be transformed into synthetic Doppler radar data that can then be used for training. This process opens up an unparalleled volume of training data for Doppler sensors, closing an important gap and elevating the feasibility of Doppler sensing for activity recognition. Results from the user study show that training a model using the proof-of-concept synthetic data output ($81.4$\% accuracy) is roughly comparable in accuracy to training with native sensor data ($90.2$\% accuracy) – a loss of around 8.8\% in accuracy in a 12-class activity recognition task. If I augment the large synthetic dataset with just a few minutes of user data captured with an in-situ sensor, accuracy jumps to $95.9$\%, suggesting a mixed approach could be successful while minimizing user burden. 

\begin{table}
  \centering
  \small
 \begin{tabularx}{\columnwidth}{XXcc}
 \toprule
  & \textbf{Training Data} &  \textbf{No. Classes} & \textbf{Accuracy} 
  \\ \midrule
 \textbf{TSN}~\cite{wang2016temporal} 
  & RGB Video       & 20     & 94.2\%  
  \\
 \textbf{TTDD\_FV}~\cite{wang2015action} 
  & RGB Video       & 20     & 90.3\%  
      \\
 \textbf{LTC}~\cite{varol2017long} 
  & RGB Video       & 20     & 91.7\%  
      \\
 \textbf{KVMF}~\cite{zhu2016key} 
  & RGB Video       & 20     & 93.1\%  
      \\
 \textbf{VideoDarwin}~\cite{fernando2015modeling} 
  & RGB Video       & 51     & 63.7\%  
  \\
 \textbf{MPR}~\cite{ni2015motion} 
  &  RGB Video       & 51     & 65.5\% 
    \\
\textbf{Zhao~\textit{et~al}.}~\cite{zhao2012combing} 
  & RGB-D video   & 12     & 89.1\% 
    \\
\textbf{Ubicoustics}~\cite{ubicoustics} 
  & Audio       & 30     & 82.1\%  
  \\
  \textbf{Liang~\textit{et~al}.}~\cite{liang2019audio} 
  & Audio       & 15     & 83.6\%  
  \\
  \textbf{Fu~\textit{et~al}.}~\cite{fu2018fitness} 
  & Doppler Ultrasound        & 3    & 92.0\% 
     \\
    \textbf{Radhar}~\cite{singh2019radhar} 
  & Doppler Radar       & 5     & 94.7\% 
     \\
 \textbf{Erol~\textit{et~al}.}~\cite{erol2019gan} 
  & Doppler Radar       & 7     & 92.8\% 
     \\
     \textbf{Kim~\textit{et~al}.}~\cite{kim2009human} 
  & Doppler Radar       & 8     & 82.6\% 
     \\
     \textbf{My Approach}
  & Doppler Radar       & 12     & 95.9\% 
  \\
 \textbf{My Approach}
  & Synthetic Doppler      & 12     & 81.4\% 
     \\
  \bottomrule
  \end{tabularx}
  \caption[External sensing solutions for Human Activity Recognition]{Activity recognition systems that make use of external sensors (i.e., not worn).}
  \label{tab:compare}
\end{table}

\section{Related Work}

I first briefly review related work on activity recognition powered many different sensing modalities, and then more specifically focus on approaches that leverage Doppler shifts induced by human motion.

\subsection{Doppler-Based Sensing}

Energy waves undergo Doppler shift when reflecting off moving objects. These waves can be sound \cite{raj2012ultrasonic, pu2013whole}, radio frequencies (RF) \cite{singh2019radhar, chen2016activity}, visible light \cite{albrecht2013laser, hecht2002optics}, or even gravitational waves \cite{barish1999ligo}. Given the focus on practical and deployable systems, the HCI community typically relies on microphone- and RF-based Doppler sensing. The ubiquity of microphones makes them extremely popular for sensing Doppler shifts (most often in ultrasonic frequency ranges so as to not interfere with human hearing). Using sound-based approaches, researchers have enabled large in-air gestures ~\cite{gupta2012soundwave, pu2013whole}, fine-grained hand gestures ~\cite{wang2016interacting, fan2016wireless}, multi-device interactions~\cite{aumi2013doplink, chen2014airlink}, and activity recognition~\cite{fu2018fitness}. 

In recent years, RF-based Doppler sensors have become significantly cheaper and more accessible. RF systems also tend to offer superior range than ultrasonic Doppler techniques, and can sometimes operate through walls. Prior work has explored through-the-wall person detection~\cite{chetty2011through, tan2016awireless, pu2013whole}, human gesture recognition~\cite{chen2016activity, li2018log, goel2015tongue}, respiratory monitoring ~\cite{li2016non}, and signs of life detection~\cite{chen2016signs}. Closer to this work are papers investigating RF-based Doppler sensing for activity recognition. Chen ~\textit{et~al}. proposed an in-home Wi-Fi signal-based activity recognition framework using passive micro-Doppler signatures~\cite{chen2016activity}. Using deep learning, Chen ~\textit{et~al}. monitored daily activities and detected falling accidents~\cite{chen2020respiration, chen2019eliminate, chen2017joint}. Similarly, Singh ~\textit{et~al}. used a sparse point cloud from a mmWave Radar sensor for recognizing five different human activities ~\cite{singh2019radhar}. A commonality in this prior work is the need for in-situ training data to develop machine learning models, which are specific to the use domain and collection environment. Given this data is manually collected, the volume of training data used in these systems is comparatively small compared to audio- and video-derived datasets. 

\subsection{Synthesizing Doppler Data}

Using data sources such as videos, motion capture, and animated 3D models, prior work has synthesized training data for IMU \cite{kwon2020imutube, huang2018deep}, audio \cite{aytar2016soundnet, zhou2018visual}, depth camera \cite{planche2017depthsynth} and human pose \cite{varol2017learning} powered-systems. The idea to specifically synthesize Doppler data to mitigate training data issues is not new either. In particular, Lin~\textit{et al}. explored using MoCap data to synthesize Doppler data for walking and running with some success~\cite{lin2017performance}. Unfortunately, MoCap data is generally sparse (often a dozen or so key joints), so researchers have also generated synthetic Doppler data using point clouds captured from depth-cameras \cite{li2019kinect, erol2015kinect, erol2014simulation}. In both MoCap and depth camera cases, I found datasets to be much smaller than video sources and missing many commonplace activities. The fact is, people capture and upload video data freely, but do not go out and capture depth-camera datasets for research use. Perhaps most similar to this work is ~\cite{erol2019gan}, which captured seed data using an actual Doppler radar sensor and then generated synthetic Doppler data using Generative Adversarial Networks (GANs). This approach is complementary to ours, and could expand the volume of training data even further. To summarize, my approach to synthetic Doppler data creation is unique in that the inputs are videos, allowing researchers to tap into a near-limitless amount of training data.

\begin{table*}
  \centering
  \tiny
 \begin{tabularx}{\textwidth}{l @{\extracolsep{\fill}}ccccccccccccc}

 \toprule & \textbf{Type}
  & \textbf{Wave} & \textbf{Climb Staircase}& \textbf{Walk} & \textbf{Squat}
  & \textbf{Run} & \textbf{Lunge} & \textbf{Jump Rope} & \textbf{Jumping  Jack}
  & \textbf{Jump} & \textbf{Cycle} & \textbf{Clean} & \textbf{Clap}\\
  \midrule
 \textbf{CMU \cite{cmu_mocap}}  & MoCap
  & ---       & ---     & \textasciitilde 300     & --- 
  & \textasciitilde 60     & ---     & ---        & --- 
  & \textasciitilde 110    & ---     & ---        & ---
  \\
  \textbf{SFU \cite{sfu_mocap}} & MoCap
  & ---       & ---     & \textasciitilde 20     & --- 
  & \textasciitilde 5     & ---     & ---        & --- 
  & \textasciitilde 10    & ---     & ---        & ---
  \\ 
\textbf{RHA}~\cite{laptev2004velocity} & RGB Video
  & \textasciitilde 100       & ---     & \textasciitilde 100     & --- 
  & \textasciitilde 100     & ---     & ---        & --- 
  & ---   & ---     & ---        & \textasciitilde 100
  \\ 
\textbf{UCF101}~\cite{soomro2012ucf101} & RGBVideo
  & ---       & ---     & ---    & \textasciitilde 115
  & ---     & \textasciitilde 130     & \textasciitilde 145       & \textasciitilde 125 
  & ---   & \textasciitilde 135     & \textasciitilde 110        & ---
  \\ 
\textbf{HMDB}~\cite{kuehne2011hmdb}& RGB Video
  & ---       & \textasciitilde 50    & \textasciitilde 155    & ---
  & ---     & ---     & ---       & --- 
  & ---   & \textasciitilde 105     & ---        & ---
  \\ 
\textbf{YouTube8M}~\cite{abu2016youtube}& RGB Video
  & ---       & ---    & 317   & 2422
  & 7628     & ---     & ---       & --- 
  & 4692   & 31080     & 206        & ---
    \\ 
\textbf{STAIR}~\cite{stairactions2018}& RGB Video
  & ---       & \textasciitilde 900    & \textasciitilde 900   & ---
  & \textasciitilde 900    & ---     & ---       & --- 
  & \textasciitilde 900   & ---     & ---        & ---
      \\ 
\textbf{ActivityNet}~\cite{caba2015activitynet}& RGB Video
  & ---       & ---    & ---   & ---
  & ---    & ---     & ---       & --- 
  & ---   & ---     & \textasciitilde 65        & ---
\\ 
\textbf{RadHAR}~\cite{singh2019radhar} & RF Doppler
  &   ---     & ---    & \textasciitilde 50   & \textasciitilde 50
  &  ---   &  ---    &   ---     & \textasciitilde 40 
  & \textasciitilde 35   & ---     & ---        & ---
 \\ 
\textbf{Gambi}~\cite{gambi2020millimeter} & RF Doppler
  & ---       & ---    & 231   & ---
  & ---    & ---     & ---       & ---
  & ---   & ---     &  ---       & ---
  \\ 
\textbf{NTU}~\cite{shahroudy2016ntu} & Depth Cam
  & \textasciitilde 1000       & ---    & \textasciitilde 1000   & ~1000
  & \textasciitilde 1000    & ---     & ---       & ---
  & \textasciitilde 1000   &---     & ---        & \textasciitilde 1000
     \\ 
\textbf{MHAD}~\cite{ofli2013berkeley} & Depth Cam
  & 50      & ---    & ---   & ---
  & ---    & ---     & ---       & 25
  & 25  & ---     & ---        & 25
      \\ 
\textbf{UTD}~\cite{chen2015utd} & Depth Cam
  & ---      & ---    & ---   & ---
  & ---    & ---     & ---       & 25
  & 25  & ---    & ---        & 25
  \\
  \textbf{YouTube} & RGB Video
  & \cmark       & \cmark     & \cmark    & \cmark 
  & \cmark     & \cmark     & \cmark        & \cmark
  & \cmark  & \cmark     & \cmark         & \cmark
  \\ 
  \bottomrule
  \end{tabularx}
  \caption[Dataset across sensing modalities]{I selected 12 activities from \cite{abu2016youtube, soomro2012ucf101, shahroudy2016ntu} to gauge data availability. This table shows multiple datasets, crossing four different categories of data (depth cameras, motion capture, Doppler radar and video). Counts are number of video snippets for that class. I use \cmark to denote classes with more data than could be practically utilized (i.e., functionally unlimited).  }\label{tab:data}
  
\end{table*}

\section{Possible Training Data Sources}
\label{sec:data_sources}

The decision to use video as the input into the synthesis pipeline was not a forgone conclusion. At the very start of the research, the preference was to use datasets that required less dramatic transformation. I now briefly describe the four main data categories I considered, and their relative pros and cons that led us to pursue a video-based approach. As a benchmark, I picked 12 activities (listed in Table~\ref{tab:data}) drawn from \cite{abu2016youtube, soomro2012ucf101, shahroudy2016ntu} as a sort of feasibility litmus test.

\textbf{Doppler Radar Datasets} - I started by surveying projects that used radar sensors for activity recognition and cataloged how they sourced their training data. For example, RadHAR \cite{singh2019radhar} manually collected data for five activities (walking, jumping, jumping jacks, squats and boxing) while Gambi~\textit{et~al}. \cite{gambi2020millimeter} collected 231 sequences of 29 people walking at different speeds. In \cite{erol2019gan}, the authors manually collected 1356 sequences of 14 people performing 8 actions at different angles. In all cases, these manually-created datasets were very small in volume and limited in their classes. Most importantly, every minute of recorded data took at least one minute of researcher time. 

\textbf{Depth Camera Datasets} - Next I considered RGBD and depth-camera video datasets (captured with sensors such as the Microsoft Kinect). I surmised the 3D point cloud of segmented users could be converted into synthetic Doppler \cite{li2019kinect, erol2015kinect}, which could then be used for training. Unlike with Doppler radar, large datasets exist for research use, which was encouraging. I surveyed 13 such datasets, but found them to be missing several of the sample activities (see Table~\ref{tab:data}), and thus several datasets (in different formats) would have to be combined. 

\textbf{MoCap Datasets} - I then looked at 3D motion capture (MoCap) datasets, digitized by professional optical tracking systems. These are very spatially accurate, but only provide a sparse 3D model of users – often just 17 key joints – and thus can only provide a very coarse synthetic Doppler signal \cite{lin2017performance}. Any Doppler-shifted reflections between e.g., the wrist and elbow must be interpolated. Additionally, of the 13 datasets I surveyed, many activities were missing, as noted in Table~\ref{tab:data}. 

\textbf{Video Datasets} - Given they contain no innate 3D data, I was initially skeptical of videos as a data source. Cameras are generally very high resolution in the plane orthogonal to the camera axis, but are largely intensive along their Z axis. However, the incredible wealth of video content – with more than 500 hours of video uploaded every minute to just YouTube alone – it soon became the clear winner. Beyond unstructured sources like Youtube, there exist scores of excellent and very large video repositories that cover all of the poses (Table~\ref{tab:data}). I was likewise encouraged by recent advances in computer vision that enabled 3D pose and even 3D meshes of users to be extracted from videos, offering the building blocks to explore synthesizing Doppler data. 

Motivated by these findings, I set out to create a software pipeline that converts videos into realistic, but synthetic Doppler radar data. If achievable, it would offer an unparalleled volume of training data for this emerging sensing modality, closing an important gap and elevating the feasibility of Doppler sensing for activity recognition. 

\section{Implementation}

\begin{figure*}
  \includegraphics[width=\textwidth]{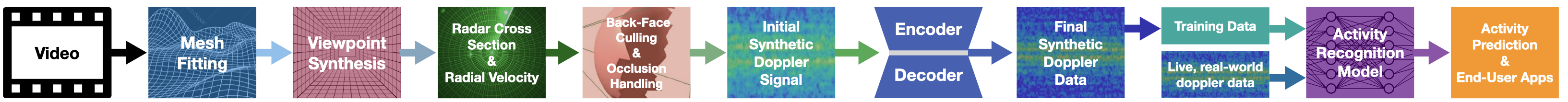}
  \caption[Vid2Doppler Pipeline Overview]{Overview of the software pipeline for synthetic Doppler data generation for training activity recognition models.}
  \label{fig:pipeline}
\end{figure*}

I now describe in detail the iterative steps of the software pipeline, illustrated in Figure~\ref{fig:pipeline}.

\subsection{Mesh fitting}

Doppler sensors measure the radial velocity of reflective surfaces in a scene. To replicate this signal, I require equivalent 3D data of a user's body against a static background. Fortunately, computer vision has made enormous strides in fitting a 3D mesh to a person's image ~\cite{kocabas2020vibe, mehta2020xnect, xu2020eventcap, habermann2020deepcap}. Hence, as a first step, I compute the position of all vertices of the human body by fitting a mesh to it. For this, I use VIBE~\cite{kocabas2020vibe}, which estimates the mesh via an adversarial learning framework for human pose estimation. Given an input video, the VIBE model estimates the human pose and outputs a human pose mesh for each frame. I track vertices across frames and also smooth their positions to increase stability.

\subsection{Viewpoint Synthesis}

Once I have the 3D mesh of a user, I can place a virtual camera in the scene to synthesize any view. Indeed, I place nine synthetic cameras in a spherical coordinate system around the user mesh to simulate different viewpoints. Specifically, I include a ``head on" view, along with views 45{\degree} to the left and right, as well as 45{\degree} up and down (forming a 3 $\times$ 3 polar coordinate grid). This simple manipulation, essentially a form of training data augmentation, has a multiplicative effect on every second of input video. Moreover, it helps to make the later machine learning model more view-invariant. 

\subsection{Radar Cross-Section \& Radial Velocity}
\label{subsec:rad_vel}

Given a viewpoint and a mesh of a user, I can compute the radar cross-section of every vertex with respect to a virtual Doppler sensor. I do this by taking the user mesh returned by VIBE \cite{kocabas2020vibe} (which uses SMPL mesh \cite{SMPL:2015}) and calculate each vertex's surface area and normal. To compute radial velocity, I look back at each vertex's movement history (previous frames), again with respect to a virtual Doppler sensor. I also further augment the training data by slightly varying the framerate of the input video (e.g., by assuming consecutive frames are not 1/30th of a second apart, but rather 1/29th or any other value) to produce realistic variations of the same activity being performed at different speeds. 

\subsection{Vertex Visibility \& Occlusion}

At this point in the pipeline, I have each vertex's contribution to the synthetic Doppler signal, assuming all were visible. Of course, there are vertices on the reverse side of the user, and also vertices that are occluded by other body parts (e.g., arms crossing the torso). Since these would not contribute to an RF Doppler signal, they must be filtered. I first perform back-face culling \cite{zhang1997fast} for each viewpoint, and then calculate if a vertex is occluded by another. Only vertices with line-of-sight to the virtual Doppler sensor are passed to the next step of the pipeline. 

\begin{figure*}
\centering
  \includegraphics[width=\textwidth]{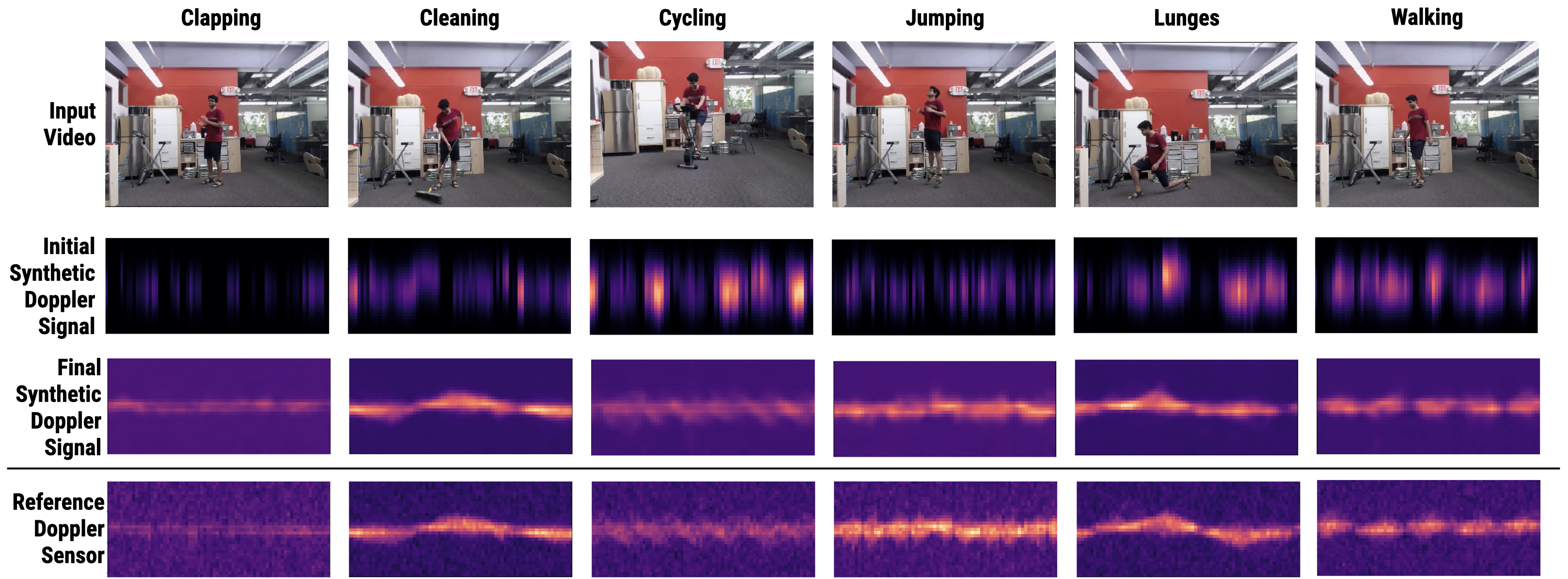}
  \caption[Synthetic and Real Doppler signal across different activities]{Top row shows a user performing different activities. Below each activity is the initial and final synthetic Doppler signal. The bottom row shows the corresponding signal captured by an actual Doppler radar sensor (positioned next to the camera that filmed the user). Note the synthetic pipeline produces comparable signal. }
  \label{fig:activity}
\end{figure*}

\subsection{Synthesizing Initial Whole-Body Doppler}
\label{subsec:initial_dop}

A useful and popular visualization of Doppler sensor data is a radial velocity profile at a given instant in time. To mimic this, I create a 32-bin histogram of the radial velocities of the visible user vertices in the velocity range of the real world sensor (in this case, -2 to 2 m/s). By stacking such signatures over time, I create a sliding, Doppler-time plot (see Figure~\ref{fig:activity}, second row). The envelope of this initial simulated Doppler signal roughly follows that of actual Doppler data, but this can be further improved as I will explain. 

\subsection{Encoder-Decoder for Domain Translation}
\label{subsec:enc_dec}

The aforementioned synthetic Doppler-time plot is very coarse, as it is heavily quantized and even small mesh fitting errors can produce big jumps in radial velocity. Additionally, it does not contain any of the characteristic noise or non-linearities found in real RF Doppler sensors. However, I found that the correspondence between synthetic and real-world Doppler can greatly be improved by making use of an encoder-decoder model. To train the encoder-decoder model, I choose a U-Net architecture \cite{ronneberger2015u}, which contains a convolutional and deconvolutional block with an embedding layer of size 128 in between. Each block has 16, 32 and 64 2D filters respectively with a kernel size of 3 $\times$ 3. A Leaky ReLU activation function and a batch normalization are applied to each convolutional layer. I take corresponding pairs of real world Doppler and synthetic Doppler to train the model for 1000 epochs with a root mean square error loss and Adam optimizer \cite{kingma2014adam} with a learning rate of 0.001.

\subsection{Final Whole-Body Doppler Signal}

The final output of the pipeline is a synthetic Doppler-time plot generated by the encoder-decoder model. This plot represents radial velocity from -2 m/s to 2 m/s (Y-axis, 32 bins) and 3.0 seconds of data (X-axis, 72 bins). As can be seen in Figure~\ref{fig:activity} (third row), this synthetic signal has a strong correspondence to real-world RF doppler data (bottom row), despite only utilizing 2D video. It is this signal that I use to train the deep learning model for activity recognition, described next.

\subsection{Activity Recognition}
As a proof-of-concept Doppler sensor, I used a Texas Instruments AWR1642 mmWave radar board \cite{ti_mmWave}, which costs around \$30~USD \cite{TiDigikey} when purchasing just the sensor. Doppler data is streamed to a MacBook Pro laptop (3.1GHz dual-core i5) over USB at 31~FPS. The machine learning model is a VGG-16-based convolutional neural network ~\cite{simonyan2014very} that ingests a real-world Doppler-time plot (32x72) and outputs one of 12 activity classes. I train the model with a categorical cross-entropy loss \cite{zhang2018generalized} and Adam optimizer \cite{kingma2014adam} with a learning rate of 0.001 for 1000 epochs. On the laptop, this model takes 47~ms of compute, allowing it to run in realtime. 

\section{Training Data}
As a proof-of-concept class set, I used the same 12 activities used to survey data sources earlier in the paper, which were drawn from the literature \cite{abu2016youtube, soomro2012ucf101, shahroudy2016ntu}. To train the activity recognition model, I aggregated $10.4$ hours of video data to serve as the input for the synthetic Doppler data generation (expanded roughly ten fold via data augmentation). Most of the video datasets that I use ($8.4$ hrs of the $10.4$ hrs) are structured (``RGB Video" datasets in Table~\ref{tab:data}), i.e., have activity labels associated with them. Similar to \cite{kwon2020imutube}, I also mined data from unstructured sources (e.g., YouTube) using queries related to the activity set and then filtered them manually. In the future, as the sophistication of vision-based Human-Activity Recognition \cite{beddiar2020vision} modules improve, I could rely on them for automatic labeling. 

To train the encoder-decoder model, I had two participants perform the 12 activities in a different room than the later study (a living room of dimension $7.2$ $\times$ $5.6$ $\times$ $3.6$ m). Each user provided one hour of data, varying their angle to the sensor. As the encoder-decoder model works with unsupervised data, collection required no labeling and motions that were not part of the activity dataset were also captured when transitioning from one activity to another. 

\section{User Study}

I now describe the physical arrangement of the study, and then walk through a series of specific experiments that I used to elucidate the feasibility of the approach. 

\subsection{Apparatus \& Location}

For this study, I used the same Texas Instruments AWR1642 RF Doppler sensor \cite{ti_mmWave} and MacBook Pro laptop as described in the previous section. I mounted the Doppler sensor to a tripod alongside a Logitech HD webcam to capture footage. I cleared a small space in my lab where users could safely perform activities I requested. 

\subsection{Procedure}

I recruited 10 participants (8 male, 2 female) with an average age of $25.3$ years. For each participant, I captured two sessions of data back-to-back in a lab space roughly $12.8$ $\times$ $6.5$ $\times$ $3.8$ m. Within each session, participants were asked to perform 12 activities (enumerated in Table~\ref{tab:data}) at 3 different angles with respect to the sensor tripod (0{\degree}, 45{\degree} and -45{\degree}), resulting in a total of 36 trials per session per participant. Thus, in total I collected: 10 participants $\times$ 2 sessions $\times$ 3 angles $\times$ 12 activities = 720 total trials (roughly $3.4$ hours of data). 
 
For each trial, I collected synchronized Doppler and video data. Note that video data was collected as a reference to benchmark the accuracy of the generated synthetic Doppler signal vs. the real world Doppler signal. Importantly, reference videos were never used to generate any synthetic data for training. As described previously, the training dataset consisted of $10.4$ hours of video data curated from various external data sources, which I processed into synthetic Doppler data.

\begin{figure*}
\centering
  \includegraphics[width=\textwidth]{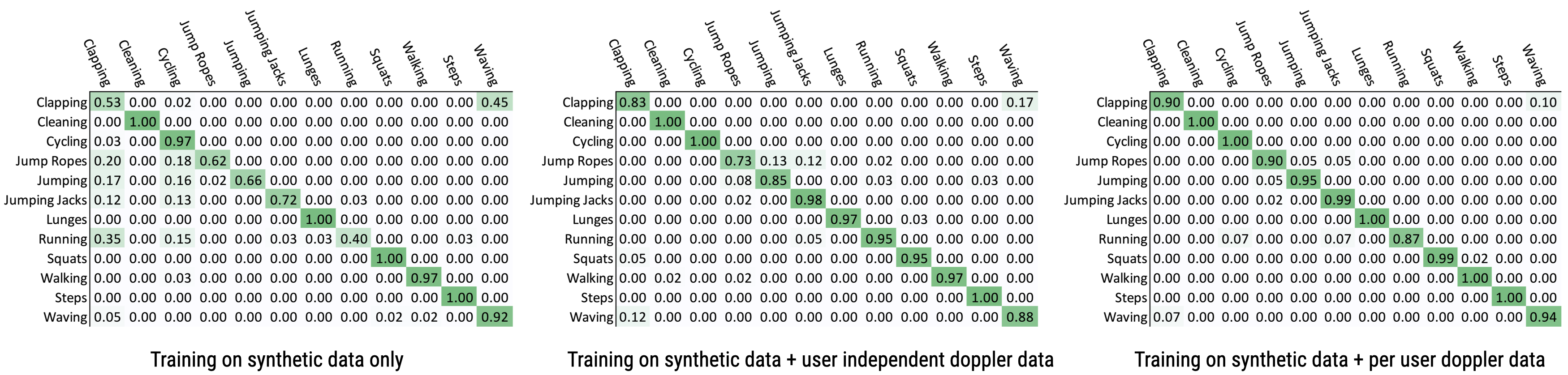}
  \caption[Vid2Doppler Confusion matrix]{Confusion matrix across different train/test conditions.}
  \label{fig:cm}
\end{figure*}

\subsection{Results}
I designed the study procedure in order to analyze and isolate different factors that affect performance. First I discuss the quality of the generated synthetic Doppler data. I then describe a series of varying train/test configurations to assess activity recognition accuracy. 

\subsubsection{Quality of Synthetic vs. Real-World Doppler Signal}

To test the efficacy of the synthetic Doppler data generation pipeline, I make use of the videos I captured in tandem with real-world Doppler data. Specifically, I run participant videos through the pipeline and compare the synthetic Doppler output to the real-world Doppler sensor signal, finding a Mean Absolute Error (MAE) of $0.09$ (SD = $0.03$) for the normalized (between 0 and 1) amplitude of Doppler shift.

\subsubsection{Recognition Accuracy: Only Synthetic Training Data}

I evaluated the performance of the model trained \textit{only} on synthetic data generated from the external video dataset (detailed in Section~\ref{sec:data_sources}) and tested using participants' real-world Doppler signals. In this configuration, the model achieved an accuracy of $81.4$\% (chance is $8.34$\%) across the 12 activities and all participants. Figure~\ref{fig:cm} (left) provides the confusion matrix. Note that in this train/test configuration, I do not train the model on \textit{any} real-world Doppler data (i.e., only synthetic Doppler data). This result can be thought of as ``out-of-the-box" accuracy, without any calibration to the local environment or user. 

\subsubsection{Recognition Accuracy: Only Real-World Training Data}

To better contextualize the model accuracies, I train a model using real-world Doppler data and then test it on real-world Doppler data. As already mentioned several times, there are not good existing datasets to run such an analysis, so instead I had to use the collected study data. Specifically, I performed a leave-one-user-out cross validation. In this process, I train on real-world Doppler data from nine of the participants and test on a tenth (all combinations, results averaged). These models achieved an average accuracy of $90.2$\% (SD = $4.8$\%). Unsurprisingly, training on data captured using the actual sensor in the same location outperforms the model trained only on synthetic data, though the difference is only $8.8$\%, which I view as a strong result for the proof-of-concept pipeline. At a high level, I believe it shows that a purely synthetic training data pipeline can be competitive with training procedures that rely on in-situ captured data. 

\subsubsection{Recognition Accuracy: Synthetic + Real-World Training Data}

It is also possible for models to leverage both synthetic and real-world Doppler data for training. This could offer the best of both worlds: a large corpus of videos for creating an even larger synthetic Doppler dataset, as well a smaller real-world dataset captured in-situ that is inherently better tuned to the local environment and physical sensor. To explore this, I again ran a leave-one-user-out cross validation. This time, I trained models using all synthetic Doppler data and real-world data from nine participants, testing on a tenth holdout user (all combinations, results averaged). In this scenario, the model achieves an accuracy of $93.4$\% (SD = $5.4$\%), a boost of $3.2$\% over using only real-world data for training. See confusion matrix in Figure~\ref{fig:cm}, center. 

\subsubsection{Recognition Accuracy: Per-User Training}

In all of the previous experiments, the model is never exposed to training data from the user it is testing. However, it is not uncommon for sensing systems to collect some training data from users, often in the guise of a calibration during setup (e.g., voice transcription systems that are trained using a large general corpus, but also ask the user to speak some phrases to calibrate). To simulate this, I performed a leave-one-session-out cross validation. Specifically, I train a model using all of the synthetic Doppler data (i.e., a large general corpus) and then add one round of a participant's real-world Doppler data, testing on the holdout round (both round combinations, for all participants, results averaged). This achieves the best accuracy of all tests: $95.9$\% (SD = $0.3$\%). The confusion matrix can be found in Figure~\ref{fig:cm}, right.

\subsubsection{Comparison to Prior Work}

The accuracy of my approach compares favorably to prior work. RadHAR makes use of point clouds generated from mmWave radar (the same sensor as ours) and achieves an accuracy of $90.5$\% across 5 activities employing a deep learning model. \cite{kim2009human} uses SVM's trained on Doppler radar to recognize 7 activities with a per-user model accuracy of $92.8$\% and a cross-user accuracy $91.9$\%. The synthetic Doppler data approach in \cite{erol2019gan} achieves an accuracy of of $82.6$\% across 8 activities on one participant. In contrast, the per-user model on 12 activities achieves an accuracy of $95.9$\%. However, it is to be noted that my goal is not to make a better framework for sensing activities via Doppler data, but rather to create a framework for synthesizing Doppler data for training a myriad of different activity recognizers. That said, higher accuracy is a nice side-effect of leveraging synthetically-created training data derived from video sources. An overview of accuracies can be found in Table~\ref{tab:compare}, though I emphasize these systems are tested on different datasets and have different applications.

\section{Limitations}

There are several key technical limitations that will need to be overcome before consumer use and widespread adoption. First, the model was trained and tested with static background scenes. The model would fail in cases where there is motion in the background or the sensor itself was in motion (i.e., the environment would create Doppler-shifted reflections in addition to a user). To overcome this in the future, it may be possible to add random Doppler signals or even simulate the physics of moving objects and walls as part as part of the synthetic data generation pipeline. Another limitation of the current system is the inability to handle multiple simultaneous users. The current approach sums all the Doppler profiles across distances to create a distance-invariant Doppler-time plot. However, some Doppler radar systems can segment and track multiple people if they are far enough apart \cite{ti_person_tracking}, so this may be overcome in the future. Lastly, the current machine learning approach makes use of convolutions on the Doppler-time plot. However, alternate feature representations that treat the Doppler-shift histograms as a time series and make use of RNN-based architectures \cite{shi2015convolutional, zaremba2014recurrent} could help take the model away from fixed window lengths. Furthermore, apart from a UNet architecture, conditional adversarial losses \cite{isola2017image} can also be explored where the encoder-decoder and discriminator (activity classification) are combined in a single step. 

\section{Discussion on Privacy and Invasiveness}
\label{discussion_privacy}

I would like to acknowledge that while RF Doppler radar and the proposed approach is privacy-preserving in comparison to cameras and microphones, the logging of activity data in itself can have significant privacy implications. This is a long standing HCI research topic, and as Doppler radar sensors become more pervasive, they too will need special scrutiny given their unique pros and cons.

Further, such an inherently privacy-aware sensing systems is another tool in the arsenal to other approaches already present. These sensing systems are not a replacement for policy based methods, differentiable privacy, edge computing or fully homomorphic encryption. Rather, they are to be used in conjunction with them to enable even higher security. 

Let's consider policy-based methods that define capture routines of high-fidelity sensors. For example, policies that ask for the camera to blur out faces or the microphone to only be activated by a wake-word. Such approaches require an aspect of trust associated with the manufacturer and developer of the sensors. There can be mal intent during data collection, false advertising of the data recording routines or even leaks of the collected data. In such cases, policy-based method would fail to curb malicious activities they are unaware of. Therefore, such privacy-aware sensing modalities can add another layer of protection in such cases. 

Further, consider the scenario in the future wherein the user is not identifiable from the raw Doppler data but a Machine Learning model can identify the user from it. In such a scenario, both the ML model and the raw Doppler data need to be compromised for the attacker to invade the privacy of the user. This is in contrast to other camera-based methods wherein raw data itself can compromise the user without the need of the model. Thus, the utility of privacy-aware sensing approaches can further boost the  capabilities of other privacy techniques already present.  

\section{Conclusion}

User digitization enables a plethora of applications, demonstrated in much prior research. With the high-fidelity, yet privacy-preserving sensing afforded by Doppler radar sensors, the ubiquity of this modality is held back by the lack of available training data. I aim to mitigate this important issue by created a software pipeline that takes videos of users performing activities and outputs realistic, synthetic Doppler data. This can then be used to train activity recognition models. Thus, in this work I have not only created a framework for activity recognition and behavior tracking that can be integrated into future consumer devices, but also an innovative way for generating Doppler data from videos, that can potentially be expanded to other human-centric applications in the future such as seizure detection, physical therapy and human-robot interactions.

\label{chaptervid2doppler}

\chapter{Passive and Longitudinal User Digitization On-the-Go}


\begin{figure}
\centering
  \includegraphics[width=\linewidth]{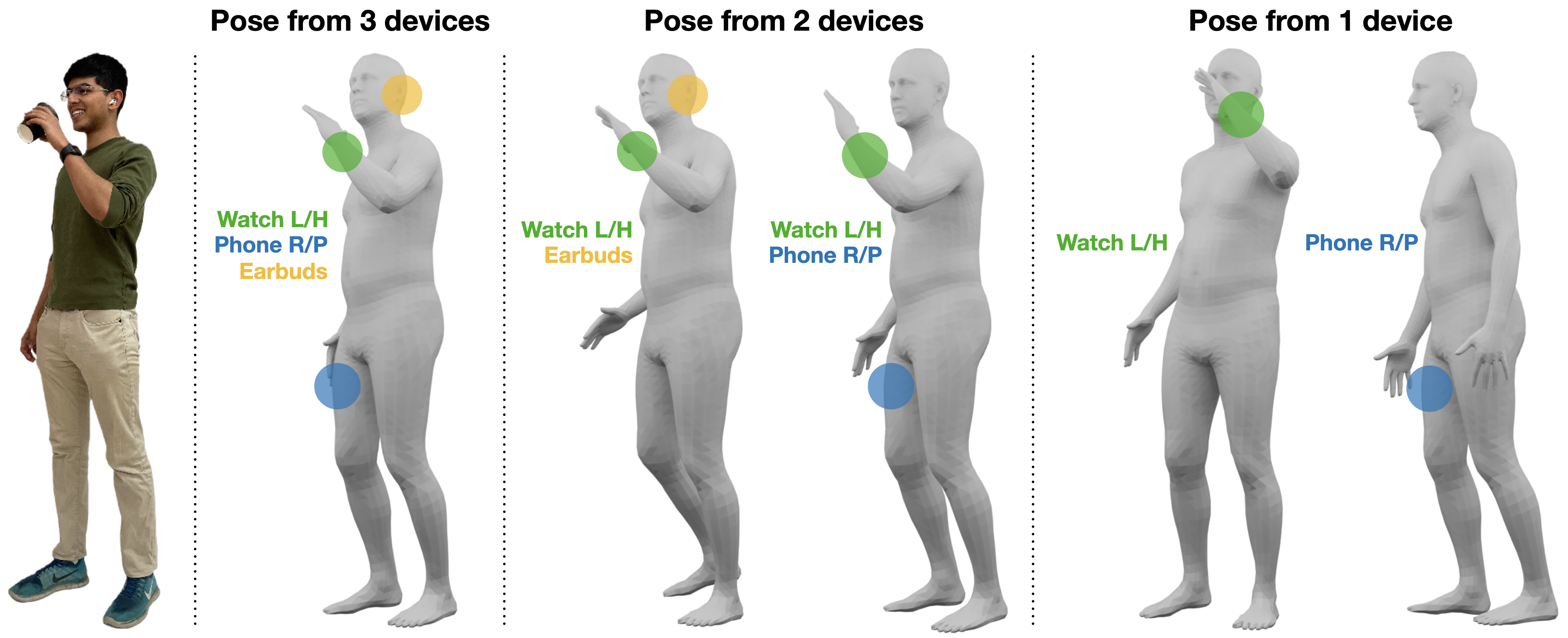}
  \vspace*{-3mm}\caption{Using whatever mobile devices a user has with them, IMUPoser estimates full-body pose. In the best case, a user can have a smartphone, smartwatch and earbuds (pose from 3 devices). Of course, the number of devices will vary over time, e.g., earbud use is intermittent and not everyone wears a smartwatch. This means IMUPoser must track what devices are present, where they are located, and use whatever IMU data is available. Abbreviation key: L-Left, R-Right, H-Hand, and P-Pocket. }
  \label{fig:teaser}
\end{figure}

\begin{figure}
    \centering
    \includegraphics[width=\textwidth]{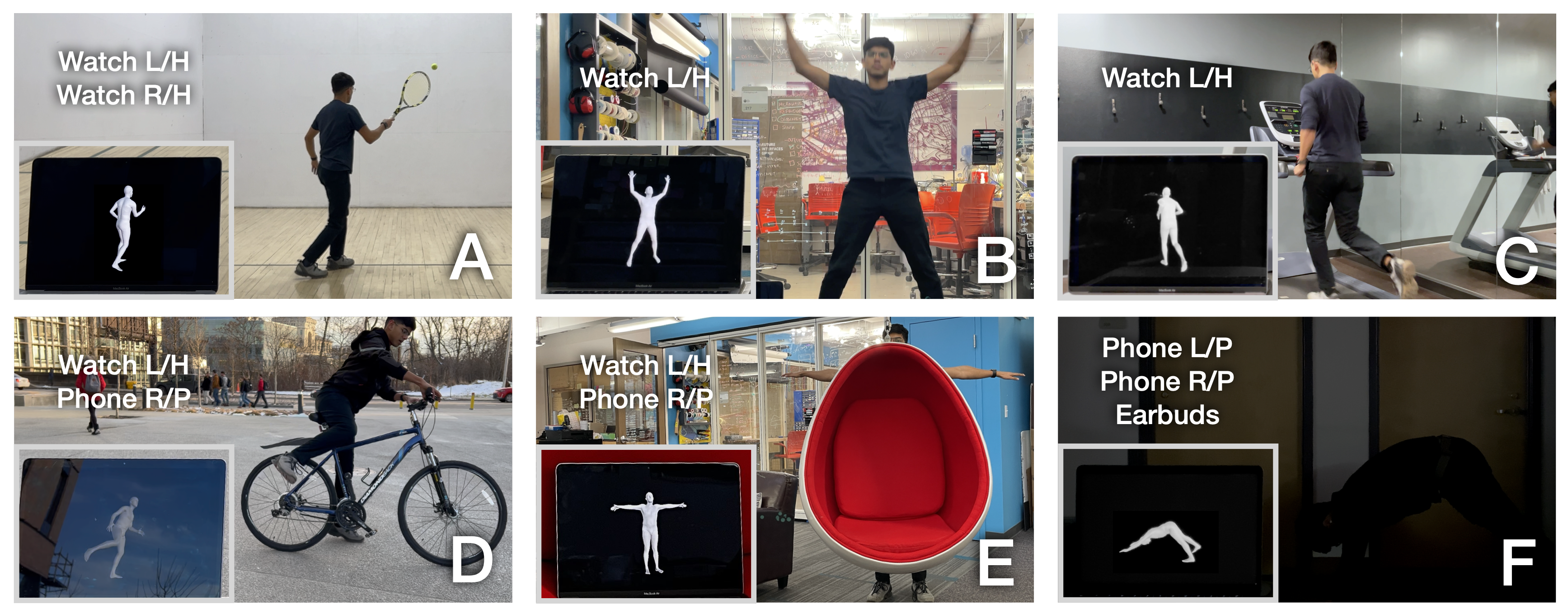}
    \caption{Real-time pose estimation (inset photos) powered by consumer mobile devices (listed in each photo) could have uses across many domains, including sports (A), rehabilitation (B), fitness (C), and transportation (D). Note also that IMUPoser is robust to occlusion (E) and lighting conditions (F). Abbreviation key: L-Left, R-Right, H-Hand, and P-Pocket.}
    \label{fig:example_applications}
\end{figure}

\section{Introduction}

Prior works such as Pose-on-the-Go \ref{chapterPOSEOTG} and BodySLAM \ref{chapterBODYSLAM} have showcased immense the utility of full-body motion capture. It has obvious applications in gaming \cite{Kinect_games}, fitness \cite{khurana2018gymcam}, rehabilitation \cite{mousavi2014review}, life logging \cite{jalal2012depth}, and context-aware interfaces \cite{dov, adeli2020socially}. For example, digital assistants with knowledge of pose could help a football player improve their form, or a patient recovering from surgery monitor changes in their gait. However, the previously described technologies either require the use of cameras for only a particular domain (e.g. BodySLAM is only for XR) or rely on the user operating a device in a particular configuration (e.g. Pose-on-the-Go's opportunistic nature relies on the user always holding the phone in their hands). In contrast, IMUPoser explores passive and longitudinal sensing of the user under different consumer device configurations using only inertial data.

Most computing devices we carry with us contain IMUs, most notably smartphones, smartwatches, and wireless earbuds. Further, in contrast to cameras and microphones, IMU's are more privacy-aware as a sensing modality and require significantly less computational overhead. In this work \cite{mollyn2023imuposer}, I study how one can use this ecosystem of worn and mobile devices to estimate a user's body pose in real-time and with no external infrastructure. This approach introduces new challenges prior sparse IMU pose models (e.g., \cite{von2017sparse, huang2018deep}) have not faced. Uniquely, the position and number of tracked body locations can change on the go. For instance, a user can take a phone from their left pocket into their right hand, or a user can add to the number of sensed points by wearing their earbuds. The model must accept various combinations of incomplete inputs and gracefully degrade as the number of active devices reduces (potentially to one). Secondarily, the system must work with IMU data received from consumer devices that are noisier than professional-grade motion capture suits (e.g., XSens \cite{xsens}) used in highly-related prior work such as Sparse Internal Poser \cite{von2017sparse}, Deep Inertial Poser \cite{huang2018deep}, and TransPose \cite{yi2021transpose}. Table~\ref{tab:imu-systems-overview} provides an overview of the most related prior work.

To evaluate our method, I created a novel dataset: professional-grade Vicon optical tracking paired with commodity device IMU data from common worn/held locations. Unsurprisingly –- given that most people have \textit{at most} three body positions to estimate hundreds of degrees of freedom in the human body -- our pose output is an approximation. However, it is rarely wildly incorrect, and most often, the main gestalt of a user's pose and locomotion mode is captured. This ``low-fi" pose output is ill-suited for high-fidelity applications, such as special effects motion capture or virtual reality avatars, where users expect a mostly-faithful body representation. Nonetheless, the adaptive and mobile nature of IMUPoser enables passive and longitudinal sensing of the user (potentially even from a single device), making it especially well-suited for health and wellness applications. For example, this low-fi body tracking could be valuable for boosting accuracy in calorie counting, tracking progress in a physical therapy regime, and monitoring exercise form and rep count. I highlight some example uses in Figure \ref{fig:example_applications}. 

\begin{table}
\centering
\small
\begin{tabularx}{\columnwidth}{Xccccc}

\toprule

\textbf{System}    & \textbf{\# Inst. Joints}  & \textbf{Sensor FPS (Hz)} & \textbf{Consumer Device} & \textbf{Real-time} & \textbf{MPJVE (cm)}   \\
\midrule

XSens\cite{xsens}  & 17  & 120 & \xmark & \xmark & -       \\
SIP\cite{von2017sparse}  & 6  & 60 & \xmark &  \xmark & 7.71         \\
DIP\cite{huang2018deep}       & 6 & 60 & \xmark & \cmark & 8.96               \\
Transpose\cite{yi2021transpose}  & 6  & 90 & \xmark & \cmark  & 7.09       \\
PIP\cite{yi2022physical}       & 6  & 60 & \xmark & \cmark    & 5.95             \\
Tautges\cite{tautges2011motion}  & 4  & 25 & \xmark & \xmark & -       \\
\rowcolor{Gray}
IMUPoser  & 1--3  & 25 & \cmark & \cmark & 12.08  \\
\bottomrule

\end{tabularx}
\caption{Comparison of worn-IMU, full-body pose estimation systems. MPJVE is calculated on the DIP-IMU dataset \cite{huang2018deep}.}
\label{tab:imu-systems-overview}
\end{table}

\section{Possible Device Combinations}\label{Possible device combos}

Smartphones, smartwatches, and earbuds have different possible body locations. For instance, a smartphone can be stored in the left or right pocket, held in the left or right hand, held to the head (to take a call), or not carried by the user at all (6 possible states). For smartwatches, they are either worn on the left or right wrist or not worn by the user (3 possible states). For earbud-like devices, they can be worn on the head, placed into a charging case and stored in the left or right pocket, or not carried by the user (4 possible states). Although at present, putting earbuds into a charging case generally puts them to sleep, I assume that in the future a firmware update could allow for continuous IMU data streaming, especially given the larger battery in the case.

Fully enumerated, this yields 72 possible device-location combinations. However, I eliminate three combinations where the user is both wearing earbuds and the phone is held to the head (to take a call), as this is not a typical use case. An additional invalid combination is no phone, smartwatch or earbuds, and thus our system would not run at all. This leaves 68 possible arrangements combinations –- 14 combinations have 1 active device, 36 combinations have 2 active devices, and 18 combinations have 3 active devices. 

\begin{figure}
    \centering
    \includegraphics[width=0.95\linewidth]{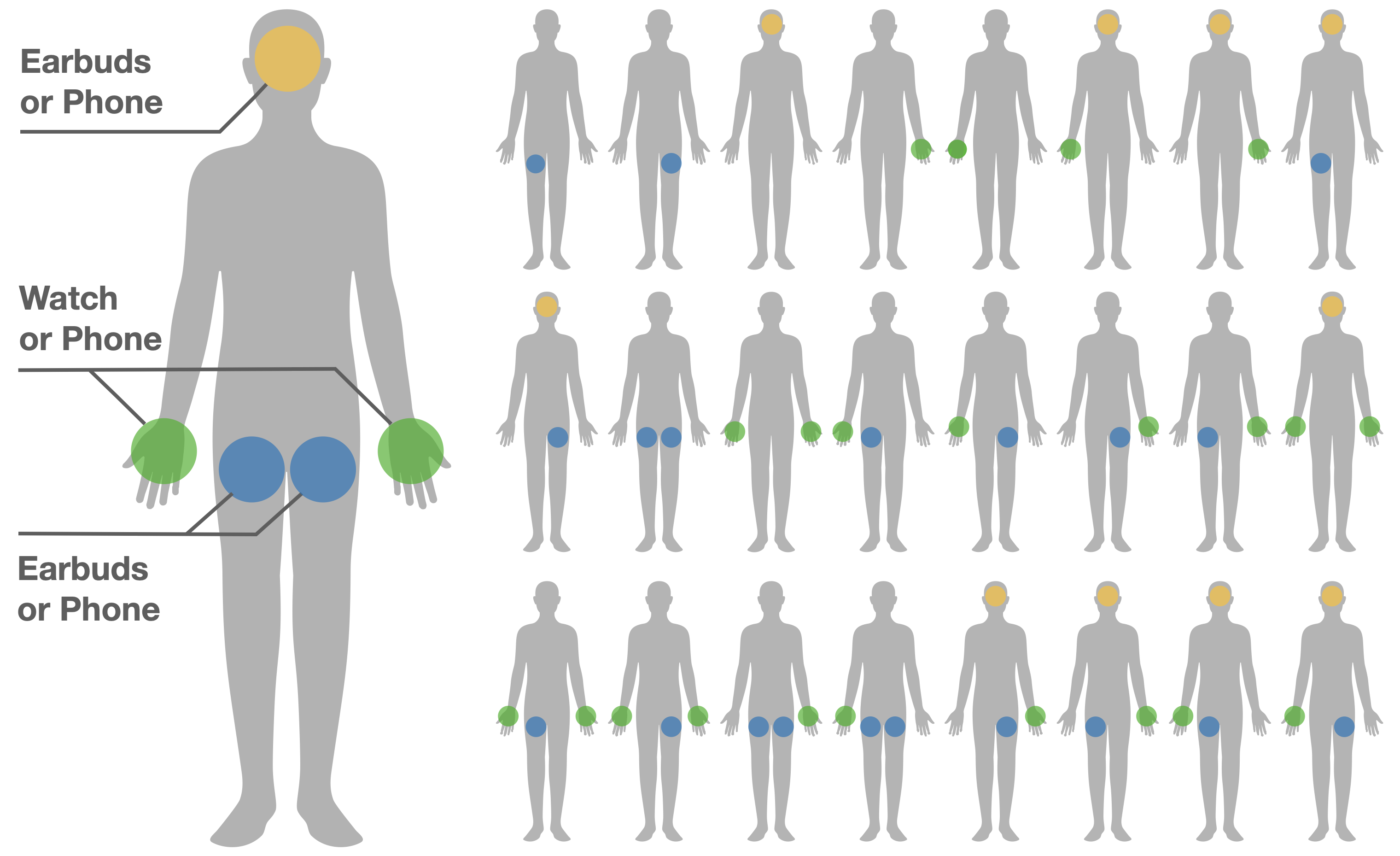}
    \caption{The 24 possible device-location combinations I support and investigate.}
    \label{fig:possible_combinations}
\end{figure}

Next, it is important to consider that some combinations of devices do not provide substantially different body data for our purposes. For instance, a user could wear a smartwatch on their left wrist and hold a smartphone in their left hand -- the IMU data would be highly correlated, and thus I treat it as a single body point. Another example is storing a smartphone in the right pocket, along with earbuds in a charging case -- again, the IMU data would be similar. Thus, what our system truly cares about are the combinations of body location enabled by the 68 possible device-position combinations -- these 24 combinations are illustrated in Figure \ref{fig:possible_combinations}. 

I note that our system makes some simplifying assumptions about body positions. For example, in order for a hand position to be considered active, it requires either a smartphone to be held in that hand or a smartwatch worn on that wrist. Even though the signal is not identical, it is highly correlated such that the information power is similar. Similarly, a smartphone held to either ear is considered to be a head location (rather than left or right ear). I made the latter simplification because Apple's AirPods (which I use in our real-time implementation) fuse their IMUs to provide a single-head 6DOF estimate, rather than provide IMU data from each Airpod individually.

\section{Implementation}

\begin{figure}
    \centering
    \includegraphics[width=\linewidth]{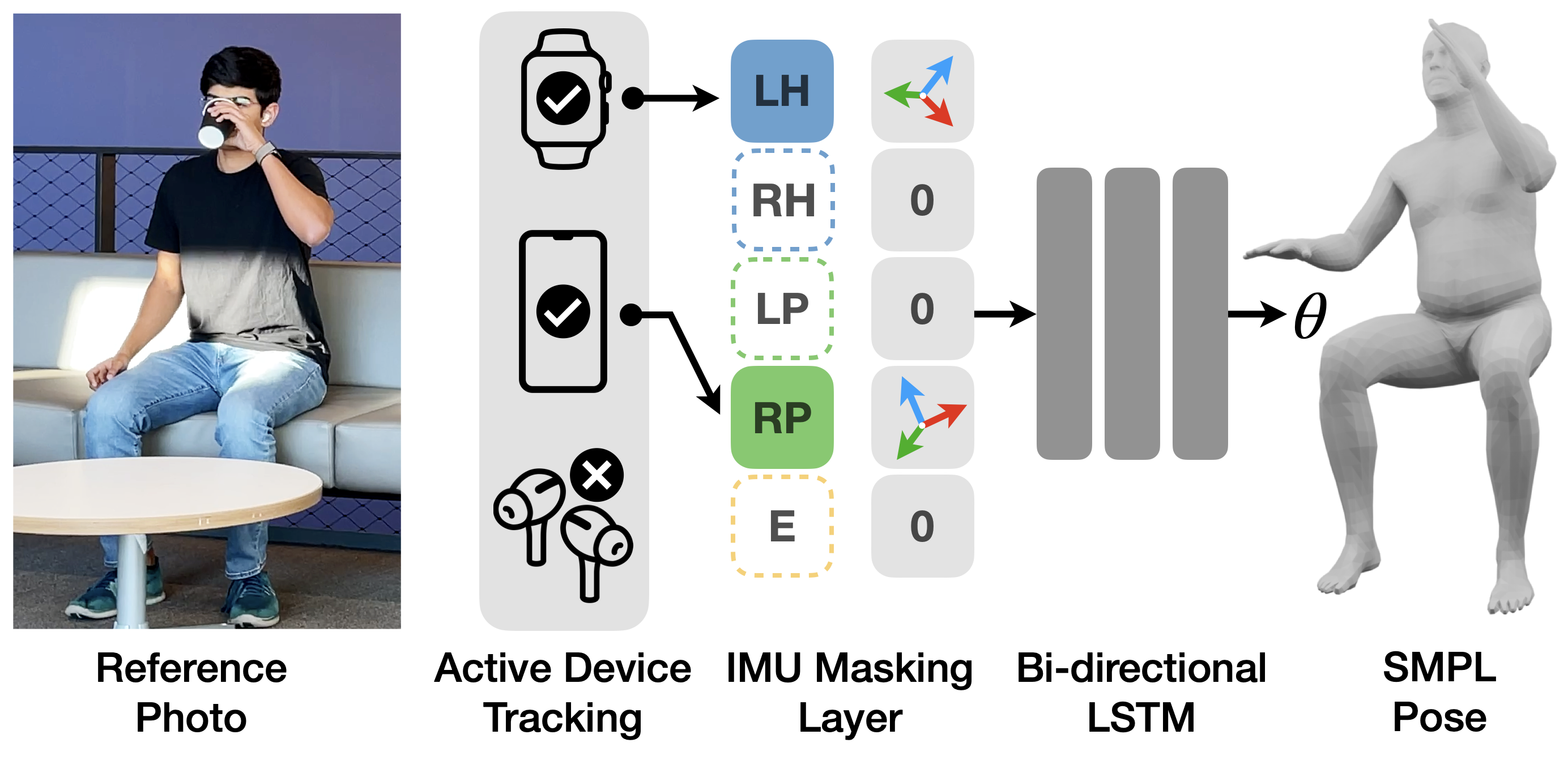}
    \caption{Overview of IMUPoser's real-time system architecture.}
    \label{fig:architecture}
\end{figure}

Figure~\ref{fig:architecture} provides an overview of our pipeline. I focus on three popular consumer devices: smartphone, smartwatch and wireless earbuds/headphones. Each of these devices contains an IMU, the ability to wirelessly transmit data, and some local compute. I envision our model executing on the most capable device carried by the user, with the other less-capable devices streaming their IMU data over e.g., Bluetooth.

\subsection{Model}
\label{ref:model_arch}

For the learning architecture, I use a two-layer Bidirectional LSTM, inspired by prior works \cite{yi2021transpose, huang2018deep}. Although I did experiment with newer architectures such as transformers, I found these models did not perform well in practice. LSTMs produced smoother output predictions than the other models I tested. For each available IMU, our system uses orientation (represented as a 3$\times$3 rotation matrix) as well as acceleration as input, both in a global coordinate frame of reference. In contrast to prior work, I do not normalize these inputs to be relative to a root IMU sensor location, such as the pelvis, as our available devices vary. I flatten and concatenate these inputs to form an input vector of size 60: 5 possible IMU locations $\times$ (3 acceleration axes + 3$\times$3 orientation rotation matrix), which I input to the model. Of note, our model can ingest any subset of the available IMU data, with absent devices masked (i.e., values set to zero).

The input vector is first transformed into an embedding of dimension 512 using a ReLU \cite{nair2010rectified} activated linear layer. Next, these embeddings are fed sequentially into a Bidirectional LSTM of hidden dimension 512. A final linear layer outputs 144 SMPL \cite{Loper2015SMPL} pose parameters --- 24 joints represented as 6D rotations (which is a smoother representation space \cite{6drotHaoLi2018}). SMPL also provides a body mesh (6890 vertices), which can be seen in Figures \ref{fig:teaser}, \ref{fig:architecture} and \ref{fig:sample_mesh}. In total, our neural network model has $10.7$M trainable parameters. During training, a forward kinematics module calculates joint positions from these pose parameters and further minimizes it with respect to the ground truth.

\subsection{IMU Dataset Synthesis}\label{ref:dataSynthsec}

Training our pose model required a significant volume of data. For this, I can leverage existing motion capture databases to generate a large synthetic corpus. Specifically, I use AMASS~\cite{mahmood2019amass}, a compilation of 24 motion capture datasets (ACCAD, BioMotion, CMU MoCap, MIXAMO, Human Eva, Human 3.6M, etc.) totaling almost 63 hours of high-quality, high-resolution motion capture data in the SMPL~\cite{Loper2015SMPL} format. A wide variety of motions and activity contexts are included, such as locomotion, sports, dancing, exercising, cooking, and freestyle interactions. For additional details on the composition of the AMASS dataset, please refer to~\cite{mahmood2019amass}. 
I note that AMASS has been used in much prior work \cite{huang2018deep, yi2021transpose, yi2022physical} as the basis for deriving synthetic datasets.

The consumer devices I use for our study and real-time demo (described in Sections \ref{sec:data_collection_apparatus} and \ref{ref:realtimeimp}) run at a common framerate of 25 FPS. Thus I resample AMASS' 60$\sim$120 FPS data to 25 FPS. I then follow the synthetic data generation process used in TransPose \cite{yi2021transpose} and DIP \cite{huang2018deep}. In short, I ``attach" virtual IMUs to specific vertices in the SMPL mesh (the left and right wrists, the left and right front pant pockets, and the scalp) and compute synthetic acceleration data using adjacent frames in the global frame of reference. To generate synthetic orientation data, I calculate joint rotations relative to the global frame by compounding local rotations starting from the joint to the pelvis (root) following the SMPL kinematic chain. I scale acceleration data ($m/s^2$) by 30 to be suitable for neural networks~\cite{yi2021transpose}. Finally, rather than adding synthetic high-frequency noise to our dataset, I instead smooth both synthetic and real-world data using an averaging window of length $5$ frames ($200$ ms), similar to \cite{transformerIntertialPoser2022}.

I use this pipeline to create 24 sets of data, one for each of our 24 device-location combinations (Figure \ref{fig:possible_combinations}), which I combine into a single dataset. I simulate missing devices by masking-out (i.e., zeroing-out) IMU data for those locations. For example, even in our best-case scenario of three devices present, this means that 2/5$^{ths}$ of the input vector is null. 63 hours of AMASS data $\times$ 25 FPS $\times$ 24 device-location combinations yields 134.8M synthetic IMU instances  with paired ground truth SMPL poses for training. 

\subsection{Training}

The model is trained end-to-end using PyTorch and PyTorch Lightning deep learning frameworks. I use a batch size of $256$ and update the weights using the Adam optimizer with a learning rate of $3e^{-4}$. While training, I use non-overlapping windows of paired IMU and pose data in 5-second (125 samples) chunks. As mentioned earlier, I train our model to regress to full-body pose and full-body joint positions using mean squared error (MSE) loss. Our total loss is the sum of these two individual losses. I train our model for 80 epochs ($22$ hours) on an NVIDIA Titan X GPU.

\subsection{Joint Rotation Refinement}

As the last step of our inference pipeline, I adopt the Inverse Kinematic refinement method presented in \cite{jiang2022avatarposer} to perform a final refinement of our output pose. Although our model predicts the rotation of legs, hands and head, it does not necessarily fully honor the absolute orientation offered by the IMUs, even when weighted heavily in our loss term. However, it is logical to take advantage of IMU orientation for limbs with devices, as it is both an absolute value and considerably less noisy than accelerometer data. More specifically, as I have absolute orientation from the IMUs, I optimize certain bone orientations for each instrumented joint. In particular, for the wrist joint (smartwatch/phone), I optimize the elbow and the shoulder orientations, and similarly for the head (earbuds/phone) and hip (smartphone/earbuds case) joints. I implement this using the PyTorch framework and optimize this error using the MSE loss and the Adam optimizer. I allow this optimization to run for 10 iterations on each frame, which I found to not impede real-time performance.

\begin{figure}
    \centering
    \includegraphics[width=0.5\linewidth]{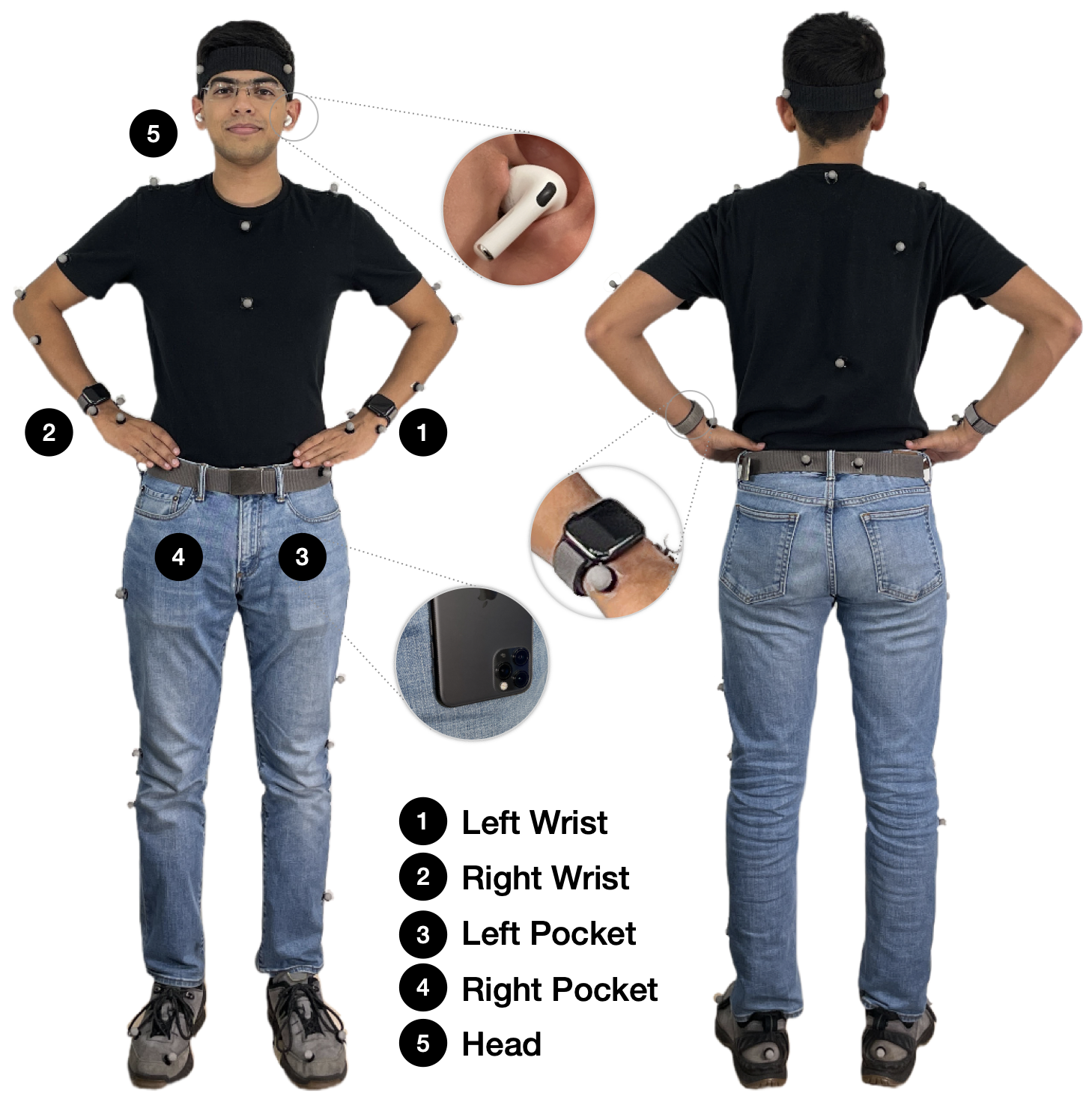}
    \caption{IMUPoser data collection setup. Participants wore 2 smartwatches, kept 2 smartphones in their front pockets, and wore wireless earbuds. 41 retroreflective motion capture markers were also placed around the body to track ground truth body pose.}
    \label{fig:ipose_data_collection_setup}
\end{figure}

\section{Evaluation}
I systematically isolate and analyze the efficacy of IMUPoser across different datasets and conditions. 

\subsection{DIP-IMU Dataset}
To test the performance of our model on real (and not synthetic) IMU data, I use DIP-IMU \cite{huang2018deep}, an IMU-based MoCap dataset. While smaller than the AMASS dataset I used for training, it offers a good variety of poses and activities across five classes: upper-body (arm raises, stretches, swings, etc.), lower-body (leg raises, squats, lunges, etc.), interaction (gestures to interact with everyday objects), freestyle (jumping jacks, punching, kicking, etc.) and locomotion (walking, side steps, etc.). A secondary benefit of using DIP-IMU is that it has been used for evaluation in other similar works \cite{yi2021transpose,huang2018deep,yi2022physical,transformerIntertialPoser2022}, permitting direct comparison. DIP-IMU used the commercially-available Xsens \cite{xsens} IMU-based system to capture data from 10 participants. The data is sampled at 60~Hz, leading to a total dataset size of approximately 90 mins.

\subsection{IMUPoser Dataset}
\begin{figure*}
    \centering
    \includegraphics[width=\linewidth]{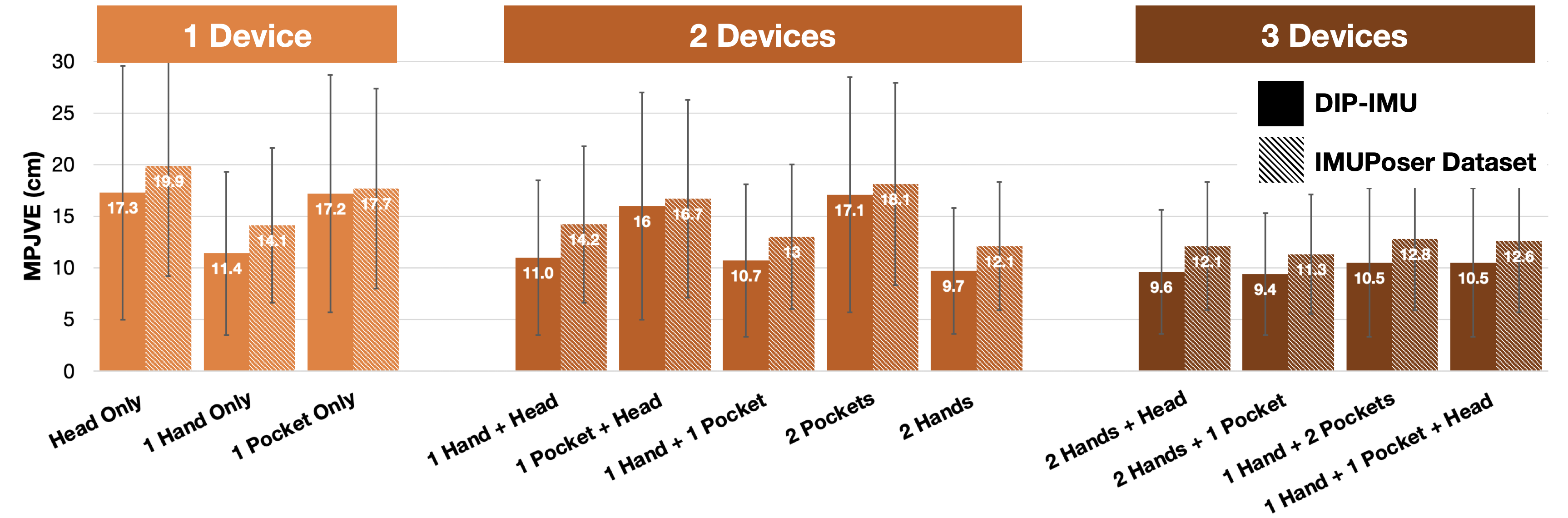}
    \caption{Accuracy across different device combinations. Error is Mean Per Joint Vertex Error (MPJVE) in cm. Note how error decreases as the number of devices increases.}
    \label{fig:sample_device_acc}
\end{figure*}

As noted above, DIP-IMU used the professional-grade XSens system for data collection, which costs approximately \$4000 USD. All of the IMUs are matched, offering similar noise and tracking performance. To complement this dataset with a \textit{consumer} device equivalent, I collected our own dataset. 

\subsubsection{Data Collection Apparatus}
\label{sec:data_collection_apparatus}
Our data collection apparatus consisted of two smartphones (Apple iPhone 11 Pro) placed in the left and right front pockets, two smartwatches (Apple Watch Series 6) placed on the left and right wrists, and one pair of Apple AirPods Pro worn in the ears (Figure \ref{fig:ipose_data_collection_setup}). The sampling rate of our system was configured to 25~Hz, the maximum sampling rate of the AirPods. The Apple Watch and AirPods communicated over Bluetooth to the iPhones, and the two iPhones relayed all IMU data to a laptop for data processing and recording. Although users had all five devices on them during data capture, I only use a subset of these devices for pose estimation, as described in Section \ref{Possible device combos}. 

For ground truth pose, I use a Vicon Motion Capture System system~\cite{Vicon} with twelve MX40 cameras and four T160 cameras capturing at 120FPS. I used Vicon Blade 3.2 for capture and data export and Vicon IQ 2.5 for data cleaning. I downsample the Vicon data and synchronize it with our collected IMU data streams. For analysis, I fit an SMPL mesh to the Vicon data using Mosh++~\cite{mahmood2019amass}.

\subsubsection{Device Calibration} \label{Calibration}
In contrast to commercial IMU-based motion capture systems like XSens, smartphones, smartwatches, and earbuds fail to provide IMU orientations in a common (global) frame of reference. If a device contains a magnetometer, the manufacturer usually provides a way to access the orientation of the device in a global frame of reference oriented with Earth's gravitation and magnetic fields. While the iPhones and Apple Watches that I used for this study contained magnetometers, I found their global orientation data to be fairly noisy. Moreover, the Apple AirPods do not contain magnetometers and hence only provide orientation relative to the initial frame of reference of the head. As a result, I opted to use the XArbirtraryCorrectedZVertical frame of reference provided by the Swift CoreMotion API \cite{CMAttitudeRef}. 

Before the study began, I aligned all the devices to a common frame of reference and recorded their orientation values over a window of three seconds. This acted as calibration data, bringing all the devices into the same global frame of reference. In practice, since the AirPods only sampled IMU data when they were in a participant's ears, the common frame of reference was set to that. In line with prior works \cite{yi2021transpose, huang2018deep}, I asked participants to make a T-pose for three seconds to calculate the orientation offsets between the device and the bone joint that it was attached to. The T-pose acts as a template pose wherein rotations are identity and thus known for each joint. This helps calibrate for users wearing the devices in different orientations, for example, a phone held in the hand vs. a watch worn on the wrist.

\subsubsection{Data Collection Procedure}

\begin{figure*}
    \centering
    \includegraphics[width=\linewidth]{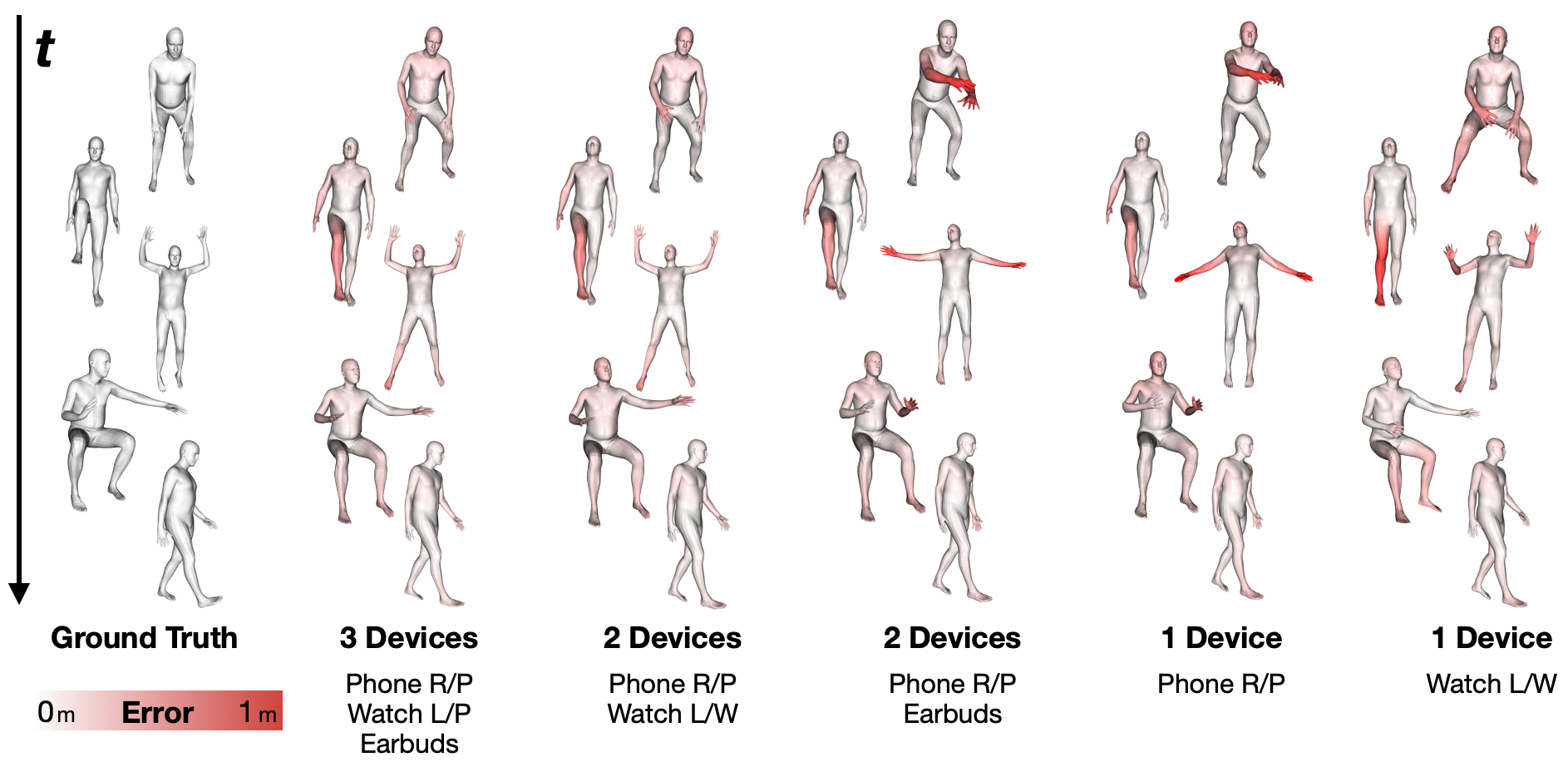}
    \caption{Sample SMPL mesh predictions for different device placements and combinations. The red color indicates the per vertex error in meters (ranging from 0 to 1 m). }
    \label{fig:sample_mesh}
\end{figure*}

For our data collection, I recruited 10 participants (5 identified as female, 5 identified as male) with a mean age of 22. The study lasted roughly 45 minutes and paid \$20 in compensation. I asked participants to wear and store the five devices in the way that felt most natural to them. Other than requesting the participants to wear pants with pockets, I did not control for differences in clothing, pocket styles, or smartwatch placement preference on the wrist, so as to get realistic real-world variation. For our Vicon-derived ground truth, I placed 41 optical markers on participants. In order to keep the markers secure, I asked participants to tuck in their shirts and provided velcro straps where needed.

Inspired by prior works \cite{huang2018deep,ahuja2021pose}, I collected our data using an ``obstacle course"-style procedure. I extended the classes in the DIP-IMU dataset and included the following motions:

\begin{itemize}[leftmargin=*]
    \item \textit{Upper Body}: Right arm raises, left arm raises, both arm raises, right arm swings, left arm swings, both arms swinging, arms crossing across the torso, and arms crossing behind the head. 
    \item \textit{Lower Body}: Right leg raises, left leg raises, squats, lunges with left leg, and lunges with right leg.
    \item \textit{Locomotion}: Walking in a straight line, walking in a figure 8, walking in a circle, sidesteps with legs crossed, and sidesteps with feet touching.
    \item \textit{Freestyle}: Jumping jacks, tennis swings, boxing with alternate arms, kicking with the dominant leg, push-ups, and dribbling a basketball.
    \item \textit{Head Motions}: Moving head up-and-down, moving head left-to-right, leaning head from shoulder-to-shoulder, and moving head in circles.
    \item \textit{Interaction}: Scrolling on a smartphone while seated in a chair. 
    \item \textit{Miscellaneous}: Waving with right arm, waving with left arm, clapping, hopping on right leg, hopping on left leg, jogging in a straight line, and jogging in a circle.
\end{itemize}

The upper body, lower body, locomotion, freestyle, head motions, interaction and miscellaneous scenarios lasted for 69.7, 43.4, 95.3, 76.2, 36.8, 19.2, and 74.3 seconds on average, respectively, resulting in roughly 7 minutes of data per participant. All the motions were continuous and data was also captured while participants were transitioning from one category to another. 

\subsection{Evaluation Protocol}
In order to compare with prior works, I follow the exact method detailed in  \cite{huang2018deep,yi2021transpose,transformerIntertialPoser2022,yi2022physical}. Specifically, I use data from the first eight participants of DIP-IMU as training data, with the last two participants used for testing. I fine-tune our AMASS-trained model using this training data downsampled to 25~Hz to match both our AMASS training data and our real-time system's capabilities. I further test this model on our IMUPoser Dataset, helping to assess real-world accuracy and performance. Our model is evaluated in an online fashion. In particular, I feed a rolling window of 125 samples (5-second history) with a 1-sample overlap, emulating real-world use. This data is smoothed using an averaging filter, as described in Section \ref{ref:dataSynthsec}. I analyze these results using different evaluation metrics across various device-location combinations. Also following prior work \cite{yi2021transpose, yi2022physical, huang2018deep}, I make use of the following evaluation metrics to quantify the performance of our full-body pose estimation pipeline:

\begin{enumerate}[leftmargin=*]

  \item \textit{Mean Per Joint Rotation Error:} MPJRE measures the mean global angular error across all joints in degrees (\textdegree). 
  
  \item \textit{Mean Per Joint Position Error:} MPJPE measures the mean Euclidean distance error of all estimated joints in centimeters (cm) with the root joint (pelvis) aligned. 
  
  \item \textit{Mean Per Joint Vertex Error:}  MPJVE measures the mean Euclidean distance error across all vertices of the estimated SMPL mesh in centimeters (cm) with the root joint (pelvis) aligned. 
 
  \item \textit{Mean Per Joint Jitter (Jitter):}  Jitter measures the average jerk of the predicted motion \cite{yi2021transpose}. A lower jerk value signifies a smoother and more natural motion.
  
\end{enumerate}

I use mesh error (MPJVE) as our primary evaluation metric for most tasks, due to its ease of understanding and its utility as a benchmark for comparison with prior work.

\section{Results}
\begin{figure*}
    \centering
    \includegraphics[width=\linewidth]{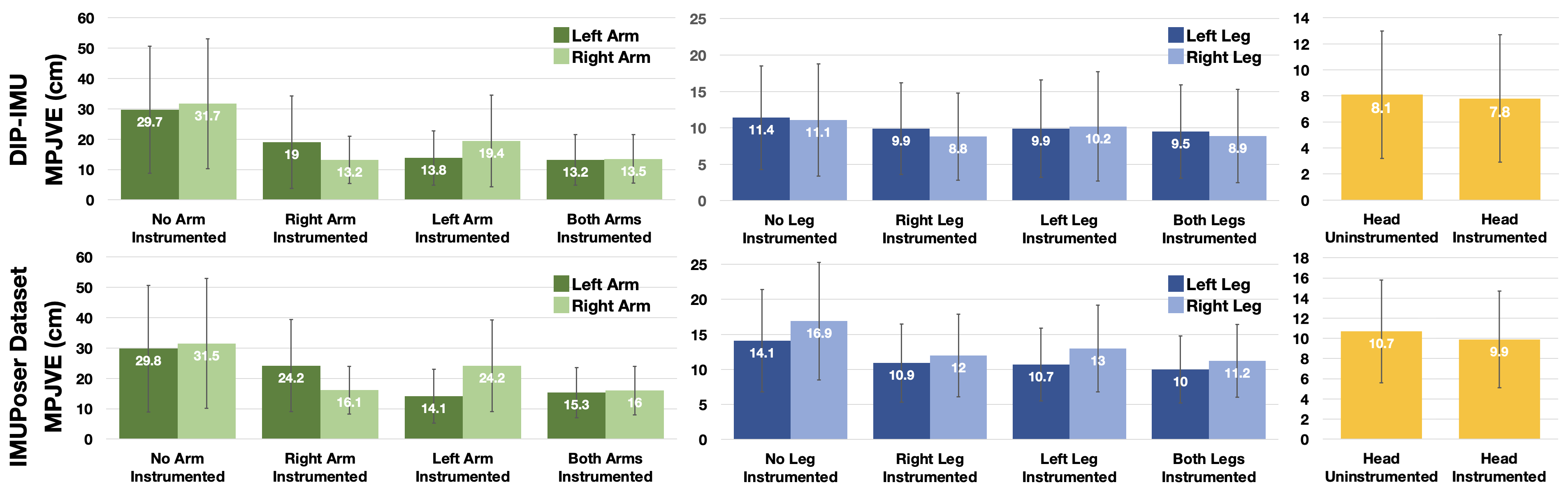}
    \caption{Summarized accuracy results across different body regions evaluated on the DIP-IMU and IMUPoser datasets.}
    \label{fig:body_region}
\end{figure*}

I first describe IMUPoser's accuracy across device-location combinations, before changing our focus to look at results by body region. I conclude this section with a comparison to other related systems. 

\subsection{Accuracy Across Device-Location Combinations}

To simplify presentation of results, I group the 24 possible device-location combinations (Figure~\ref{fig:possible_combinations}) into 12 supersets based on the number of devices present and their body locations (ignoring left/right placements). Figure~\ref{fig:sample_device_acc} presents the results for the IMUPoser and DIP-IMU datasets. Note, that our model has not been fine-tuned on the IMUPoser Dataset. Across all device combinations, I find a MPJVE of 14.1~cm on the IMUPoser Dataset and a MPJVE of 12.1~cm with DIP-IMU. When averaging the results across both datasets, having one device on the user results in a MPJVE of 16.27~cm (SD=9.93~cm), which decreases to 13.9~cm (SD=8.36~cm) when a second device is present. The lowest error, unsurprisingly, is when three devices are present -- a MPJVE of 11.1~cm (SD=6.51~cm) across all possible three-device combinations.


Figure~\ref{fig:sample_mesh} offers example mesh predictions across different device-location combinations. As expected, accurate head orientation estimation is only plausible when earbuds are present. Other times the head regresses to the most natural orientation given where the body is facing. Global body orientation works best when at least two devices are present. Lastly, motions that have a characteristic cadence, such as walking, work well across all combinations. Similarly, activities with symmetric limb motions, such as jumping jacks, work fairly well even with no sensor data from important limbs. On the other hand, activities with uncorrelated limb motions fail unless limbs are instrumented. 

\subsection{Accuracy Across Body Regions}

Figure~\ref{fig:body_region} provides a breakdown of system accuracy across different body regions for the IMUPoser and DIP-IMU datasets. I note that the accuracy for a limb with an instrumented point is always greater than that of an uninstrumented one. For example, averaging across both datasets and with an IMU present on the right hand, the MPJVE is 14.65~cm for the right arm (right hand = 17.2~cm)  vs. 21.6~cm for the left arm (left hand = 26.9~cm). Unsurprisingly, the highest error is when none of the limbs in a particular body region have IMU data.

Also unsurprising is that the lowest error is achieved when both left and right limbs have IMUs present. For example, with only one IMU on the arms, the MPJVE for both arms is 18~cm (right hand = 22.17~cm; left hand = 20.4~cm). Whereas with both arms having IMUs, the MPJVE is 14.5~cm (right hand = 17.35~cm; left hand = 16.9~cm). A partial exception to this trend is the legs. Unlike the arms, which can move independently, legs tend to move in tandem (out of phase when walking, or in phase for activities such as jumping). This means that even one IMU on the legs is still highly effective at predicting both legs, and two IMUs offer just a modest gain. Looking at our results, the MPJVE for the left leg is 10.3~cm (left foot = 15.65~cm) when the IMU is in the left pocket, and the error for the right leg is 10.4~cm (right foot = 16~cm) when the IMU is in the right pocket. When both IMUs are present (i.e., left and right pockets), the error of the left and right legs drop very modestly to 10.05~cm and 9.75~cm, respectively. 

I note that error accumulates along the kinematic chain (see Figure \ref{fig:sample_mesh}). Across all conditions, the average error of the end-effectors (left hand, right hand, left foot, right foot, head) is 20.28~cm and 17.29~cm on the IMUPoser Dataset and DIP-IMU Dataset, respectively (vs. 12.92~cm and 11.23~cm for joints that are not end-effectors). 

\subsection{Comparison to Prior Work}

\begin{table}
\small
\centering
\begin{tabular}{l c c c c c}
\toprule
\textbf{System} & \textbf{\# Inst. Joints}    & \textbf{MPJRE (\textdegree)}  & \textbf{MPJPE (cm)} & \textbf{MPJVE (cm)} & \textbf{Jitter ($10^2m/s^3$)}\\
\midrule

SIP (offline)   & 6 & 8.7  & 6.7 & 7.7 & 3.8 \\
DIP (online)   & 6 & 15.1  & 7.3 & 8.9 & 30.13 \\
TransPose (online)  & 6 &  8.8 & 5.9 & 7.1 & 1.4 \\
PIP (online)  & 6   & -  & - & 5.9 & 0.24 \\

\rowcolor{Gray}
IMUPoser (online)  & 1--3  & 23.9  & 9.7 & 12.1 & 1.9 \\

\bottomrule
\end{tabular}
\vspace*{2mm}\caption{Comparison of IMUPoser to key prior work, all evaluated on the DIP-IMU Dataset \cite{huang2018deep}.}
\label{tab:comparison}
\end{table}

To the best of our knowledge, no prior research has investigated deriving full-body pose from such a sparse set of consumer-grade devices equipped with IMUs. Table~\ref{tab:comparison} offers a quantitative comparison against key prior work, all evaluated on the same DIP-IMU Dataset \cite{huang2018deep}. 

Unsurprisingly, for a system that uses between 1 and 3 IMUs, our model is less accurate than those utilizing 6 sensors (i.e., IMUs placed on each limb). However, compared to DIP \cite{huang2018deep} and Transpose \cite{yi2021transpose}, our MPJVE is only worse by 3.2~cm and 5.0~cm, respectively. It is interesting to note that the Jitter of our system is in line with prior work (1.9 vs. 1.4 of TransPose). At a high level, even with an impoverished sensing configuration, I are able to produce natural, realistic and smooth pose estimation sequences. In the future, I hope to combine physics-backed models (as in PIP) to further improve the pose estimation of our system. 

\section{Active Device Tracking}
A crucial piece of information our pose model needs before it can run is: 1) What devices are present on the user? And 2) where these devices are located on the body? For this, I created a separate piece of software, which runs in parallel with our pose model. 


\subsection{Implementation}
To determine where devices are located on the body, I require three pieces of information from the user, which I envision being collected when a user first purchases a device. 1) In which pocket do they typically store their phone? 2) In which hand do they typically hold their phone? And 3) On which arm would they wear a smartwatch? After this basic initialization, I use a series of automated heuristics.


I make the assumption a smartphone is held in the hand if the screen is on and the IMU is reporting even slight motion. If the user is wearing a smartwatch, I can use the distance between the watch and phone (provided by Apple's NINearbyObject API \cite{NINearbyObject}, which uses UWB) to guess the holding hand automatically (see Figure~\ref{fig:active_device}). If no smartwatch is worn, our system falls back on the hand specified by the user during setup. If the smartphone screen is off and the IR proximity sensor is triggered, I assume the phone is in a pocket. If the user has a smartwatch, I can similarly use UWB-derived distance to guess the pocket. If no smartwatch is worn, I default to the user-specified pocket.

As most users wear watches in a consistent location, the logic for smartwatches is simpler. If it is connected to the iPhone and moving, I assume it is worn on the user-specified hand. Similarly, for Airpods, if they are connected to the iPhone, I know they are in the ear. 
When in their charging case, Airpods go to sleep and stop transmitting IMU data. However, I believe that Apple could modify the Airpods firmware such that in the future they could continue to transmit IMU data even when stored in a pocket.
\subsection{Evaluation}

\begin{figure}
    \centering
    \includegraphics[width=0.8\linewidth]{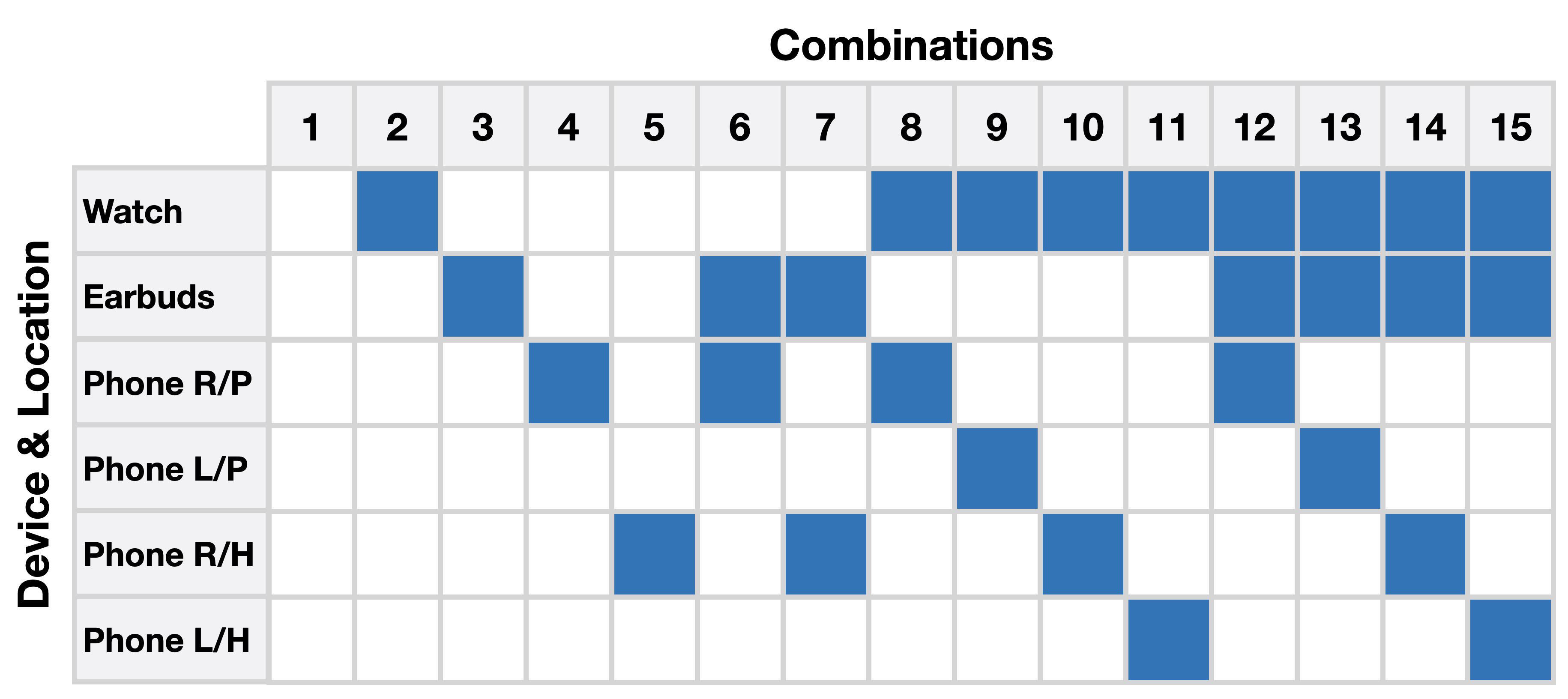}
    \caption{Device combinations tested as part of our active device tracking study. Blue denotes presence in the set. }
    \label{fig:active_device_combos}
\end{figure}

\begin{figure}
    \centering
    \includegraphics[width=\linewidth]{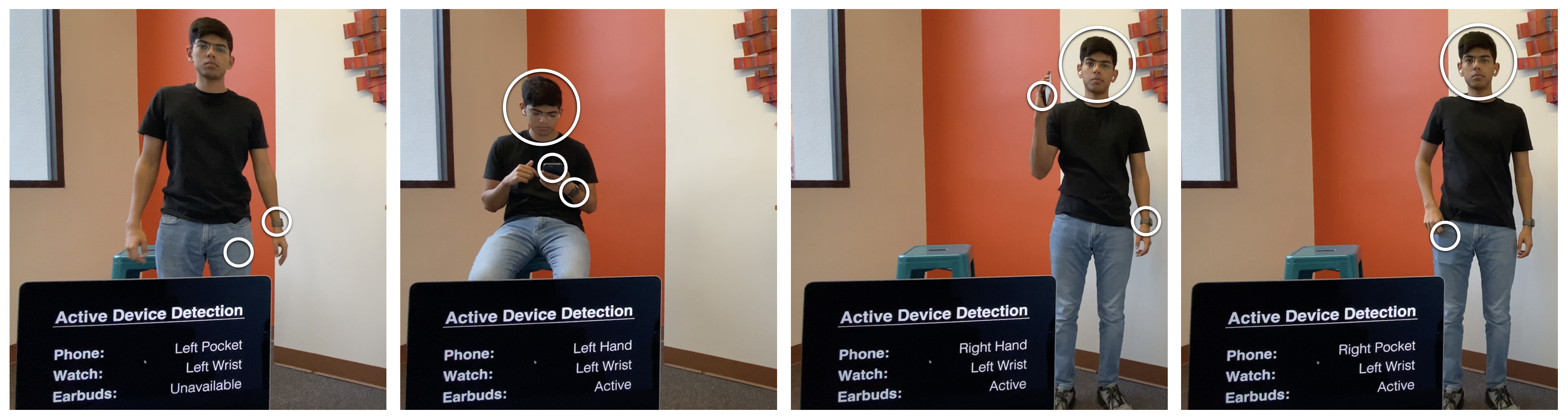}
    \caption{Active device tracking across different device combinations. Active devices are highlighted with a white circle for illustration and the foreground laptop shows the live tracking result.}
    \label{fig:active_device}
\end{figure}

As a preliminary evaluation of our active device tracking prediction, I ran a user study with 7 participants (5 identified as male, 2 identified as female; mean age 27.8; all right-handed with a preference for wearing watches on the left wrist). The study lasted approximately 15 minutes and paid \$5. To initialize our system, I recorded participants' answers to the three preference questions listed in the previous section. I then asked users to transition between 15 device combinations, in a random order, documented in Figure \ref{fig:active_device_combos}. When a device was not in a requested set, it was set aside on a nearby table. For each requested device-location combination, I asked participants to walk around for about 10 seconds, then sit down briefly, rise to stand again, and lastly return to the starting position. Before the next trial began, the necessary devices were given or taken from the participants. Throughout the study, our active device tracking process ran, making live predictions about what devices were active and where they were located. A trained experimenter conducted the study, marking the start and stop of each device combination trial, alongside ground truth labels. 

Across all participants and all data instances, the accuracy of earbuds and smartwatch tracking was 100\%, owing to their known locations and very reliable detection of worn vs. not worn. Smartphone tracking is the most challenging, with five possible states (not present, left pocket, right pocket, left hand, right hand). I found the instance-wise accuracy for smartphone tracking was 90.8\%. 


\section{Real-Time Implementation}
\label{ref:realtimeimp}

To help demonstrate the imminent feasibility of our approach, I created a real-time implementation of IMUPoser, which can be seen in our Video Figure. It is comprised of two main processes working together. First is active device tracking, which monitors what devices are available to provide IMU data and predicts where they are located on the body. Second is our pose model, which is passed the location inferences and IMU data. 

\subsection{Proof-of-Concept Device Ecosystem}
As a proof-of-concept implementation, I use an Apple iPhone 11 Pro, Apple Watch Series 6, and AirPods Pro. Apple offers a mature inter-device API that allows these devices to exchange data. Each device reports 6DOF IMU data at different rates, with the slowest being AirPods at roughly 25 FPS. I also note that although AirPods come as a pair, they fuse their individual IMUs into a single 6DOF head estimate.

\subsection{Output}\label{real time pose prediction}
As a proof of concept, I use an iPhone optionally connected to an Apple Watch and AirPods. The iPhone streams all available IMU data back to a MacBook Air (2021), which runs our active device tracking and pose estimation processes, with a mean inference time of 26.8~ms. I believe our model could be run on a mobile phone with additional engineering effort. Regardless of where the model runs, it is capped at 25 FPS, the reporting frequency of our slowest IMU (Airpods). Before running our system, I must perform the same calibration as in data collection (see Section \ref{Calibration}). For real-time output, I visualize the SMPL mesh.

\section{Open Source}

To enable other researchers and practitioners to build upon our system, I have made our dataset, architecture, trained models, and visualization tools freely available at \url{https://github.com/FIGLAB/IMUPoser} with the gracious permission of our participants.

\section{Limitations}

While IMUPoser enables pathways to full-body pose estimation with minimal user instrumentation, it has pros and cons like any other technical approach. While IMUPoser can glean insights about the pose of limbs for which it has no direct sensor data, it is important to note that such a pose is only an approximate result. For cases where the motion of the instrumented joint is completely independent of that of the uninstrumented one, IMUPoser tends to regress to the mean pose. IMUPoser can support the incorporation of new joint locations by using the corresponding SMPL mesh vertices for training. Thus, in the future, IMUPoser can potentially support and track new device placements, such as a phone in a back pants pocket, coat pocket, armband, etc. The fidelity of the system could also be improved by integrating additional consumer devices (e.g., smart shoes, eye-wear, rings) into the ecosystem. This would help expand the range of poses supported by IMUPoser, allowing it to track dynamic activities such as cycling, kayaking, skiing, etc. 

Unlike Transpose~\cite{yi2021transpose} and PIP~\cite{yi2022physical}, the current implementation of IMUPoser does not predict global root translation. In the future, using better learning methods and multimodal cues when available (e.g., visual odometry from the smartphone \cite{ahuja2021pose}) could help predict translation. Another limitation of our system is active device tracking. Currently, this is a basic, proof-of-concept implementation that needs further refinement before it can be deployed for consumer use. Furthermore, all of the devices need to be in a homogeneous ecosystem (e.g., Apple) to work effectively. In the future, the use of a common industry-wide standard to connect and network between different consumer devices can help mitigate this issue.

\section{Conclusion}

I have presented a user digitization system that is both practical and rich. It enables real-time, full-body pose estimation using IMUs present in consumer devices such as phones, smartwatches and earbuds. The approach automatically tracks devices that are available and where they are currently located on the body, and uses their IMU data to estimate pose. It can contend with the noisy signals of consumer IMUs and produce natural and temporally-coherent pose estimates with as little as one device. This opens up new and interesting long-term whole-body digitization applications with no additional user instrumentation. 
\label{chapterIMUPoser}

\chapter{Summary}

\begin{figure}
\centering
  \includegraphics[width=0.5\columnwidth]{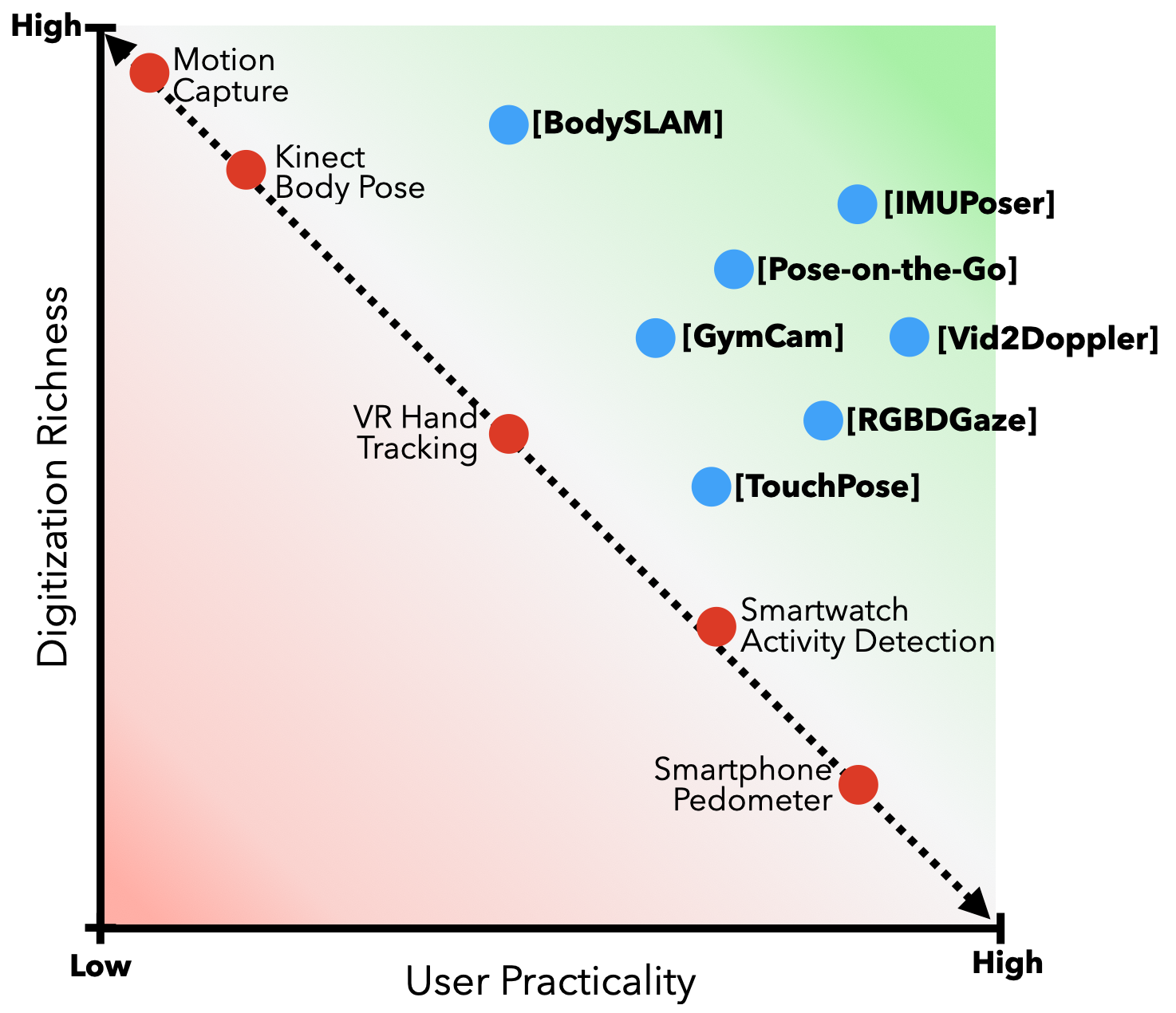}
  \caption[Design Space]{Design space plotting digitization richness vs. user practicality.}
  \label{fig:summary_design_space}
\end{figure}

Through my research efforts, I have focused on creating user digitization systems that are both practical and rich. With my research (Figure \ref{fig:summary_design_space}, blue dots), I have started to considerably pull away from the axis occupied by conventional approaches, into the area highlighted in green. I draw on three pillars of strength to build towards this vision. First, I use my knowledge in human factors to understand people; and more importantly, the problems they face. Second, I build new sensors and sensing approaches to capture how they behave, move within and interact with the world around them. Finally, I use my background in AI and ML to make sense of these captured signals, to model and derive utility from them for these human-centric tasks. This enables consumer devices to calculate a continuous comprehensive representation of the user. To achieve this research vision, I develop sensing systems that make lower fidelity systems higher fidelity, by adding layers of intelligence over devices, to create new and powerful computing experiences; while retaining or even improving consumer practicality and accessibility, for instance by making them less invasive or more privacy aware.

Such digital representations go beyond a simple step count and capture a user’s gait cycle, dexterity, balance, posture, and active range of motions of joints \ref{chaptergymcam}, providing contextualized behavior and pose insights in every day practical scenarios that even a once a day visit to a Motion Capture studio would fail to provide. Using this holistic continuous information, a user can get fine-grained wellness data and even enable doctors to remotely monitor their patients \ref{lemurdx}, for example, tracking a user’s recovery after surgery, and even chances of injury and relapse. This will also inform the design of future consumer devices that enable full-body avatars for telepresence \ref{chapterPOSEOTG}, interactions in extended reality \ref{chapterBODYSLAM} and to create more productive and collaborative work environments. To summarize, my research has led to the creation of powerful user digitization systems that are minimally invasive \ref{chaptervid2doppler}, mobile and deployable in-the-wild \ref{chapterIMUPoser}. 

\section{Key Takeaways}

Here are the key takeaways summarized. First off, framing user digitization as a richness vs. user practically continuum helps answer what the technology enables and where it can be used. My research has shown that higher-order full-body digitization can be enabled on mobile consumer devices. They are interpret able, easily visualizable, and transferable to other tasks. Further, user practicality should be a key consideration as it can lead to wide-spread adoption. Lastly, different applications demand different accuracies, and thus even approximate user digitization can be immensely useful.    

Furthermore, here are key takeaways that can not only be used for creating user digitization, but can also transfer to other domains which essentially try to do ``more from less". To combat less training data - synthetic data generation, data augmentation and contextualization can play a key role. To make most of the few points of instrumentation, we can combine multiple sensor modalities and locations. Lastly, to increase fidelity of systems, we can incorporate physics based modeling as we saw with human kinematics, and also employ data-driven machine learning as an effective learning tool.

\section{Vision for the Future}

\begin{figure}
\centering
  \includegraphics[width=\columnwidth]{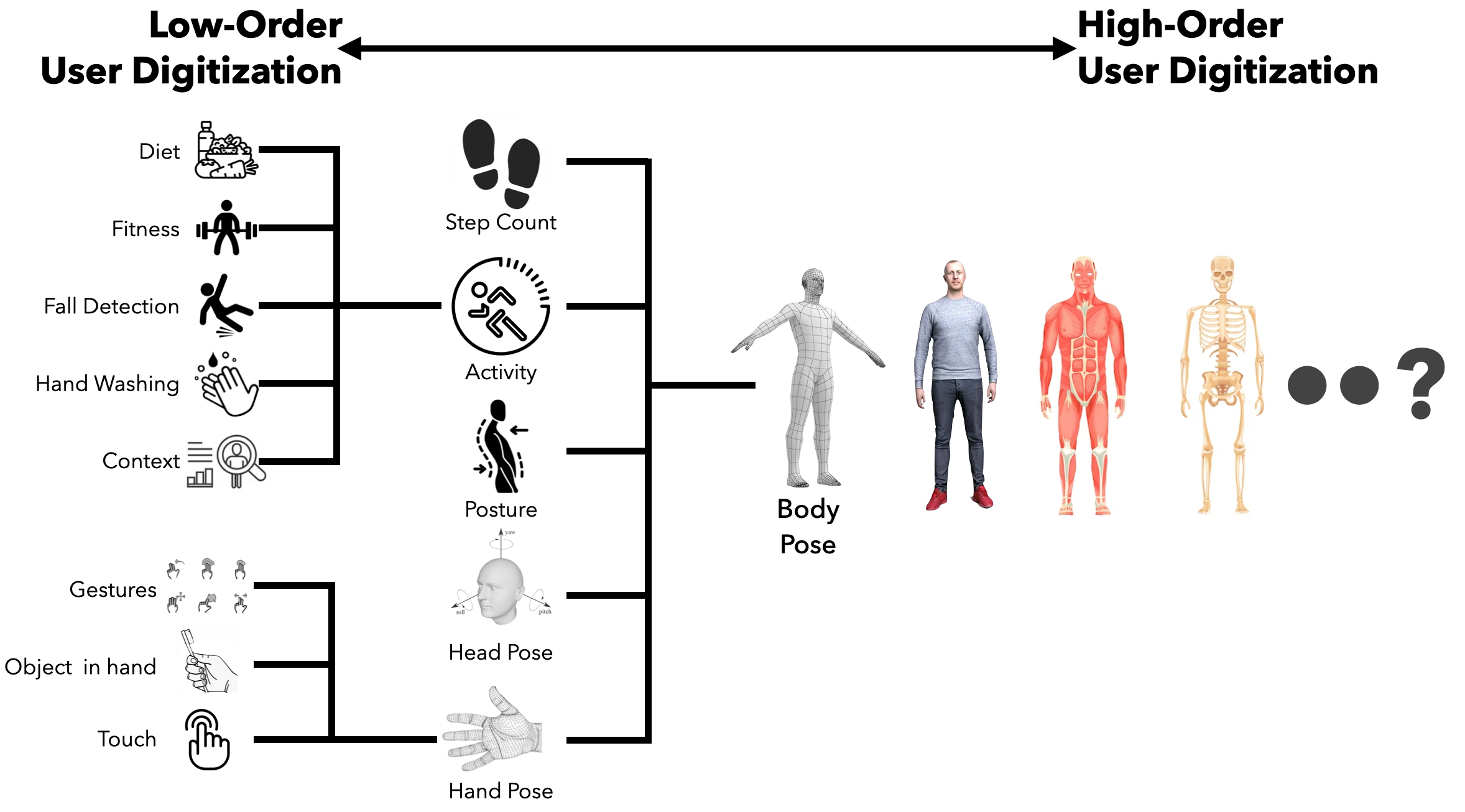}
  \caption[Future Vision]{Future of Higher-Order User Digitization.}
  \label{fig:future_vision}
\end{figure}

While I have made significant progress, there is still work to be done to enable “motion capture” quality resolution on consumer devices. We have seen that these higher order user digitization have immense utility for improving their users interactions, productivity and wellness. Thus, in the future, I plan to explore even higher order representations (Figure \ref{fig:future_vision}) that go beyond the users activity and pose and towards capturing their appearance, muscle structure, and even fine grained anatomy in a practical, minimally-invasive manner. These will have a similar trickle-down effect to unlock entirely new applications and augment existing ones in interesting ways. 

For example, in addition to using images or optical motion capture, muscle maps can directly capture fine-grained facial action units that can be used to create photo-realistic avatars for telepresence and character animation. Imagine context-aware fitness assistants that not only know what exercise you are doing, what your pose is, but also which muscles you are activating, which muscles you ideally should activate and how do change your posture do do that. Further, this will even reduce chances of injury, improve caloric estimates and help with better diet, nutrition and overall well-being. 

Finally, I believe the holy grail of practical and rich user digitization lies not in looking outwards but inwards. Let's take the example of detecting arthritis. The gold standard for diagnosing it is an X-ray. The problem is, when you see arthritis on X-rays, the damage has already been done. Imagine medical quality resolution of imaging our bones and muscles but with devices you carry with you everyday. This will create digitization that can go beyond monitoring illnesses and injuries to taking proactive steps to reduce their damage or in some cases prevent them. With novel sensing technologies, coupled with advances in AI, big data and medical imaging, I fully believe as a field we can work towards enabling this future of creating a holistic human digital twin.

\backmatter



\bibliographystyle{plainnat}
\bibliography{sample} 

\end{document}